\pdfoutput=1

\documentclass[11pt]{article}

\newif\ifNoAnonymous
\newif\ifNoComments
\newif\ifProduction
\newif\ifOnlyArxiv

\Productiontrue
\NoCommentstrue
\NoAnonymoustrue

\newcommand{\tablecustomsmall}{\fontsize{8.25pt}{9pt}\selectfont}

\newcommand{\ttt}{\texttt}

\newcommand{\clstok}{\ttt{[CLS]}}
\newcommand{\septok}{\ttt{[SEP]}}

\usepackage{color}

\usepackage[normalem]{ulem}

\ifProduction
    \NoCommentstrue
    \NoAnonymoustrue
\fi

\ifOnlyArxiv
    \newcommand{\onlyarxiv}[1]{#1}
\else
    \newcommand{\onlyarxiv}[1]{}
\fi

\ifNoAnonymous

    \newcommand{\nmslib}{NMSLIB \cite{boytsov2013engineering}}
    \newcommand{\anonurl}[1]{\url{#1}}
    
    \newcommand{\revision}[1]{#1}
\else

    \newcommand{\nmslib}{an \textbf{anonymous} toolkit for k-NN search}
    \newcommand{\anonurl}[1]{\textbf{anonymous} URL.}
    
    \newcommand{\revision}[1]{{\color{red} #1}}
\fi

\ifNoComments
\newcommand{\mynote}[1]{}
\newcommand{\davidnote}[1]{}
\newcommand{\strikethrough}[1]{}

\newcommand{\lb}[1]{}
 
\else
\newcommand{\mynote}[1]{\textbf{\color{red}\small #1}}
\newcommand{\davidnote}[1]{\textbf{\color{green}\small #1}}
\newcommand{\strikethrough}[1]{\sout{#1}}

\newcommand{\lb}[1]{\textcolor{orange}{{\textbf{LB:} #1}}}

\fi

\ifNoAnonymous
\usepackage[final]{acl2023}
\else
\usepackage[review]{acl2023}
\fi

\usepackage{times}
\usepackage{latexsym}

\usepackage[T1]{fontenc}

\usepackage[utf8]{inputenc}

\usepackage{microtype}

\usepackage{inconsolata}

\usepackage{graphicx}

\usepackage{multirow}
\usepackage{balance}
\usepackage{caption}
\usepackage{placeins}
\usepackage{float}
\usepackage{graphicx}
\usepackage{subcaption}
\usepackage{wrapfig}
\usepackage{hyperref}
\usepackage{rotating}
\usepackage{lscape}
\usepackage{amsfonts}       
\usepackage{amsmath}
\usepackage{amssymb}
\usepackage{nicefrac}       
\usepackage{microtype}      
\usepackage{booktabs}
\usepackage{mdframed}
\usepackage[para,online,flushleft]{threeparttable}
\usepackage{algorithm}

\usepackage{algpseudocode}

\ifNoAnonymous
\fi
\newmdenv[
  backgroundcolor=gray!10,
  linecolor=black,
  linewidth=0.5pt,
  innertopmargin=6pt,
  innerbottommargin=6pt,
  innerleftmargin=6pt,
  innerrightmargin=6pt,
  skipabove=10pt,
  skipbelow=10pt
]{textenv}

\title{Positional Bias in Long-Document Ranking: Impact, Assessment, and Mitigation}

\newcommand{\betterthanks}[1]{\thanks{\ \ #1\ \ }}

\author{Leonid Boytsov \betterthanks{Work done outside the scope of Amazon employment.}\hspace{0.3em}\betterthanks{Equal contribution.}\\
  Independent Researcher\\ Pittsburgh, USA\\
  \texttt{leo@boytsov.info} \\ \And
  David Akinpelu \footnotemark[2]\\ Louisiana State University \\ Baton Rouge, USA \\
  \texttt{akinpeluakorede01@gmail.com} \\ \AND
  Nipun Katyal \betterthanks{Work done while studying at Carnegie Mellon University.} \and
  Tianyi Lin \footnotemark[3] \and 
  Fangwei Gao \footnotemark[3] \and 
  Yutian Zhao \footnotemark[3] \and 
  Jeffrey Huang \footnotemark[3] 
   \AND Eric Nyberg \\
  Carnegie Mellon University \\ Pittsburg, PA, USA }

\newcommand{\ourcode}{\footnote{\url{https://github.com/searchivarius/long_doc_rank_model_analysis_v2/.}}}

\begin{document}
\maketitle
\begin{abstract}
We tested over 20 Transformer models for ranking long documents (including recent \emph{LongP} models  trained with FlashAttention and RankGPT models ``powered'' by OpenAI and Anthropic cloud APIs).
We compared them with the simple \emph{FirstP} baseline,  which applied the \emph{same} model to truncated input (up to 512 tokens).
On MS MARCO, TREC DL, and Robust04 no long-document model outperformed \emph{FirstP} by more than 5\% (on average). 
We hypothesized that this lack of improvement is not due to inherent model limitations, 
but due to benchmark positional bias (most relevant passages tend to occur early in documents), 
which is known to exist in MS MARCO.
To confirm this, we analyzed positional relevance distributions across four \emph{long-document} corpora (with six query sets) and observed the same early-position bias.
Surprisingly, we also found bias in six BEIR collections, which are typically categorized as
\emph{short-document} datasets.
We then introduced a new diagnostic dataset, MS MARCO FarRelevant, where relevant spans were deliberately placed beyond the first 512 tokens.
On this dataset, many long-context models---including RankGPT---performed at random-baseline level, suggesting overfitting to positional bias.
We also experimented with debiasing training data, but with limited success.
Our findings (1) highlight the need for careful benchmark design in evaluating long-context models for document ranking, (2) identify model types that are more robust to positional bias, and (3) motivate further work on approaches to debias training data.
We release our code and data to support further research.
\end{abstract}

\begin{figure*}[tb]
    \centering  
    \begin{subfigure}{0.3\textwidth}
        \centering
        \includegraphics[width=\linewidth]{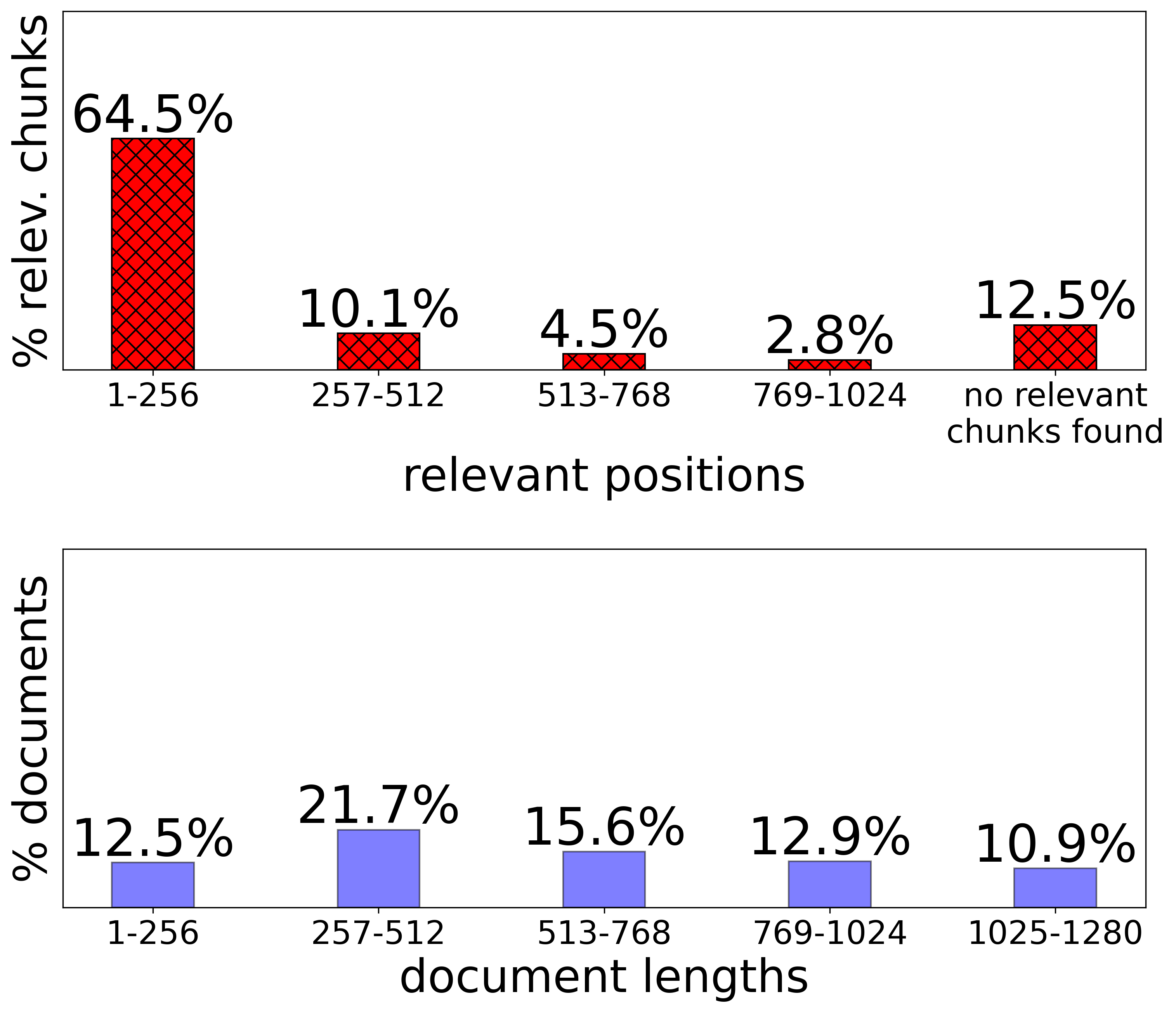}
        \caption{Robust04}
    \end{subfigure}    
    \begin{subfigure}{0.3\textwidth}
        \centering
        \includegraphics[width=\linewidth]{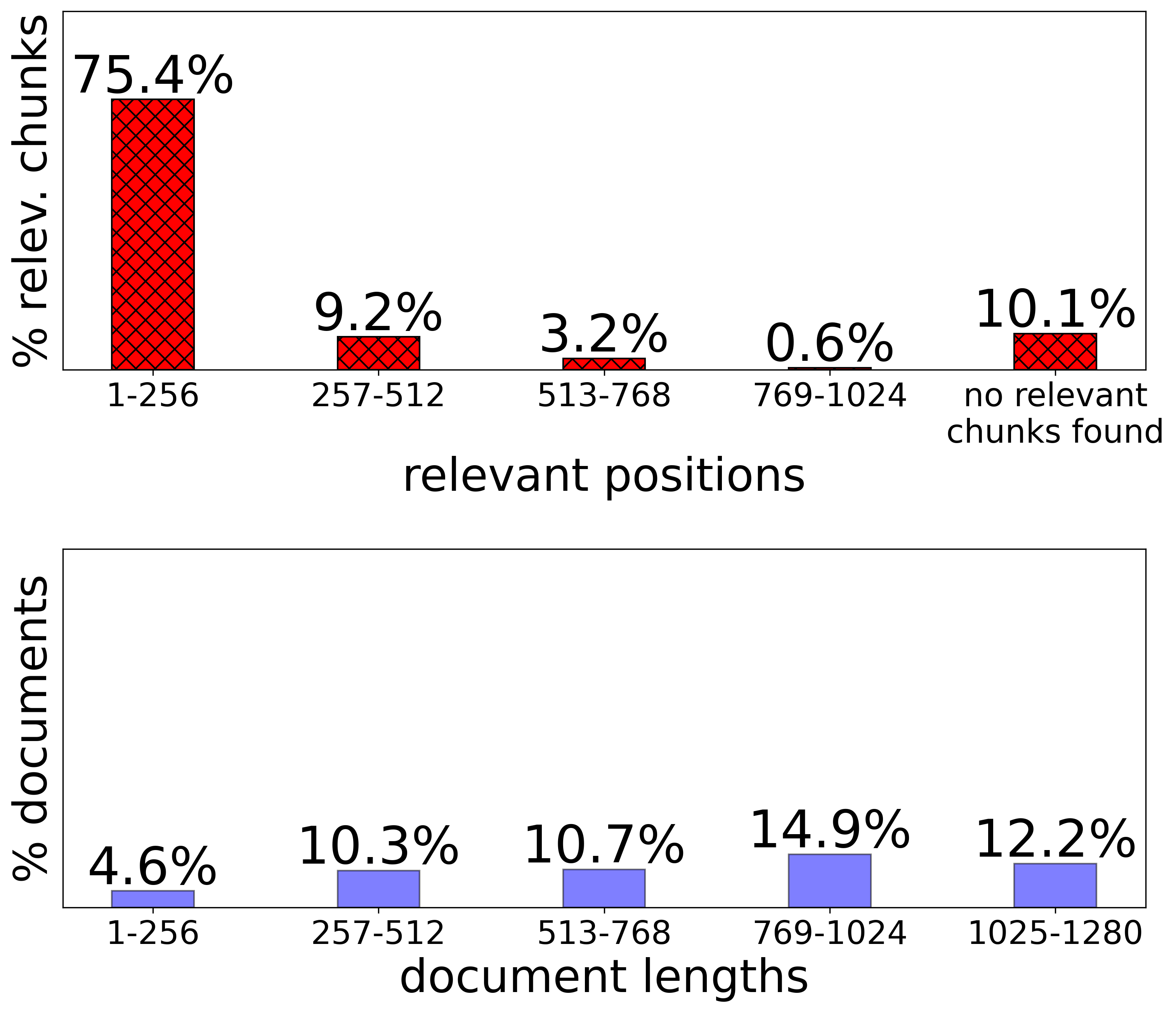}
        \caption{ClueWeb12 (WebTrack)}
    \end{subfigure}
    \begin{subfigure}{0.3\textwidth}
        \centering
        \includegraphics[width=\linewidth]{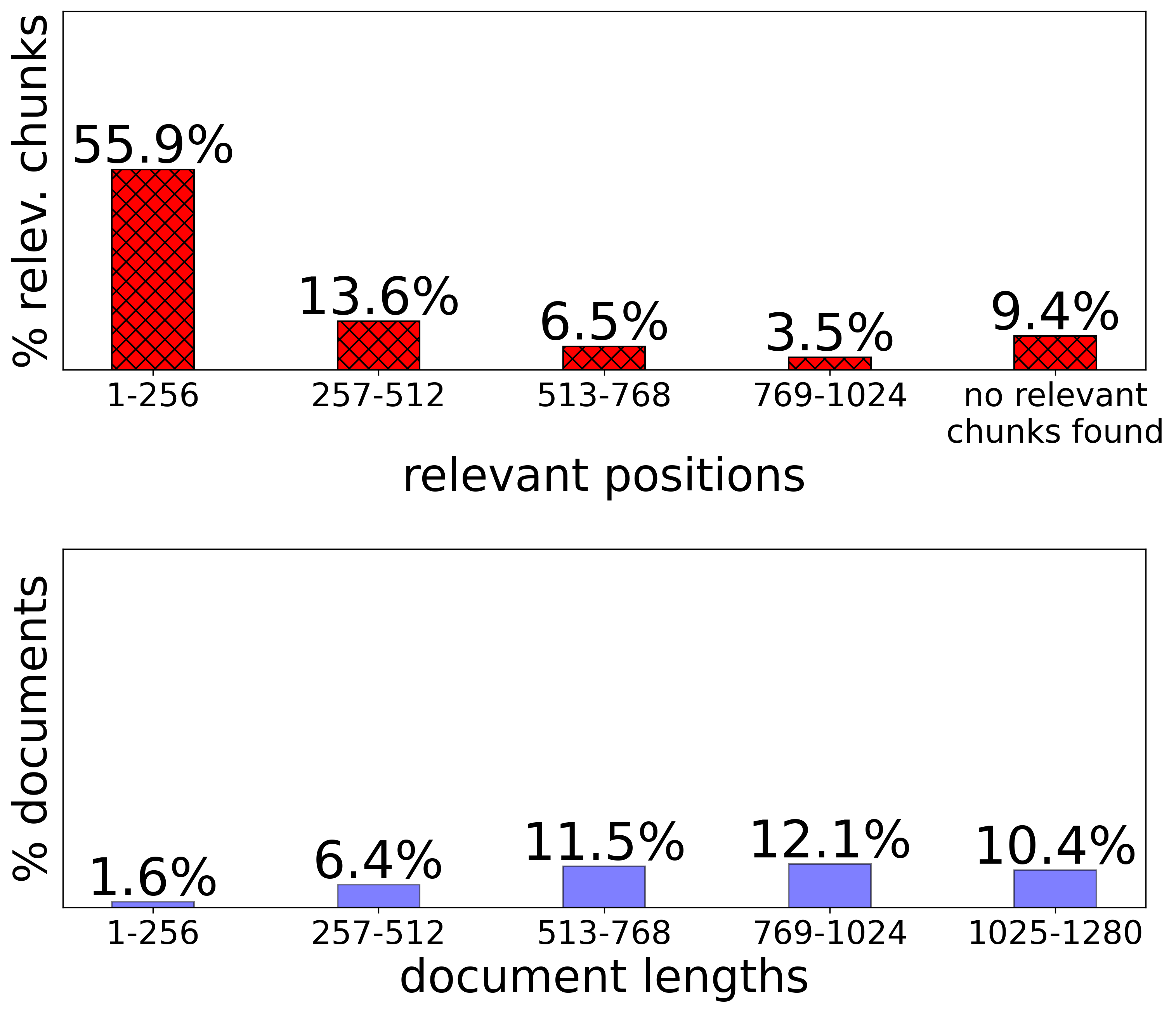}
        \caption{TREC DL 2019-2021 (combined)}
    \end{subfigure}
    
    \caption{\revision{Positional relevance bias for three long-document collections (best viewed in color). 
    We show a distribution of first relevant passage positions (red bars) vs. relevant document lengths (blue bars). Lengths and offsets are measured in the number of subword tokens (BERT-base tokenizer). See more results (including BEIR) in Figures~\ref{fig:relev_match_full_plot}~and~\ref{fig:relev_match_full_beir_plot} in Appendix~\ref{sec:data_details}.
    }
    }
    \label{fig:relev_match_plot}
\end{figure*}

\section{Introduction}\label{sec:intro}
Various advances in Transformer architectures---including sparse attention \cite{BigBird2020,longformer2020} and FlashAttention \cite{FlashAttention2022}---have motivated a growing interest in long-document ranking and retrieval. However, despite the ability of these models to process substantially more text,
on \emph{popular retrieval benchmarks} the improvements by these models over simpler truncation-based approaches remain surprisingly modest 
\cite{DaiC19,luyu2022moresplus,boytsov2022understanding,DBLP:conf/acl/CoelhoMMCX24}.
A widely used truncation-based \emph{FirstP} baseline \cite{DaiC19}---where models score only the first 512 tokens of each document---often performs competitively or sometimes even better than long-context counterparts (see, e.g., Table~\ref{tab:main_results_real_short}). 

Despite anecdotal knowledge about the presence of this phenomenon in the MS MARCO document-retrieval collection among TREC Deep Learning track participants 
and in some early reports \cite{FIRA2020,DBLP:conf/ecir/HofstatterLAZH21,boytsov2022understanding} the available evidence has been scattered and incomplete.
In particular, despite the track's five-year history, none of the track's overview papers have mentioned this issue \cite{trecdl2019overview,trecdl2020overview,trecdl2021overview,trecdl2022overview,trecdl2023overview}.

Moreover, it remains unclear whether these limitations stem from model deficiencies or from characteristics of the benchmarks themselves. In this paper, we initially hypothesized that both factors---model robustness and benchmark design---may be responsible for the limited gains achieved by long-document models over \emph{FirstP} baselines. However, our findings suggest that benchmark design, particularly positional relevance bias, is the dominant factor.

To verify our research hypotheses, we first conducted a large-scale, systematic study of over 20 Transformer-based ranking models \cite{devlin2018bert,vaswani2017attention} for long-document retrieval. This was done using three popular document collections---MS MARCO Documents v1/v2~\cite{trecdl2020overview} and Robust04~\cite{Terabyte2004}---along with diverse query sets (both large and small), several Transformer backbones, and multiple training seeds.  In addition to locally trained models, we also assessed a listwise LLM ranker RankGPT \cite{RankGPT2023} ``powered'' by OpenAI \cite{openaiGPT2023} and Anthropic \cite{anthropic2024claude3} cloud APIs. 

Despite the increased context capacity of long-document models, we found that none of them consistently outperformed their \emph{FirstP} baselines by more than 5\% on average. 
An ablation experiment showed that limited gains over \emph{FirstP} were not attributable to the choice of a Transformer backbone model (see Table~\ref{tab:first_p_backbone_ablation}). 

Next, we estimated positional relevance bias across five document collections and more than six query sets. 
As can be seen from Fig.~\ref{fig:relev_match_plot}, in the vast majority of cases, the first relevant passage occurred within the initial 512 tokens.
In contrast, the distribution of relevant passage positions is more uniform, with a pronounced long tail. 
This confirms that positional bias is substantial not only in MS MARCO, but also in other TREC collections.
Surprisingly, we also found bias in several BEIR collections \cite{BEIR}, which are typically categorized as 
\emph{short-document} datasets (a complete set of plots is provided in Fig.~\ref{fig:relev_match_full_plot} and~\ref{fig:relev_match_full_beir_plot} of Appendix \S~\ref{sec:positional_bias_indent}). 

Our initial exploration prompted two \emph{broad} research questions:
\begin{itemize}
    \item \textbf{RQ1:} How robust are long-document models to the positional bias of relevant passages?
    \item \textbf{RQ2:} 
    To what extent has the research community advanced the state of long-document ranking models? Specifically, do current approaches yield substantial improvements over \emph{FirstP} baselines? Considering that all existing long-document models are at least 2$\times$ slower than their respective \emph{FirstP} counterparts (see Figure~\ref{fig:tradeoff_plot_rel}, \S~\ref{sec:efficiency}), it is reasonable to question the practicality of such models and to consider whether \emph{FirstP} variants might be preferable in real-world applications.
\end{itemize}

To answer these questions,
we constructed a new \emph{diagnostic} synthetic collection \emph{MS MARCO FarRelevant}
where relevant passages were not present among the first 512 tokens.
On this dataset, many long-context models---including RankGPT \cite{RankGPT2023}---failed to generalize and performed at a random baseline level, suggesting overfitting to positional bias.
Poor performance of models on MS MARCO FarRelevant prompted another important question \textbf{RQ3:} Can debiasing of training data mitigate model overfitting to positional bias?
We addressed this question by evaluating an existing debiasing approach by \citet{DBLP:conf/ecir/HofstatterLAZH21}. 

Our paper makes the following contributions:
\begin{itemize}
    \item We re-examined the issue of positional relevance bias, gathered extensive evidence confirming its presence in both
    long- and short-document datasets, showed that it negatively affects all models including recent LLM-based rankers, and evaluated the robustness of ranking models against this bias using a new diagnostic dataset MS MARCO FarRelevant.

    \item Our work highlights the need for careful benchmark design in evaluating long-context models for document ranking, so that benchmarks do not mask the benefits of long-context models, and identify model types that are more robust to positional bias;
    \item We performed an extensive reproduction study of over 20 ranking models using established benchmark collections for long-document retrieval and ranking: MS MARCO Documents v1 and v2~\cite{trecdl2020overview} and Robust04~\cite{Terabyte2004};
    \item We experimented with an existing approach to debiasing training data \cite{DBLP:conf/ecir/HofstatterLAZH21} and motivated further research in this area.
\end{itemize}
Our code and data are available.\ourcode

\section{Related Work}\label{sec:related_work}
\textit{Neural Ranking} models have been a widely studied topic in recent years \cite{guo2019deep}, though the success of early approaches was debated \cite{lin2019neural}. This changed with the introduction of BERT, a bidirectional encoder-only Transformer model \cite{devlin2018bert}, which significantly outperformed previous methods in both NLP \cite{devlin2018bert} and information retrieval (IR) tasks \cite{nogueira2019document,trecdl2020overview}.

Several Transformer-based models, such as ELECTRA~\cite{ELECTRA2020} and DEBERTA~\cite{he2021debertav3}, have improved upon BERT through different training strategies and datasets. However, due to their architectural similarities, we---following~\citet{2021LinNY}---refer to these collectively as BERT models.

Despite their strong performance, neural models are vulnerable to distribution shifts, often relying on superficial features 
and exhibiting various \emph{biases}.
They do not consistently outperform BM25 on out-of-domain data \cite{Boytsov_Pseudo_2021, BEIR}, can be misled by minor text modifications and distractor sentences~\cite{DBLP:journals/tacl/MacAvaneyFGDC22}, or reformulated queries \cite{DBLP:conf/ecir/PenhaCH22}.
They also struggle to effectively utilize information located in the middle of long input contexts, in particular
for retrieval-augmented generation \cite{DBLP:journals/tacl/LiuLHPBPL24}.
Recent work showed that this can be an \emph{intrinsic architectural} artifact such that
the model gives disproportionately high attention to tokens at the beginning or end of the input context, regardless of their actual relevance, 
which can be partially mitigated by re-calibration \cite{DBLP:conf/acl/HsiehCL0LKGRLKP24}.
However, \citet{DBLP:conf/nips/AnML0LC24} argue that architectural biases can be exacerbated by insufficient 
explicit supervision during long-context training,
which causes models to assume that important information is concentrated near context-window edges.

A related but distinct issue---\emph{positional relevance bias}---which this study attributes primarily to the characteristics of supervised training data, has been identified in information-retrieval settings, particularly within the MS MARCO document-retrieval collection
\cite{FIRA2020,boytsov2022understanding,DBLP:conf/acl/CoelhoMMCX24}.
Some studies have reported the strong performance of \emph{FirstP} baselines on long-document retrieval collections, 
interpreting this as evidence of benchmark-induced positional bias \cite{boytsov2022understanding,E52024,DBLP:journals/tois/RauDK24}.
Note, however, that strong performance of \emph{FirstP} baseline is only indirect evidence, 
potentially resulting from implementation bugs, suboptimal training methods, or model-inherent biases.

The earliest study on this topic obtained direct evidence of bias only for a small set of TREC DL
queries, without studying how bias affects model performance \cite{FIRA2020}. 
\citet{boytsov2022understanding} tested a variety of long-document ranking models (on MS MARCO and Robust04 datasets) and found them
to be only marginally better than the \emph{FirstP} baseline, which they attributed to positional relevance bias.
\citet{DBLP:conf/acl/CoelhoMMCX24} found that two embedding models trained on MS MARCO ``dwell'' in the beginning and are 
 less effective when relevant information is present elsewhere in a document. 
These studies, however, (1) had \emph{limited} evidence that bias was attributable to data, (2) did not evaluate
key modern long-context models, and (3) did not assess a debiasing strategy.

Because positional relevance bias can ``obscure'' benefits of long-document ranking models,
\citet{DBLP:journals/tois/RauDK24} proposed to compare these models with \emph{RandP} baselines, 
which score a randomly selected passage. However, we believe this approach is problematic for two reasons.
First, since a \emph{RandP} model often fails to score an entire relevant passage, 
it artificially underestimates a model's accuracy, making comparisons with \emph{RandP} unfair. Second, this approach does not address the core problem of biased benchmarks, which still allow models to exploit the shortcut of focusing only on the document’s initial portion.

Interestingly, \citet{DBLP:journals/tois/RauDK24} argue that positional bias is strong only in MS MARCO, but not in Robust04. 
We believe this is because they only compared \emph{FirstP} and \emph{MaxP} models in a zero-shot setting (trained on MS MARCO and tested on Robust04), whereas we fine-tuned all models on Robust04 as well. 
The relatively strong performance of \emph{MaxP} compared to \emph{FirstP} in their study may be due to the generally stronger zero-shot performance of \emph{MaxP} models.
We observe similarly \emph{small} gains over \emph{FirstP} for both MS MARCO and Robust04,
and our passage-relevance analysis confirms that they have comparable levels of positional bias.
Thus, despite their widespread use, Robust04 and MS MARCO are not particularly useful for benchmarking of long-document models.

To address the issue with existing benchmarks, \citet{E52024} proposed a \emph{LongEmbed} benchmark with two synthetic tasks where relevant mini-passages were scattered uniformly across documents with lengths varying from 256 to 32768 tokens. 
However, as discussed in \S \ref{sec:comp_synthetic}, these synthetic sets are quite unnatural and lack diversity.
Furthermore, \citet{E52024} provide only small query sets and no in-domain training data, making it difficult to assess the upper performance bound that models can achieve on this dataset.
As another important limitation \citet{E52024}, only explored training-free extensions of positional encoding and did not investigate methods to debias training data. 
In contrast, \citet{DBLP:conf/ecir/HofstatterLAZH21} proposed to debias training data using a simple yet effective approach.
However, they did \emph{not} evaluate it on \emph{challenging} long-document datasets. 

In this paper, we aim to provide stronger evidence through collecting more comprehensive statistics of relevant passages in documents, 
experimenting with a diverse set of models, and applying the diagnostic synthetic set,
which also demonstrates models' abilities to adapt to different positional biases.

Due to the quadratic complexity of the Transformer's attention mechanism \cite{vaswani2017attention,BahdanauCB14}, 
early Transformer models restricted input length to a maximum of 512 (subword) tokens. Until around 2022, two main strategies were used to process long documents: (1) localizing attention and (2) splitting documents into smaller, independently processed chunks. Attention-localization methods apply a limited-span (sliding window) attention and selective global attention. Given the vast number of such approaches (see  \citealt{Tay2020}), evaluating all of them is impractical. Therefore, we focus on two popular models: Longformer \cite{longformer2020} and BigBird \cite{BigBird2020}. More recently, it has also become feasible to train long-context models with an IO-efficient FlashAttention algorithm without sparsifying attention \cite{FlashAttention2022}. 
\onlyarxiv{
In our work, we use three such models: JINA \cite{gunther2023jina}, MOSAIC \cite{portes2023mosaicbert} and TinyLLAMA \cite{TinyLLAMA2024}. 
}

In summary, methods to tackle longer documents are divided into \emph{LongP} methods---where longer document are ``natively'' supported and \emph{SplitP} methods---where a longer document cannot be processed as a whole and needs to be processed in chunks. 
The results of each chunk are aggregated together using various aggregation techniques, including computation of 
a maximum or a weighted-sum prediction score \cite{YilmazWYZL19,DaiC19,CEDR2019}.
This includes MaxP \cite{DaiC19}, AvgP, SumP \cite{CEDR2019}, as well as PARADE Avg and PARADE Max models \cite{CEDR2019}. 
MaxP is an important baseline that computes relevance scores for each chunk independently and takes their maximum.

Some \emph{SplitP} approaches aggregate using simple neural networks. This includes all CEDR \cite{CEDR2019}
models, the Neural Model 1 \cite{BoytsovK21}, and the PARADE Attention model \cite{Parade2020}.
In contrast, PARADE Transformer \cite{Parade2020} models' aggregator network is an additional Transformer model. 
Due to space constraints, a detailed description of document-splitting (\emph{SplitP}) approaches is provided in the Appendix~\S~\ref{sec:methods_detailed}. 

The recent success of decoder-only models---commonly known as LLMs---has led to a new generation of cross-encoding \emph{LongP} models
that natively support longer contexts. First, a pre-trained cross-encoding decoder-only model can be directly fine-tuned on a ranking or embedding task \cite{DBLP:journals/corr/abs-2310-08319}. 
Second, the RankGPT approach~\cite{RankGPT2023} formulates document ranking as a generation task: The model is prompted
with a list of documents and an instruction to generate their ranking---an ability made possible through instruction-tuning and/or alignment \cite{FLAN22,InstructGPT2022}.
When the combined length of concatenated documents exceeds the input context size, RankGPT employs an overlapping
sliding window strategy, followed by aggregation of the results.

\begin{table*}[!tb]
\tablecustomsmall
\scriptsize
\centering
\begin{threeparttable}
\setlength{\tabcolsep}{0.5em}
\begin{tabular}{l|l|l|ll|l}\toprule
Retriever / Ranker  &  \multicolumn{1}{c}{MS MARCO}  & \multicolumn{1}{|c|}{TREC DL}   & \multicolumn{2}{|c|}{Robust04}    &  \multicolumn{1}{c}{Avg. gain } \\
      &  \multicolumn{1}{|c|}{dev}         &  \multicolumn{1}{c|}{(2019-2021)} & \multicolumn{1}{c}{title} & \multicolumn{1}{c|}{description}    & \multicolumn{1}{c}{over FirstP} \\\midrule
      & \multicolumn{1}{|c|}{\scriptsize\textbf{MRR}} &  
      \multicolumn{1}{|c|}{\scriptsize\textbf{NDCG@10}} &  \multicolumn{2}{c|}{\scriptsize\textbf{NDCG@20}} &   \\\midrule
\revision{BM25} & \revision{0.274} & \revision{0.545} & \revision{0.428} & \revision{0.402} & -- \\
retriever (if different from BM25) & 0.312 & 0.629 & -- & -- & -- \\ \midrule
FirstP (BERT) & 0.394 & 0.632 & 0.475 & 0.527 & -- \\
FirstP (Longformer) & 0.404 & 0.643 & 0.483 & 0.540 & -- \\
FirstP (ELECTRA) & 0.417 & 0.662 & 0.492 & 0.552 & -- \\
FirstP (DEBERTA) & 0.415 & 0.672 & 0.534 & 0.596 & -- \\
FirstP (Big-Bird) & 0.408 & 0.656 & 0.507 & 0.560 & -- \\
FirstP (JINA) & 0.422 & 0.654 & 0.488 & 0.532 & -- \\
FirstP (MOSAIC) & 0.423 & 0.643 & 0.453 & 0.538 & -- \\ 
FirstP (TinyLLAMA) & 0.395 & 0.615 & 0.431 & 0.473 & -- \\
FirstP (E5-4K) \textbf{zero-shot} & 0.380 & 0.641 & 0.438 & 0.429 &  -- \\
FirstP RankGPT (GPT-4o-mini) & -- & \textbf{0.708} & -- & 0.562 &  \\
\midrule
AvgP & 0.389 $(- 1.3\%)$ & 0.642 $(+ 1.5\%)$ & 0.478 $(+ 0.5\%)$ & 0.531 $(+ 0.9\%)$ & +$0.4\%$ \\\midrule
MaxP & 0.392 $(- 0.4\%)$ & 0.644$^{a}$ $(+ 1.9\%)$ & 0.488$^{a}$ $(+ 2.6\%)$ & 0.544$^{a}$ $(+ 3.3\%)$ & +$1.9\%$ \\
MaxP (ELECTRA) & 0.414 $(- 0.6\%)$ & 0.659 $(- 0.5\%)$ & 0.502 $(+ 2.0\%)$ & 0.563 $(+ 2.1\%)$ & +$0.8\%$ \\
MaxP (DEBERTA) & 0.402$^{a}$ $(- 3.2\%)$ & 0.671 $(- 0.1\%)$ & 0.535 $(+ 0.2\%)$ & 0.609$^{a}$ $(+ 2.2\%)$ & -$0.2\%$ \\
SumP & 0.390 $(- 1.0\%)$ & 0.639 $(+ 1.0\%)$ & 0.486 $(+ 2.2\%)$ & 0.538 $(+ 2.1\%)$ & +$1.1\%$ \\ \midrule
CEDR-DRMM & 0.385$^{a}$ $(- 2.3\%)$ & 0.629 $(- 0.5\%)$ & 0.466 $(- 2.0\%)$ & 0.533 $(+ 1.3\%)$ & -$0.9\%$ \\
CEDR-KNRM & 0.379$^{a}$ $(- 3.8\%)$ & 0.630 $(- 0.3\%)$ & 0.483 $(+ 1.7\%)$ & 0.535 $(+ 1.7\%)$ & -$0.2\%$ \\ 
CEDR-PACRR & 0.395 $(+ 0.3\%)$ & 0.643$^{a}$ $(+ 1.6\%)$ & 0.496$^{a}$ $(+ 4.3\%)$ & 0.549$^{a}$ $(+ 4.2\%)$ & +$2.6\%$ \\\midrule
Neural Model1 & 0.398 $(+ 0.9\%)$ & 0.650$^{a}$ $(+ 2.8\%)$ & 0.484 $(+ 1.8\%)$ & 0.537 $(+ 1.9\%)$ & +$1.8\%$ \\ \midrule
PARADE Attn & 0.416$^{a}$ $(+ 5.5\%)$ & 0.652$^{a}$ $(+ 3.1\%)$ & 0.503$^{a}$ $(+ 5.7\%)$ & 0.556$^{a}$ $(+ 5.6\%)$ & \textbf{+5.0\%} \\
PARADE Attn (ELECTRA) & 0.431$^{a}$ $(+ 3.3\%)$ & 0.680$^{a}$ $(+ 2.7\%)$ & 0.523$^{a}$ $(+ 6.4\%)$ & 0.581$^{a}$ $(+ 5.3\%)$ & +$4.4\%$ \\
PARADE Attn (DEBERTA) & 0.422$^{a}$ $(+ 1.6\%)$ & 0.688$^{a}$ $(+ 2.4\%)$ & \textbf{0.549$^{a}$ $(+ 2.9\%)$} & \textbf{0.615$^{a}$ $(+ 3.2\%)$} & +$2.5\%$ \\
PARADE Avg & 0.392 $(- 0.6\%)$ & 0.646$^{a}$ $(+ 2.1\%)$ & 0.483 $(+ 1.5\%)$ & 0.534 $(+ 1.5\%)$ & +$1.1\%$ \\
PARADE Max & 0.405$^{a}$ $(+ 2.7\%)$ & 0.655$^{a}$ $(+ 3.5\%)$ & 0.489$^{a}$ $(+ 2.8\%)$ & 0.548$^{a}$ $(+ 4.0\%)$ & +$3.3\%$ \\\midrule
PARADE Transf-RAND-L2 & 0.419$^{a}$ $(+ 6.3\%)$ & 0.655$^{a}$ $(+ 3.6\%)$ & 0.488$^{a}$ $(+ 2.8\%)$ & 0.548$^{a}$ $(+ 4.1\%)$ & +$4.2\%$ \\
PARADE Transf-RAND-L2 (ELECTRA) & \textbf{0.433$^{a}$ $(+ 3.9\%)$} & 0.670 $(+ 1.2\%)$ & 0.523$^{a}$ $(+ 6.3\%)$ & 0.574$^{a}$ $(+ 3.9\%)$ & +$3.8\%$ \\
PARADE Transf-PRETR-L6 & 0.402$^{a}$ $(+ 1.9\%)$ & 0.643 $(+ 1.6\%)$ & 0.494$^{a}$ $(+ 4.0\%)$ & 0.554$^{a}$ $(+ 5.1\%)$ & +$3.2\%$ \\
\midrule
LongP (Longformer) & 0.412$^{a}$ $(+ 1.9\%)$ & 0.668$^{a}$ $(+ 3.9\%)$ & 0.500$^{a}$ $(+ 3.6\%)$ & 0.568$^{a}$ $(+ 5.1\%)$ & +$3.6\%$ \\
LongP (Big-Bird) & 0.397$^{a}$ $(- 2.9\%)$ & 0.651 $(- 0.7\%)$ & 0.452$^{a}$ $(-10.9\%)$ & 0.477$^{a}$ $(-14.9\%)$ & -$7.3\%$ \\
LongP (JINA) & 0.416$^{a}$ $(- 1.5\%)$ & 0.665$^{a}$ $(+ 1.7\%)$ & 0.503$^{a}$ $(+ 2.9\%)$ & 0.558$^{a}$ $(+ 4.9\%)$ & +$2.0\%$ \\
LongP (MOSAIC) & 0.421 $(- 0.4\%)$ & 0.664$^{a}$ $(+ 3.3\%)$ & 0.456 $(+ 0.6\%)$ & 0.570$^{a}$ $(+ 6.0\%)$ & +$2.4\%$ \\
LongP (TinyLLAMA) & 0.402$^{a}$ $(+ 1.7\%)$ & 0.608 $(- 1.1\%)$ & 0.452$^{a}$ $(+ 4.8\%)$ & 0.505$^{a}$ $(+ 6.7\%)$ & +$3.0\%$ \\
LongP (E5-4K) \textbf{zero-shot} & 0.353$^{a}$ $(- 7.1\%)$ & 0.649 $(+ 1.3\%)$ & 0.439 $(+ 0.1\%)$ & 0.434 $(+ 1.1\%)$ & -$1.1\%$ \\
LongP RankGPT (GPT-4o-mini) & -- & 0.706 $(- 0.3\%)$ & -- & 0.562 $(+ 0.0\%)$ & -$0.1\%$ \\
\bottomrule
\end{tabular}

    \begin{tablenotes}
        \small
{\tablecustomsmall  
  In each column we show a relative gain with respect model's respective \emph{FirstP} baseline: The last column shows the average relative gain over \emph{FirstP} baselines. Best numbers are in \textbf{bold}: Results are averaged over three seeds. Unless specified explicitly, the backbone is \textbf{BERT-base}.  
  Statistical significant differences with respect to this baseline are denoted using the superscript \textbf{a}. 
  $p$-value threshold is 0.01 for an MS MARCO development collection and 0.05 otherwise. \\
 }
    \end{tablenotes}
\end{threeparttable}
\caption{\revision{Comparison between long-document models and respective \emph{FirstP} (truncation) baselines. Results for MS MARCO, TREC DL, and Robust04.} \label{tab:main_results_real_short}}
\end{table*}

\section{Experiments}

\subsection{Data}
Our primary datasets, used for both training and evaluation, consist of several realistic collections (along with their respective query sets) and synthetic data. All datasets are in \emph{English}. Document and query statistics are provided in Appendix \S~\ref{sec:data_details}; see Tables \ref{tab:queries} and \ref{tab:docs}.

The realistic datasets include three long-document collections: MS MARCO Documents v1 and v2 \cite{msmarco,trecdl2019overview,trecdl2021overview}, {Robust04} \cite{Robust04},
and several short-document collections: {MS MARCO Passages}  (v1) \cite{msmarco,trecdl2019overview,trecdl2021overview}, 
and seven BEIR datasets (see \S~\ref{sec:data_details}). 
Several BEIR datasets and the MS MARCO Documents collections 
include document titles, which we prepend to the main document text. 
Following \citet{DBLP:conf/ecir/HofstatterLAZH21}, we created a debiased version of MS MARCO by randomly splitting documents at word boundaries and  concatenating reordered segments (see Algorithm \ref{DebiasAlgo} in \S~\ref{sec:training_setup_expanded}). 
\onlyarxiv{This debiasing process is only partial, as shorter documents remain more frequent. To address this imbalance, we experimented with oversampling longer documents, but this approach did not yield improvements.}

Our synthetic data consists of two subsets from LongEmbed \cite{E52024} and our newly created MS MARCO FarRelevant collection.
All these can be considered variants of the needle-in-the-haystack test, where an informational ``nugget'' is randomly embedded  within unrelated text \cite{LoCo2024,E52024,DBLP:journals/tacl/LiuLHPBPL24}.
We use two LongEmbed subsets: {Needle} and {Passkey}. 
Each subset has 800 question-document pairs with document lengths varying from approximately 256 to 32768 tokens.

MS MARCO FarRelevant was created by randomly mixing relevant and non-relevant passages from the MS MARCO \emph{Passage} collection \cite{trecdl2019overview} in such a
way that (1) each document contains exactly one relevant passage, (2) this passage does not
start before token 512, and (3) less than 1\% of documents have more than 1500 tokens (see an algorithm in Appendix \S~\ref{sec:farrelevant_algo}).
It has about 0.5 million documents with an average length of 1.1K tokens.
Due to the {MS MARCO} datasets' \textbf{non-commercial license}, {MS MARCO FarRelevant} has the same licensing restriction.
In Appendix \S~\ref{sec:comp_synthetic}, we present dataset examples and argue that---while all these collections share the limitation of not resembling natural documents---{MS MARCO FarRelevant} offers greater diversity and serves as a more suitable benchmark for evaluating text retrieval systems.

{Robust04} is another relatively small dataset containing 0.5 million documents, comprising a mix of news articles and government records, some of which are quite lengthy. However, it includes only a limited number of queries (250), making it a challenging benchmark for training models in low-data scenarios. Each query has a title and a description. The title expresses a concise information need, while the description provides a more detailed request, often written in proper English prose.
We use {Robust04} in a cross-validation setting with folds created by \citet{huston2014comparison} and provided via IR-datasets~\cite{irds2021}.\footnote{In this setting we do not train Robust04 models from scratch, but instead fine-tune models trained on the MS MARCO Documents collection.}

MS MARCO v1 was created from the MS MARCO reading comprehension dataset \cite{msmarco} and consists of two \emph{related} collections: {MS MARCO Passages} and {MS MARCO Documents}.
MS MARCO v1 comes with \emph{large} query sets, which is particularly useful for training and testing models in the big-data regime.
These query sets include question-like queries selected from the Bing search engine logs \cite{trecdl2021overview}.
Note that queries are not necessarily proper English questions, e.g., ``lyme disease symptoms mood'', but they are answerable by a short passage retrieved from a set of about 3.6M Web documents \cite{msmarco}. 
MS MARCO v1 test sets were created in two stages. First, relevance judgments were created for the passage variant of the dataset.
Then, document-level relevance labels were created by transferring passage-level relevance to original documents from which passages were extracted.

The MS MARCO v2 collection was created for the TREC 2021 Deep Learning (DL) track \cite{trecdl2021overview}. It is an expanded version of MS MARCO v1 and incorporates a subset of sparse relevance judgments from MS MARCO v1. In the training set, newly added documents lack both positive and negative judgments, introducing a bias where many relevant documents are mistakenly considered non-relevant. As a result, we do not train on v2 data and only use it for testing.

Relevance labels in the training and development sets are ``sparse'':
There is about one positive example per query without explicit negatives.
In addition to sparse relevance judgments---separated into 
training and development subsets---there is a small number of (about 150) test queries from the TREC Deep Learning~(DL) track.

\subsection{Setup}\label{sec:exper_setup}
We focus on cross-encoding rankers, which process queries concatenated with documents \cite{nogueira2019passage}.
This includes various \emph{SplitP} and \emph{LongP} models discussed in \S~\ref{sec:related_work} and 
in the Appendix~\S~\ref{sec:methods_detailed}.
As a reference point, we also tested
a bi-encoder embedding E5-4K model, which has strong performance on the LongEmbed benchmark with
context sizes under 4K tokens \cite{E52024}. 
E5-4K was tested only in zero-shot mode (without fine-tuning)---as a ranking model.
Except for LongEmbed subsets {Needle} and {Passkey}, a ranker is applied to a top-$k$ set produced by a retriever.
Each LongEmbed subset contains only 800 documents, and we re-rank them all without using a first-stage retriever.

Nearly all rankers are based on BERT models (bidirectional encoder-only Transformer)
with 100-200M parameters (see Table~\ref{tab:model_params}).
Additionally, we evaluated two types of LLM rankers:
(1) a fine-tuned TinyLLAMA model, which delivers strong performance relative to its
compact size \cite{TinyLLAMA2024} and~(2) generative black-box LLMs.
For (2), we used OpenAI's GPT-4o-mini \cite{openaiGPT2023} and Anthropic's Claude Haiku-3 \cite{anthropic2024claude3}, both of which
support at least a 128K-token input context.

Unless explicitly specified, the backbone
Transformer model for \emph{SplitP} methods is BERT-base \cite{devlin2018bert}. Although
using other backbones such as ELECTRA \cite{ELECTRA2020} and DEBERTA \cite{he2021debertav3}
can improve overall accuracy, we observe larger gains compared to the \emph{FirstP}
baseline when \emph{we use BERT-base} (see \S~\ref{sec:repro_and_backbone} in the Appendix).

Except for E5 \cite{E52024} and RankGPT \cite{RankGPT2023}, 
which were evaluated only in  zero-shot mode, we trained each model using \emph{three} seeds.
Due to the high evaluation cost (more than \$1 per 1000 query-document pairs)\footnote{\url{https://openai.com/api/pricing/}}, we also did not test RankGPT on some query sets, in particular, we excluded the large MS MARCO development set.

We measure effectiveness of the models using the mean reciprocal rank (MRR),
 the normalized discounted cumulative gain at rank $k$ (NDCG@k) \cite{DBLP:journals/tois/JarvelinK02}, precision at rank $k$ (P@k),
 and the mean average precision (MAP).
To assess statistical significance, we averaged per-query metric values across three seeds.

\begin{figure}[!tb]
    \centering
    \vspace{-1em}
    \includegraphics[width=0.5\textwidth]{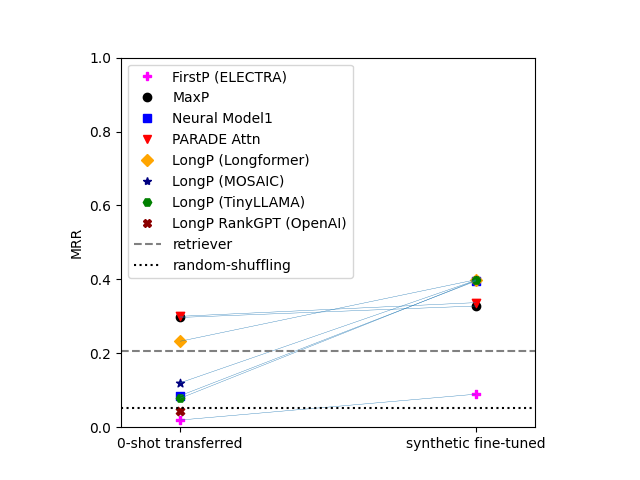}
     \vspace{-1em}
    \caption{Zero-shot vs. fine-tuned performance on MS MARCO FarRelevant. \revision{This figure shows results for a representative set of models.}}
    \label{fig:result_synthetic}
\end{figure}

\subsection{Results}\label{sec:results}
\paragraph{Realistic Data.} Our main experimental results for MS MARCO, TREC DL 2019-2021, and Robust04 are presented 
in Table~\ref{tab:main_results_real_short}.
There we show only a single aggregate number for all TREC DL query sets,
which is obtained by combining all the queries and respective relevance judgments (i.e., we
post an overall average rather than an average over the mean values for 2019, 2020, and 2021).
More detailed results, including Anthropic-based RankGPT, 
are available in Appendix~\ref{sec:detailed_exper_results}, specifically in Tables~\ref{tab:msmarco} and~\ref{tab:robust04}.
Efficiency evaluation is presented in \S~\ref{sec:efficiency} (see Fig.~\ref{fig:tradeoff_plot_rel}).
In \S~\ref{sec:repro_and_backbone}
we also show that we can match or outperform key prior results, which---we believe---boosts the trustworthiness of our experiments.

We abbreviate names of some PARADE models:
Note that \textsc{PARADE Attn} denotes a PARADE Attention model.
The \textsc{PARADE Transf} or \textsc{P.~Transf} prefix denotes PARADE Transformer models
where an aggregator Transformer can be either trained from scratch (\textsc{Transf-RAND-L2}) 
 or initialized with a pretrained model (\textsc{Transf-PRETR-L6}).
 L2 and L6 denote the number of aggregating layers (two and six, respectively).\footnote{
 Note, however, that \textsc{Transf-PRETR-L2} has only four attention heads.}

From Table~\ref{tab:main_results_real_short} and Fig.~\ref{fig:tradeoff_plot_rel}   we learn that the maximum average
gain relative to respective \emph{FirstP} baselines is only 5\% (when measured using MRR or NDCG@k).
Gains are much smaller for a number of models,
which sometimes even \emph{underperform} their \emph{FirstP} baselines on one or more datasets.
In particular, this is true for RankGPT \cite{RankGPT2023}, CEDR-DRMM, CEDR-KNRM \cite{CEDR2019}, JINA
\cite{gunther2023jina}, and MOSAIC \cite{portes2023mosaicbert}. 

\begin{table*}[!tb]
\centering
\setlength{\tabcolsep}{3pt}
\scriptsize
\begin{threeparttable}
\begin{tabular}{l|ll|ll|ll} \toprule
Ranker  &  \multicolumn{1}{c}{\textbf{MS MARCO}}  & \multicolumn{1}{c|}{\textbf{TREC DL}} &
       \multicolumn{2}{|c|}{\textbf{FarRelevant}}   &  
       \multicolumn{2}{|c}{\textbf{LongEmbed}}     \\
      &  \multicolumn{1}{c}{dev}       &  
        \multicolumn{1}{c|}{(2019-2021)} & 
        \multicolumn{1}{c}{zero-shot transf.}       &
        \multicolumn{1}{c|}{fine-tuned}     &
        \multicolumn{1}{c}{Needle}         &
        \multicolumn{1}{c}{Passkey}         \\ \midrule
      & \multicolumn{1}{|c}{\scriptsize\textbf{MRR}} &  
      \multicolumn{1}{c|}{\scriptsize\textbf{NDCG@10}} &
       \multicolumn{1}{c}{\scriptsize\textbf{MRR}} &   
       \multicolumn{1}{c|}{\scriptsize\textbf{MRR}} &          
       \multicolumn{1}{c}{\scriptsize\textbf{MRR}} &   
       \multicolumn{1}{c}{\scriptsize\textbf{MRR}}                
      \\ \midrule
\revision{BM25} & \revision{0.274} & \revision{0.545} & 
\revision{ 0.207 } &
\revision{ 0.207 } &
\revision{ 0.305 } &
\revision{ 0.339 } 
\\ \midrule
\multicolumn{7}{c}{\textbf{Original MS MARCO training set}} \\ \midrule
FirstP (ELECTRA) & 0.417 & 0.662 & 0.019 & 0.089 & 0.205 & 0.235 \\ \midrule
MaxP (ELECTRA) & 0.414 & 0.659 & 0.328 & 0.349 & \textbf{0.331} & \textbf{0.338} \\
PARADE Attn (ELECTRA) & 0.431 & \textbf{0.680} & \textbf{0.338} & 0.354 & 0.270 & 0.334 \\
PARADE Transf-RAND-L2 (ELECTRA) & \textbf{0.433} & 0.670 & 0.229 & \textbf{0.432} & 0.321 & 0.333 \\
CEDR-KNRM & 0.379 & 0.630 & 0.055 & 0.382 & 0.129 & 0.166 \\\midrule
\multicolumn{7}{c}{\textbf{Debiased MS MARCO} \cite{DBLP:conf/ecir/HofstatterLAZH21}} \\ \midrule
MaxP (ELECTRA) & 0.377$^{a}$ $(- 9.1\%)$ & 0.665 $(+ 0.8\%)$ & 0.321 $(- 2.1\%)$ & 0.349 & 0.316 $(- 4.6\%)$ & 0.325 $(- 3.9\%)$ \\
PARADE Attn (ELECTRA)  & 0.390$^{a}$ $(- 9.4\%)$ & 0.653$^{a}$ $(- 3.9\%)$ & 0.326 $(- 3.6\%)$ & 0.354 & 0.251 $(- 7.1\%)$ & 0.330 $(- 1.0\%)$ \\
PARADE Transf-RAND-L2 (ELECTRA)  & 0.410$^{a}$ $(- 5.4\%)$ & 0.677 $(+ 1.0\%)$ & 0.328$^{a}$ $(+43.5\%)$ & \textbf{0.432} & 0.259$^{a}$ $(-19.4\%)$ & 0.331 $(- 0.7\%)$ \\
CEDR-KNRM  & 0.269$^{a}$ $(-29.0\%)$ & 0.503$^{a}$ $(-20.2\%)$ & 0.202$^{a}$ $(+268.8\%)$ & 0.382 & 0.121 $(- 5.8\%)$ & 0.181 $(+ 8.6\%)$ \\
\bottomrule
\end{tabular}
\begin{tablenotes}
  \tablecustomsmall
{\tablecustomsmall  
  Except \emph{FirstP} we train each model using the original and the debiased MS MARCO. 
  For each model trained on the debiased dataset, we compute a gain (or loss) compared to the same model trained on the original 
  training set. Statistical significance of the respective difference is denoted using the superscript superscript \textbf{a} ($p$-value threshold is 0.05
  for TREC DL and 0.01 for other collections). 
 Best numbers are in \textbf{bold}: Results are averaged over three seeds. \\
 }
\end{tablenotes}
\end{threeparttable}
\caption{\revision{Debiasing effectiveness. Performance of (selected) rankers trained on original and debiased MS MARCO.}\label{tab:main_debias_results}}
\end{table*}

We can also see that the \emph{LongP} variant of the Longformer model appears to have a relatively strong performance, but so does the \emph{FirstP} version of Longformer.
Thus, we think that a good performance of Longformer on MS MARCO and 
Robust04 collections can be largely explained by better pretraining compared to the original BERT-base model rather than by its ability to process long contexts.
Moreover, FirstP (ELECTRA) and FirstP (DEBERTA) are even more accurate than FirstP (Longformer) and perform comparably well to (or better than) some chunk-and-aggregate document models that use BERT-base as the backbone model. 
This is a fair comparison aiming to demonstrate that---on a typical
test collection---the benefits of long-context models are so small that comparable benefits can be obtained by finding or training a more effective
\emph{FirstP} model. 
\emph{FirstP} models are more efficient during inference and can be pretrained on a larger number of tokens for the same cost, 
so they may perform better (see Fig.~\ref{fig:tradeoff_plot_rel} in \S~\ref{sec:efficiency}).

\paragraph{Synthetic Data.} 
Based on our analysis of positions of first relevant passages, 
we hypothesized that limited benefits of long-context models are not due to inability to process long contexts, but rather due to a positional bias of relevant passages, which tend to be among the first 512 document tokens
(see Figure~\ref{fig:relev_match_plot} and Figure~\ref{fig:relev_match_full_plot} in Appendix \ref{sec:positional_bias_indent}).
To support this hypothesis, 
we carried out two sets of experiments using  MS MARCO FarRelevant,
where a relevant passage did not start until token 512. 
We carried out both a zero-shot experiment (evaluation of the model trained on MS MARCO) and a fine-tuning 
experiment using 50K in-domain queries (from the MS MARCO FarRelevant). 

Results for key models are shown in Fig.~\ref{fig:result_synthetic} and more detailed results can be found 
in Table \ref{tab:main_results_synthetic} of the Appendix \ref{sec:detailed_exper_results}.
The \emph{FirstP} models performed roughly at the random-baseline level in both zero-shot and fine-tuning modes (\textbf{RQ1}). 
Because of this, our main baselines here are Longformer and \emph{MaxP} models. 
For models with ELECTRA and DEBERTA backbones, we compare with MaxP (ELECTRA) and MaxP (DEBERTA), respectively. Otherwise, the baseline is MaxP~(BERT).

Surprisingly, E5-4K performance is also at a random-baseline level
despite its competitive performance on LongEmbed benchmark \cite{E52024},
MS MARCO, and Robust04 (see Table~\ref{tab:main_results_real_short}).
Both GPT-4o-mini and Claude Haiku-3 RankGPT perform at the random-baseline level as well! As a sanity check,
and to verify if more accurate and expensive LLMs could do better, we assessed performance of GPT-4o for a sample of 100 queries. The respective RankGPT \cite{RankGPT2023} ranker was still not better than a random baseline (\textbf{RQ1}).

Simple aggregation models including MaxP and PARADE Attention had good zero-shot accuracy, but benefited little from fine-tuning on MS MARCO FarRelevant (\textbf{RQ1}).

In contrast, other long-document models had poor zero-shot performance (sometimes at a random baseline level), but outstripped \emph{respective} MaxP baselines  by as much as \mbox{9.2\%-27.7\%} after fine-tuning, see Table \ref{tab:main_results_synthetic} and Fig.~\ref{fig:result_synthetic} (\textbf{RQ1} and \textbf{RQ2}). 

With the exception of RankGPT \cite{RankGPT2023} on TREC DL 2019-2021, 
PARADE Transformer models were more effective than other models on standard collections, but their advantage was small (a few~\%). 
In contrast, on MS MARCO FarRelevant,
PARADE Transformer (ELECTRA) outperformed the next competitor
Longformer by 8\% and PARADE Max (ELECTRA)---an early chunk-and-aggregate approach---by as much as 23.8\%, 
thus showcasing substantial advances in long-document
ranking modeling and answering \textbf{RQ2}.

\paragraph{Bias Mitigation.} To address \textbf{RQ3}, we trained four representative models on the debiased MS MARCO 
 and tested them on MS MARCO FarRelevant as well as Needle/Passkey subsets of LongEmbed  \cite{E52024}.\footnote{(see Appendix~\S~\ref{sec:exper_setup_misc} for model selection rationale)}
We also tested four models fine-tuned on MS MARCO FarRelevant on TREC DL 2019-2021 query sets.
Due to the substantial NDCG@10 drop (0.1–0.15) observed for PARADE Transformer and CEDR-KNRM, 
we concluded that fine-tuning on purely synthetic data is not viable and did not pursue it further.

According to Table~\ref{tab:main_debias_results},
debiasing improved the performance of CEDR-KNRM and PARADE Transformer on MS MARCO FarRelevant.
Yet, it mostly caused performance degradation on the original MS MARCO dataset and on LongEmbed subsets. 
It did not benefit the MaxP and PARADE Attention models, which were the most robust to positional bias.
\revision{We believe this result---along with the strong performance of all models after fine-tuning on MS MARCO FarRelevant---supports the conclusion that the models are not inherently biased toward the start of the document (or at least this is not a key factor). Instead, the primary cause appears to be positional bias in the training data.}

We further tested debiased models on short-document collections: the MS MARCO Passage
collection and seven BEIR collections \cite{BEIR}.
According to Table~\ref{tab:main_debias_short_collect_results} in Appendix~\ref{sec:exper_appendix},
for three out of four models, debiasing has either a positive or small negative effect (at most 1\% degradation). 
In particular, on the BEIR NQ subset, PARADE Transformer and CEDR-KNRM improved by 2.5\% and 11\%, respectively.

Despite the positive average improvement, we also observe substantial degradation for several datasets.
For example, the effectiveness of CEDR-KNRM decreases by 12.4\% and 4.6\% on SciFact and  SciDocs, respectively.
One possible reason is that short-document collections can have positional relevance bias as well
(see Figure \ref{fig:relev_match_full_beir_plot} in Appendix~\ref{sec:data_details}).
Thus, training on the original (biased) MS MARCO collection teaches the model to rely on a helpful 
shortcut---to focus more on the beginning of a document---whereas training on debiased data removes this beneficial behavior.

In summary, debiasing results are promising, but they also suggest that mitigating positional bias remains a challenging problem (\textbf{RQ3}).

\paragraph{Key Findings.} 
We observe the following:
\begin{itemize}

\item We confirmed the presence of positional relevance bias in both  long- and short-document datasets.

\item Not only did positional relevance bias diminish the benefits of processing longer document contexts, 
but it also led to model overfitting to the bias and performing poorly in a zero-shot setting when the distribution of relevant passages changes substantially.

\item We found that debiasing training data had mixed success:  
Although it improved the effectiveness of some models on some long-document ranking tasks without
substantial degradation on short-document collections, 
it degraded performance in several cases---especially for MaxP and PARADE Attention models---which were already intrinsically more robust to positional relevance bias---possibly because many short-document collections also have positional relevance bias. 
 
\item It is also worth highlighting the consistently strong performance of PARADE models on both standard long-document collections and MS MARCO FarRelevant. The best PARADE models substantially outperformed the best \emph{LongP} models in both zero-shot and fine-tuning settings, although the specific models leading in each setting may differ.

\end{itemize}

\section{Conclusion}
In this work, we revisited the problem of positional relevance bias in long-document retrieval and presented extensive evidence of its \emph{widespread} presence across existing benchmarks,
including several BEIR datasets \cite{BEIR}, which contain relatively short documents. Using both real and synthetic datasets---including our new \emph{diagnostic} dataset MS MARCO FarRelevant---we evaluated the effectiveness of over 20 ranking models as well as their robustness to positional relevance bias.

Our findings highlight the importance of a benchmark design that does not obscure the benefits of long-context modeling. We identified model families (i.e., PARADE Attention and MaxP) that are more robust to positional bias, and confirmed the strong performance of PARADE models \cite{Parade2020}, which remain competitive even against recent long-context architectures.

Finally, our debiasing experiments yielded limited and inconsistent gains, motivating further research into more effective mitigation strategies, including combining debiasing with training on well-designed synthetic data.

\section{Limitations}
Our paper has several limitations related primarily to the choice of datasets, models,
and strength of evidence for positional relevance bias.

First of all, our evaluation uses only cross-encoding ranking models. 
With the exception of E5-4K model, which is used in the zero-shot ranking mode,
we do not train or evaluate bi-encoding models (typically used to
create query and document embeddings for the first-stage retrieval). 
We nonetheless believe that---given a large number of proposals for long-document
ranking---a reproduction and evaluation of cross-encoding long-document rankers is a sufficiently important topic that alone warrants a publication. 

Moreover, we also use cross-encoding rankers as a tool to 
detect and expose positional relevance bias. 
In this regard, cross-encoders are easier to train using standard (rather than high-memory) GPUs 
with a  mini-batch size of one  and gradient accumulation. 
They also typically require only one epoch to converge (only a few models need two or three epochs). 
In contrast, bi-encoders are trained using large batches with in-batch negatives for multiple epochs (e.g., \citet{DPR2020} report using at least 40 epochs).

Second, the bulk of our ranking experiments uses only two \emph{English} document collections: MS MARCO Documents v1 and v2~\cite{trecdl2021overview} and Robust04~\cite{Terabyte2004}.
However, we have to restrict the choice of datasets to make multi-seed evaluations of 20+ models feasible. 
Thus, to corroborate the \emph{widespread} presence of positional bias we used two additional popular long-document collections: Gov2 \cite{DBLP:conf/trec/AllanAPKC08} and ClueWeb12 \cite{Web2012}
as well as seven BEIR datasets \cite{BEIR}.
For the study on the robustness of models to positional bias and its mitigation, we used two additional synthetic collections: 
MS MARCO FarRelevant and two subsets from LongEmbed \cite{E52024} together with eight
short-document collections (see Table~\ref{tab:main_debias_short_collect_results}).

One could argue that the limited improvements over \emph{FirstP} baselines result from the models' inability to handle long contexts. 
To address this concern, we trained and evaluated a diverse set of cross-encoding ranking models, including both split-and-aggregate models and models explicitly designed for long input sequences. Additionally, we assessed cloud-based RankGPT rankers, which have shown strong performance in recent research~\cite{RankGPT2023}.

However, we can still test only a limited number of models.
In this regard, one might always argue that there are untested architectures that would 
outperform \emph{FirstP} baselines by a much larger margin.
To demonstrate that selected models can, in principle, benefit from long
contexts and decisively outperform simple baselines such as \emph{FirstP} and
even \emph{MaxP} models, we trained and/or evaluated them on a synthetic
collection MS MARCO FarRelevant, which can be seen as a challenging version of a needle-in-the-haystack test. 

Admittedly, this is still a limiting experiment, because synthetic collections---with documents 
composed of randomly selected passages---do not represent complete and coherent documents.
In \S~\ref{sec:comp_synthetic} we discuss this limitation in detail; nevertheless, we argue that MS MARCO FarRelevant
is a more suitable synthetic benchmark for evaluating text retrieval systems compared to LongEmbed subsets Needle and Passkey \cite{E52024}.

In summary, we provided three types of evidence of positional bias in relevant passages:
strong performance of \emph{FirstP} models on standard collections,
direct estimation of the distribution of relevant passages using substring matching and LLM relevance judges \cite{DBLP:journals/corr/abs-2406-06519}, 
as well as experimentation with the synthetic collection MS MARCO FarRelevant 
where the distribution of relevant passages was not skewed towards the beginning of a document. 
Each experiment provided \emph{imperfect} or \emph{limited} evidence on its own, 
but \emph{together they strongly} support the existence of positional relevance bias.

While our analysis confirms a strong early-position relevance bias across multiple retrieval benchmarks, we acknowledge that this pattern may not generalize to all domains. For example, prior work has shown that in scientific abstracts, both the first and the last sentences tend to be crucial \cite{DBLP:conf/acl/RuchTGA06}. Investigating positional relevance patterns in such domains is an important direction for future work.

In our experiments with Robust04 and MS MARCO, we truncated documents to a maximum of 1431 BERT tokens. 
However, this constraint did not hinder our ability to address key research questions. As detailed in Appendix~\S~\ref{sec:chunk_ablation}, using larger inputs led to only marginal improvements.

\onlyarxiv{
\section{Ethics Statement} 
We believe our study does not pose any ethical concerns. We do not collect any new data with the help of human annotators and we do not use human or animal subjects in our study. Although we do discover
a positional bias in existing retrieval collections, 
we are not aware of any potential risks or harms that can be caused by the
exposure of this bias.

In terms of the environmental impact, our computational requirements are rather modest,
because we only fine-tuned our models rather than trained them from scratch.
These models were also rather small by modern standards.
Except 1B-parameter TinyLLAMA \cite{TinyLLAMA2024}, 
each model has about 100M parameters (see Table~\ref{tab:model_params} for details).
}

\bibliographystyle{acl_natbib}

\appendix

\begin{table*}[!htbp]
\centering
\setlength{\tabcolsep}{2.5pt}
\scriptsize
\begin{threeparttable}
\begin{tabular}{l|l|l|lllllll} \toprule
Ranker  &    \multicolumn{1}{|c}{\textbf{TREC DL}} &
             \multicolumn{1}{|c}{\textbf{NQ}} &
             \multicolumn{7}{|c}{\textbf{BEIR (without NQ)}}              \vspace{0.25em} \\
        &    \multicolumn{1}{|c|}{(2019-2020)} &
              &
             \multicolumn{1}{|c}{Touche} &
             \multicolumn{1}{c}{COVID} &
             \multicolumn{1}{c}{NFC} &
             \multicolumn{1}{c}{DBP} &
             \multicolumn{1}{c}{SciFact} &             
             \multicolumn{1}{c}{SciDocs} &                          
             \multicolumn{1}{l}{average}
       \\ 
       \midrule
      &         
       \multicolumn{1}{|c}{\scriptsize\textbf{NDCG@10}} &
       \multicolumn{1}{|c}{\scriptsize\textbf{NDCG@10}} &
       \multicolumn{6}{|c}{\scriptsize\textbf{NDCG@10}}          
      \\ \midrule

\revision{
BM25 } & \revision{ 0.519 } &\revision{  0.325 } &  \revision{ 0.336 } & \revision{ 0.677 } & \revision{\textbf{0.326} } & \revision{0.339 } & \revision{0.655 } & \revision{\textbf{0.160} } & \revision{0.417\phantom{$(- 99.9\%)$} }\\\midrule

\multicolumn{10}{c}{\textbf{Original MS MARCO training set}} \\ \midrule

MaxP (ELECTRA) & 0.715 & 0.514 & 0.314 & 0.744 & 0.312 & 0.404 & \textbf{0.659} & 0.153 & \textbf{0.477}\phantom{$(- 99.9\%)$} \\
PARADE Attn (ELECTRA) & 0.710 & 0.496 & \textbf{0.356} & 0.743 & 0.266 & 0.386 & 0.606 & 0.140 & 0.463\phantom{$(- 99.9\%)$} \\
PARADE Transf-RAND-L2 (ELECTRA) & 0.703 & 0.474 & 0.293 & 0.728 & 0.297 & 0.380 & 0.658 & 0.154 & 0.461\phantom{$(- 99.9\%)$} \\
CEDR-KNRM & 0.599 & 0.318 & 0.242 & 0.710 & 0.268 & 0.305 & 0.478 & 0.127 & 0.381\phantom{$(- 99.9\%)$} \\

\midrule
\multicolumn{10}{c}{\textbf{Debiased MS MARCO} \cite{DBLP:conf/ecir/HofstatterLAZH21}} \\ \midrule

MaxP (ELECTRA)  & \textbf{0.716 $(+ 0.1\%)$} & \textbf{0.516 $(+ 0.6\%)$} & 0.309 & 0.742 & 0.296$^{a}$ & \textbf{0.408} & 0.646 & 0.150 & 0.473 $(- 0.8\%)$ \\
PARADE Attn (ELECTRA)  & 0.675$^{a}$ $(- 4.9\%)$ & 0.427$^{a}$ $(-14.0\%)$ & 0.335 & 0.742 & 0.226$^{a}$ & 0.353$^{a}$ & 0.522$^{a}$ & 0.122$^{a}$ & 0.425 $(- 8.2\%)$ \\
PARADE Transf-RAND-L2 (ELECTRA)  & 0.706 $(+ 0.4\%)$ & 0.485$^{a}$ $(+ 2.5\%)$ & 0.299 & \textbf{0.753$^{a}$} & 0.288$^{a}$ & 0.393$^{a}$ & 0.643 & 0.153 & 0.465 $(+ 0.9\%)$ \\
CEDR-KNRM  & 0.604 $(+ 0.7\%)$ & 0.353$^{a}$ $(+11.0\%)$ & 0.231 & 0.683$^{a}$ & 0.265 & 0.340$^{a}$ & 0.419$^{a}$ & 0.122$^{a}$ & 0.377 $(- 1.0\%)$ \\

\bottomrule
\end{tabular}
\begin{tablenotes}
  \tablecustomsmall
{\tablecustomsmall  
  We train each model using the original and the debiased MS MARCO. 
  For each model trained on the debiased dataset, we compute a gain (or loss) compared to the same model trained on the original 
  training set. Statistical significance of the differences  are denoted using the superscript superscript \textbf{a} (except TREC DL and TREC COVID that have $p$-value threshold of 0.05, the $p$-value threshold is 0.01). 
 Best numbers are in \textbf{bold}: Results are averaged over three seeds. \\
 }

\end{tablenotes}
\end{threeparttable}
\caption{Debiasing impact on short-document collection performance. The table shows effectiveness of (selected) rankers trained on original and debiased MS MARCO (and tested on short-document collections).
\label{tab:main_debias_short_collect_results}
}
\end{table*}

\begin{algorithm}[H]
\begin{algorithmic}[1]
\Require Document text $D$
\Ensure Debiased document $D'$
\State Remove trailing whitespace from $D$ and find word-boundary positions $B = \{b_1,\dots,b_n\}$
\State Sample $b^\star \sim \mathrm{Unif}(B)$
\State $D_{\text{left}} \gets D[0{:}b^\star]$
\State $D_{\text{right}} \gets D[b^\star{:}]$
\State $D' \gets \text{concat}(D_{\text{right}}, \texttt{" "}, D_{\text{left}})$

\State \Return $D'$
\end{algorithmic}
\caption{Document debiasing via random word-boundary rotation proposed by \citet{DBLP:conf/ecir/HofstatterLAZH21} \label{DebiasAlgo}}
\end{algorithm}

\section{Experimental Addendum: Training/Evaluation Setup, Ablations, and Detailed Results}\label{sec:exper_appendix}

\subsection{Detailed Training and Evaluation Setup}\label{sec:training_setup_expanded}
\subsubsection{General Setup}
Except for LongEmbed \cite{E52024}, we applied a ranker to the output of the first-stage retrieval model,
also known as a candidate generator. 
However, each LongEmbed subset contains only 800 documents, and we re-rank them all without using a first-stage retriever.

Depending on the experiment and the dataset, we used different candidate generators.
For MS MARCO v1 (training and development sets) and Robust04, we used a BM25 ranker \cite{Robertson2004}.
Note that for MS MARCO v1, the ranker was applied to documents expanded using the doc2query approach \cite{Nogueira2019FromDT}.

For MS MARCO v2, we used a hybrid retriever where candidate records were first
produced using a k-NN search and subsequently re-ranked using a linear combination of BM25 scores and the cosine similarity between query and
document embeddings. 
We also used k-NN search with neural embeddings for TREC DL queries.
Embeddings were generated using ANCE \cite{ANCE2021}.

\begin{table}[!tb]
\setlength{\tabcolsep}{1pt}
\scriptsize
\begin{threeparttable}
\begin{tabular}{lll} \toprule
Retriever / Ranker  & zero-shot  & fine-tuned \\        
                    & transferred &           \\
                    \midrule
 Random shuffling of top-100 & 0.052 & 0.052 \\
 \revision{Retriever (BM25)} & 0.207$^{\phantom{b}}$ & 0.207$^{\phantom{b}}$ \\\midrule
 FirstP (BERT) & 0.016$^{b}$ & 0.090$^{b}$ \\
 FirstP (Longformer) & 0.017$^{b}$ & 0.091$^{b}$ \\
 FirstP (ELECTRA) & 0.019$^{b}$ & 0.089$^{b}$ \\
 FirstP (Big-Bird) & 0.021$^{b}$ & 0.089$^{b}$ \\
 FirstP (JINA) & 0.018$^{b}$ & 0.088$^{b}$ \\
 FirstP (MOSAIC) & 0.018$^{b}$ & 0.089$^{b}$ \\
 FirstP (TinyLLAMA) & 0.020$^{b}$ & 0.079$^{b}$ \\
 FirstP (E5-4K)        & 0.015$^{ab}$ & -- \\
 \midrule
 AvgP & 0.154$^{ab}$ {  $(-48.1\%)$} & 0.365$^{ab}$ {  $(+11.4\%)$} \\\midrule
 MaxP & 0.297$^{b}$ & 0.328$^{b}$ \\
 MaxP (ELECTRA) & 0.328$^{b}$ & 0.349$^{b}$ \\
 MaxP (DEBERTA) & 0.298$^{b}$ & 0.332$^{b}$ \\
 SumP & 0.211$^{ab}$ {  $(-28.8\%)$} & 0.327$^{b}$ {  $(- 0.4\%)$} \\\midrule
 CEDR-DRMM & 0.157$^{ab}$ {  $(-47.3\%)$} & 0.372$^{ab}$ {  $(+13.3\%)$} \\
 CEDR-KNRM & 0.055$^{ab}$ {  $(-81.5\%)$} & 0.382$^{a\phantom{b}}$ {  $(+16.4\%)$} \\
 CEDR-PACRR & 0.209$^{ab}$ {  $(-29.6\%)$} & 0.393$^{a\phantom{b}}$ {  $(+19.9\%)$} \\\midrule
 Neural Model1 & 0.085$^{ab}$ {  $(-71.3\%)$} & 0.396$^{a\phantom{b}}$ {  $(+20.6\%)$} \\\midrule
 PARADE Attn & 0.300$^{b}$ {  $(+ 1.0\%)$} & 0.337$^{b}$ {  $(+ 2.8\%)$} \\
 PARADE Attn (ELECTRA) & \textbf{0.338$^{b}$ {  $(+ 3.3\%)$}} & 0.354$^{b}$ {  $(+ 1.6\%)$} \\
 PARADE Attn (DEBERTA) & 0.307$^{b}$ {  $(+ 3.2\%)$} & 0.343$^{b}$ {  $(+ 3.4\%)$} \\
 PARADE Avg & 0.274$^{ab}$ {  $(- 7.6\%)$} & 0.322$^{b}$ {  $(- 1.7\%)$} \\
 PARADE Max & 0.291$^{b}$ {  $(- 2.1\%)$} & 0.330$^{b}$ {  $(+ 0.6\%)$} \\\midrule
 PARADE Transf-RAND-L2 & 0.243$^{a\phantom{b}}$ {  $(-18.2\%)$} & 0.419$^{ab}$ {  $(+27.7\%)$} \\
 P. Transf-RAND-L2 (ELECTRA) & 0.229$^{a\phantom{b}}$ {  $(-30.2\%)$} & \textbf{0.432$^{ab}$ {  $(+23.8\%)$}} \\
 PARADE Transf-PRETR-L6 & 0.267$^{ab}$ {  $(-10.0\%)$} & 0.413$^{a\phantom{b}}$ {  $(+26.0\%)$} \\
 P. Transf-PRETR-LATEIR-L6 & 0.244$^{a\phantom{b}}$ {  $(-18.0\%)$} & 0.358$^{ab}$ {  $(+ 9.2\%)$} \\\midrule
 LongP (Longformer) & 0.233$^{a\phantom{b}}$ {  $(-21.7\%)$} & 0.399$^{a\phantom{b}}$ {  $(+21.7\%)$} \\
 LongP (Big-Bird) & 0.126$^{ab}$ {  $(-57.4\%)$} & 0.401$^{a\phantom{b}}$ {  $(+22.1\%)$} \\
 LongP (JINA) & 0.069$^{ab}$ {  $(-76.9\%)$} & 0.372$^{ab}$ {  $(+13.4\%)$} \\
 LongP (MOSAIC) & 0.120$^{ab}$ {  $(-59.6\%)$} & 0.397$^{a\phantom{b}}$ {  $(+21.2\%)$} \\
LongP (TinyLLAMA) & 0.078$^{ab}$ $(-73.6\%)$ & 0.397$^{a\phantom{b}}$ $(+21.1\%)$ \\
LongP (E5-4K) & 0.057$^{ab}$ $(-80.7\%)$ & N/A (zero-shot only) \\
LongP RankGPT (GPT-4o-mini) & 0.043$^{b}$ & N/A (zero-shot only) \\
LongP RankGPT (Claude-3-haiku) & 0.051$^{b}$ & N/A (zero-shot only) \\
\bottomrule
\end{tabular}
\begin{tablenotes}
  \tablecustomsmall
  In each column we show a relative gain over models' respective \emph{MaxP} baseline. For \emph{LongP} models, the gain is over \emph{MaxP} (BERT). \\
  Statistically significant differences from a respective \emph{MaxP} baseline are denoted with the superscript \textbf{a}. \\
  Statistical significant differences with respect to \emph{Longformer} are denoted with the superscript  \textbf{b} ($p$-value $<0.01$). 
\end{tablenotes}
\end{threeparttable}
\caption{
\revision{
Comparison between long-document models and respective \emph{FirstP} (truncation) baselines. Results on MS MARCO FarRelevant.} \label{tab:main_results_synthetic}}
\end{table}

Depending on the collection we computed a subset of the following metrics:
the mean reciprocal rank (MRR), the normalized discounted cumulative gain at rank $k$ (NDCG@k) \cite{DBLP:journals/tois/JarvelinK02}, 
the mean average precision (MAP), and precision at rank (P@k), $k \in \{10,20\}$. 
Due to space constraints,
we included results with MAP and P@K only in the Appendix (see \S~\ref{sec:detailed_exper_results}).
Note that for test sets with sparse labels (MS MARCO development set and MS MARCO FarRelevant) we computed only MRR.

All experiments were carried out using the FlexNeuART framework \cite{FlexNeuART},
which employed Lucene and \nmslib\   to provide retrieval capabilities.
Deep learning support was provided via PyTorch \cite{paszke2019pytorch} and HuggingFace Transformers library  \cite{Wolf2019HuggingFacesTS}.
The instructions to reproduce our key results are publicly available.\footnote{\url{https://github.com/searchivarius/long_doc_rank_model_analysis_v2/}}

\subsubsection{Model Training}
A ranker was trained to distinguish between positive examples (known relevant documents)
and hard negative examples (documents not marked as relevant)
sampled from the set of top-$k$ candidates returned by the candidate generator.
Based on preliminary experiments, we chose $k=100$ for MS MARCO and MS MARCO FarRelevant.
For Robust04 we used $k=1000$.

Each model was trained using \emph{three} seeds.
To compute statistical significance, we averaged query-specific metric values over these seeds. 
All models except MOSAIC were trained using half-precision.
MOSAIC models were trained using full-precision. 
MOSAIC training was unstable (even though we used the full precision) and 
often resulted in close-to-zero performance. 
For this reason we continued training with \emph{more} seeds until we obtained three models with reasonable performance. This seed selection strategy could potentially have biased (up) MOSAIC results.

\begin{table*}[tb]
    \centering
\scriptsize
\begin{threeparttable}
\setlength{\tabcolsep}{0.25em}
    \begin{tabular}{l|l|lll|ll}\toprule
\multicolumn{1}{c|}{Model} & \multicolumn{1}{c}{MS MARCO}  & \multicolumn{3}{|c|}{TREC DL }   & \multicolumn{2}{|c}{Robust04}   \\
      &  \multicolumn{1}{c|}{dev}         &  \multicolumn{1}{c}{2019} & \multicolumn{1}{c}{2020} & \multicolumn{1}{c|}{2021}  &   \multicolumn{1}{c}{title} & \multicolumn{1}{c}{description}    \\ \midrule
      & \multicolumn{1}{c}{\scriptsize\textbf{MRR}} &  \multicolumn{3}{|c|}{\scriptsize\textbf{NDCG@10}}  & \multicolumn{2}{|c}{\scriptsize\textbf{NDCG@20}}  \\ \midrule
\revision{ BM25 } & 
\revision{ 0.274 } &
\revision{ 0.548 } & \revision{ 0.538 } & \revision{ 0.549 } &
\revision{0.428} & \revision{0.402}
\\ \midrule
\multicolumn{7}{c}{\textbf{Prior work (FirstP, MaxP), Zhang et al.~\cite{DBLP:conf/ecir/ZhangYL21} }} \\ \midrule
FirstP (BERT)        & --  &  -- & -- & --      & 0.449 & 0.510 \\
MaxP (BERT)         & --  &  -- & -- & --      & 0.477 { $(+6.2\%)$}& 0.530 { $(+3.9\%)$}  \\
MaxP (ELECTRA)        & --  &  -- & -- & --      & 0.523 & 0.574 \\\midrule
\multicolumn{7}{c}{\textbf{Prior work (PARADE) Li et al.~\cite{Parade2020} }} \\\midrule
PARADE Attn (ELECTRA)               & --  &  -- & -- & --      &  0.527  &  0.587  \\
PARADE Max (ELECTRA)                & --  &  0.679 & 0.613 & --   & 0.544   &  0.602  \\
PARADE Transf-RAND (ELECTRA)        & --  &  0.650 & 0.601 & --   & \textbf{0.566}  &  0.613 \\\midrule
\multicolumn{7}{c}{\textbf{Our results}} \\ \midrule

FirstP (BERT) & 0.394 & 0.631 & 0.598 & 0.660 & 0.475 & 0.527 \\
MaxP (BERT) & 0.392 { $(- 0.4\%)$} & 0.648 { $(+ 2.6\%)$} & 0.615 { $(+ 2.8\%)$} & 0.665 { $(+ 0.8\%)$} & 0.488$^{a}$ { $(+ 2.6\%)$} & 0.544$^{a}$ { $(+ 3.3\%)$} \\
PARADE Attn & 0.416$^{a}$ { $(+ 5.5\%)$} & 0.647 { $(+ 2.5\%)$} & 0.626$^{a}$ { $(+ 4.6\%)$} & 0.677 { $(+ 2.5\%)$} & 0.503$^{a}$ { $(+ 5.7\%)$} & 0.556$^{a}$ { $(+ 5.6\%)$} \\\midrule
FirstP (ELECTRA) & 0.417 & 0.652 & 0.642 & 0.686 & 0.492 & 0.552 \\
MaxP (ELECTRA) & 0.414 { $(- 0.6\%)$} & 0.659 { $(+ 1.0\%)$} & 0.630 { $(- 1.9\%)$} & 0.683 { $(- 0.5\%)$} & 0.502 { $(+ 2.0\%)$} & 0.563 { $(+ 2.1\%)$} \\
PARADE Attn (ELECTRA) & \textbf{0.431$^{a}$ { $(+ 3.3\%)$}} & 0.675$^{a}$ { $(+ 3.5\%)$} & 0.653 { $(+ 1.8\%)$} & 0.705 { $(+ 2.8\%)$} & 0.523$^{a}$ { $(+ 6.4\%)$} & 0.581$^{a}$ { $(+ 5.3\%)$} \\\midrule
FirstP (DEBERTA) & 0.415 & 0.675 & 0.629 & 0.702 & 0.534 & 0.596 \\
MaxP (DEBERTA) & 0.402 { $(- 3.2\%)$} & 0.679 { $(+ 0.6\%)$} & 0.620 { $(- 1.4\%)$} & 0.705 { $(+ 0.4\%)$} & 0.535 { $(+ 0.2\%)$} & 0.609 { $(+ 2.2\%)$} \\
PARADE Attn (DEBERTA) & 0.422$^{a}$ { $(+ 1.6\%)$} & \textbf{0.685 { $(+ 1.4\%)$}} & \textbf{0.659$^{a}$ { $(+ 4.8\%)$}} & \textbf{0.713 { $(+ 1.4\%)$}} & 0.549$^{a}$ { $(+ 2.9\%)$} & \textbf{0.615}$^{a}$ { $(+ 3.2\%)$} \\\midrule
FirstP (Longformer) & 0.404 & 0.657 & 0.616 & 0.654 & 0.483 & 0.540 \\
LongP (Longformer) & 0.412$^{a}$ { $(+ 1.9\%)$} & 0.676$^{a}$ { $(+ 2.9\%)$} & 0.628 { $(+ 2.0\%)$} & 0.693$^{a}$ { $(+ 6.0\%)$} & 0.500$^{a}$ { $(+ 3.6\%)$} & 0.568$^{a}$ { $(+ 5.1\%)$} \\\midrule
FirstP (Big-Bird) & 0.408 & 0.663 & 0.620 & 0.679 & 0.507 & 0.560 \\
LongP (Big-Bird) & 0.397$^{a}$ { $(- 2.9\%)$} & 0.655 { $(- 1.1\%)$} & 0.618 { $(- 0.3\%)$} & 0.675 { $(- 0.5\%)$} & 0.452$^{a}$ { $(-10.9\%)$} & 0.477$^{a}$ { $(-14.9\%)$} \\\midrule
FirstP (JINA) & 0.422 & 0.658 & 0.618 & 0.679 & 0.488 & 0.532 \\
LongP (JINA) & 0.416$^{a}$ { $(- 1.5\%)$} & 0.670$^{a}$ { $(+ 1.8\%)$} & 0.632 { $(+ 2.1\%)$} & 0.689 { $(+ 1.4\%)$} & 0.503$^{a}$ { $(+ 2.9\%)$} & 0.558$^{a}$ { $(+ 4.9\%)$} \\\midrule
FirstP (MOSAIC) & 0.423 & 0.654 & 0.607 & 0.662 & 0.453 & 0.538 \\
LongP (MOSAIC) & 0.421 { $(- 0.4\%)$} & 0.660 { $(+ 0.9\%)$} & 0.630$^{a}$ { $(+ 3.7\%)$} & 0.694$^{a}$ { $(+ 4.9\%)$} & 0.456 { $(+ 0.6\%)$} & 0.570$^{a}$ { $(+ 6.0\%)$} \\
\bottomrule
      
\end{tabular}

    \begin{tablenotes}
        \tablecustomsmall
  In each column we show a relative gain over model's respective \emph{FirstP} baseline: The last column shows the average relative gain over \emph{FirstP}. Best numbers are in \textbf{bold}: Our results  are averaged over three seeds. Prior-art seed strategy is unknown. \\
  Statistical significant differences with respect to this baseline are denoted using the superscript superscript \textbf{a}. 
  $p$-value threshold is 0.01 for an MS MARCO development collection and 0.05 otherwise.
    \end{tablenotes}
\end{threeparttable}
    \caption{
    \revision{Comparison between long-document models and respective \emph{FirstP} (truncation) baselines for several backbone Transformer models as well as comparison to \emph{prior art}. Results for MS MARCO, TREC DL, and Robust04.}\label{tab:first_p_backbone_ablation}
    }
\end{table*}

All MS MARCO models were trained from scratch. Afterward, these models were fine-tuned on Robust04. Note that except for the 
aggregator Transformers (see \ref{sec:split_p_desc} for architectural details) and TinyLLAMA, we use a \emph{base}, i.e.,  a 12-layer Transformer \cite{vaswani2017attention} models. TinyLLAMA has 22 layers and about 1B parameters.

BERT-base is more practical than a 24-layer BERT-large and performed on par with BERT-large on MS MARCO Documents and Robust04 \cite{HofstatterZH20,Parade2020}.
In our own preliminary experiments, we observed that the 24-layer BERT-large model performed much better
on the MS MARCO Passage collection, 
but we were not able to outperform 12-layer BERT-base models on the MS MARCO Documents collection.
Note that Longformer \cite{longformer2020}, BigBird \cite{BigBird2020}, 
 DEBERTA-base \cite{he2021debertav3}, JINA \cite{gunther2023jina}, and MOSAIC \cite{portes2023mosaicbert} all have 12 layers, but a larger embedding matrix compared to BERT-base. 

One training epoch consisted in iterating over all queries and sampling one positive and one negative example
with a subsequent computation of a pairwise margin loss.
We used the mini-batch size one with gradient accumulation over 16 steps.
The learning rates are provided in the model configuration files (in the code repository).\ourcode\  
We used the AdamW optimizer \cite{loshchilov2017decoupled} 
and  a constant learning  rate with a 20\% linear warm-up \cite{Mosbach2020-kn}.

We have learned that---unlike neural \emph{retrievers}---cross-encoding rankers \cite{nogueira2019passage} are relatively insensitive to learning rates, their schedules, and the choice of loss functions. 
We were sometimes able to achieve better results using multiple negatives per query
and a listwise margin loss (or cross-entropy).
However, the gains were small and not consistent compared to a simple pairwise margin loss used in our work 
(in fact, using a listwise loss function sometimes led to overfitting).
Note again that this is different from neural \emph{retrievers}
where training is sometimes difficult without using a listwise loss 
and/or batch-negatives \cite{DPR2020,ANCE2021,RocketQA2021,Zerveas2021CODERAE,SPLADEv2}.

For MS MARCO, all models except PARADE-Transf-Pretr-LATEIR-L6 and PARADE-Transf-RAND-L2 were trained for one epoch:
Further training did not improve (and sometimes degraded) accuracy.
However, PARADE-Transf-RAND-L2 and PARADE-Transf-Pretr-LATEIR-L6 required two-to-three epochs to reach the maximum accuracy.
For training with the debiased MS MARCO, we used only one epoch.
In the case of Robust04, each model was fine-tuned for 100 epochs, but 
all epochs were short, so the overall training and evaluation time was comparable to that of fine-tuning on MS MARCO for a single epoch.

Except 1B-parameter TinyLLAMA \cite{TinyLLAMA2024}, 
each model has about 100M parameters (see Table~\ref{tab:model_params} for details).
Despite training and testing 20+ models with three seeds, 
we estimate to have spent only about 6400 GPU hours for our main experiments.
96\% of the time we used NVIDIA A10 (or similarly-powerful) RTX 3090 GPUs
and 4\% of the time we used NVIDIA A6000. 

We believe this is roughly equivalent to pre-training a single
1B-parameter TinyLLAMA model with a next-token prediction objective, 
which required about 3400 GPU hours using a more powerful NVIDIA A100. 
This, in turn, this is only a tiny fraction  of compute required to train LLAMA2 models (2\% compared to a 7B LLAMA2 model).\footnote{\url{https://github.com/microsoft/Llama-2-Onnx/blob/main/MODEL-CARD-META-LLAMA-2.md}}

From our experience, models trained on MS MARCO v2 performed  worse on 
TREC 2021 queries compared to models trained on MS MARCO v1.
This may indicate that models \emph{implicitly} learned to distinguish between
original MS MARCO v1 documents and newly added ones (which did not have positive judgements in the training sets).
As a result, these models are biased and tend to rank  newly added documents
poorly even when they are considered to be relevant by NIST assessors.
For this reason, we used MS MARCO~v2 data \emph{only} in a zero-shot transfer mode.
To this end, ranking models trained on MS MARCO v1 were evaluated using TREC DL 2021 queries.

\begin{table*}[tb]
\tablecustomsmall
\setlength{\tabcolsep}{0.5em}
\centering
\begin{tabular}{l|l|l|ll|l}\toprule
Retriever / Ranker  &  \multicolumn{1}{c}{MS MARCO}  & \multicolumn{1}{|c|}{TREC DL}   & \multicolumn{2}{|c|}{Robust04}    &  \multicolumn{1}{c}{Avg. gain } \\
      &  \multicolumn{1}{|c|}{dev}         &  \multicolumn{1}{c|}{(2019-2021)} & \multicolumn{1}{c}{title} & \multicolumn{1}{c|}{description}    & \multicolumn{1}{c}{Over FirstP} \\\midrule
      & \multicolumn{1}{|c|}{\scriptsize\textbf{MRR}} &  
      \multicolumn{1}{|c|}{\scriptsize\textbf{NDCG@10}} &  \multicolumn{2}{c|}{\scriptsize\textbf{NDCG@20}} &   \\\midrule
\revision{BM25} & \revision{0.274} & \revision{0.545} & \revision{0.428} & \revision{0.402} & -- \\\midrule
Retriever (if different from BM25) & 0.312 & 0.629 & -- & -- & -- \\\midrule
PARADE Attn (1 chunk) & 0.401 & 0.637 & 0.476 & 0.527 & --  \\
PARADE Attn (2 chunks) & 0.408$^{a}$ $(+ 1.8\%)$ & 0.653$^{a}$ $(+ 2.7\%)$ & 0.499$^{a}$ $(+ 4.9\%)$ & 0.544$^{a}$ $(+ 3.3\%)$ & +$3.2\%$ \\
PARADE Attn (3 chunks) & 0.406$^{a}$ $(+ 1.3\%)$ & 0.648$^{a}$ $(+ 1.7\%)$ & \textbf{0.505$^{a}$ $(+ 6.1\%)$} & 0.557$^{a}$ $(+ 5.7\%)$ & +$3.7\%$ \\
PARADE Attn (4 chunks) & \textbf{0.412$^{a}$ $(+ 2.9\%)$} & \textbf{0.654$^{a}$ $(+ 2.7\%)$} & 0.504$^{a}$ $(+ 5.9\%)$ & \textbf{0.558$^{a}$ $(+ 5.9\%)$} & \textbf{+4.3\%} \\
PARADE Attn (5 chunks) & 0.409$^{a}$ $(+ 2.0\%)$ & 0.652$^{a}$ $(+ 2.4\%)$ & 0.502$^{a}$ $(+ 5.6\%)$ & 0.556$^{a}$ $(+ 5.5\%)$ & +$3.9\%$ \\
PARADE Attn (6 chunks) & 0.411$^{a}$ $(+ 2.4\%)$ & 0.653$^{a}$ $(+ 2.6\%)$ & 0.504$^{a}$ $(+ 5.9\%)$ & 0.554$^{a}$ $(+ 5.2\%)$ & +$4.0\%$ \\\bottomrule
\end{tabular}
\caption{Effectiveness of the PARADE Attention model for different input truncation thresholds. \revision{Results for MS MARCO, TREC DL, and Robust04.}\label{tab:chunk_ablation}}
\end{table*}

\subsubsection{Miscellaneous Notes}\label{sec:exper_setup_misc}

To enable efficient training and evaluation of the large number of models,
both Robust04 and \emph{original} MS MARCO documents were truncated to have at most 1431 BERT tokens. 
Thus, for \emph{SplitP} approaches,
queries were padded to 32 BERT tokens and long documents were split into at most three chunks, each containing 477 document tokens.
However, for training on \emph{debiased} MS MARCO, the truncation threshold was much higher: 8109 tokens.

In \S~\ref{sec:chunk_ablation} (see Table~\ref{tab:chunk_ablation}) we  show 
that for our top-performing model PARADE Attention 
\cite{Parade2020} using a larger number of chunks only marginally improves outcomes.
Depending on a dataset, the highest accuracy is achieved using either three or four chunks. 

We evaluated over 20 models, but we had to exclude two LongT5 variants \cite{guo2022longt5}
and RoFormer (with ROPE embeddings) \cite{RoFORMER} due to poor convergence and/or
crashes. 
Specifically, even after 10 epochs of training LongT5 models were $\approx 10\%$ less accurate than BERT-base \emph{FirstP} trained for one epoch.
Given the uncertainty regarding the possible convergence of models as well as the need to train
these for three epochs, we had to abandon this experiment as overly expensive.
RoFormer models were failing due to CUDA errors when the context length exceeded 512: 
We were not able to resolve this issue.

For bias-mitigation experiments, for several reasons, we used only a subset of models. First, we had to exclude all \emph{LongP} models since none of them supported a context longer than 8192 tokens. 
In contrast, we trained our chunk-and-aggregate models on documents limited to 8109 tokens 
and then extrapolated their use to rank documents up to 32768 tokens long. 

Second, we chose representative models with vastly different generalization properties. 
MaxP and PARADE Attention models performed well on MS MARCO FarRelevant in the zero-shot setting, but did not benefit much from in-domain fine-tuning. PARADE Transformer MRR dropped from 0.433 to 0.229 in the zero-shot setting, but increased up to 0.432 after in-domain fine-tuning. 
CEDR-KNRM also benefited a lot from fine-tuning on MS MARCO FarRelevant, but its zero-shot performance was at the level of the random-baseline.

\section{Additional Experimental Results}\label{sec:detailed_exper_results}

This section presents supplementary experimental results.
In particular, we compute additional effectiveness metrics for MS MARCO, TREC DL, and Robust04.
MS MARCO and TREC DL results are shown in Table~\ref{tab:msmarco}.
Results for Robust04  datasets are presented in Table~\ref{tab:robust04}.
Evaluation results for rankers trained on the debiased MS MARCO and tested on short-document collections are shown in Table \ref{tab:main_debias_short_collect_results}.
Furthermore, we provide detailed results for MS MARCO FarRelevant in Table~\ref{tab:main_results_synthetic}.

To justify the use of BERT-base \cite{devlin2018bert} as the main model backbone,
we carry out an ablation study in \S~\ref{sec:repro_and_backbone}.
Moreover, in this section we also compare our results to prior art,
with the objective to boost trustworthiness of our experiments.

This is followed by 
a miscellaneous ablations subsection (\S~\ref{sec:misc_ablations}), 
where we carry out an efficiency evaluation and assess
if truncating input to at most 1431 BERT tokens affected our experimental outcomes.

\subsection{Backbone Selection for \emph{SplitP} Models and Prior Art Comparison}\label{sec:repro_and_backbone}
\subsubsection{Choice of a Backbone}

To understand if using BERT-base as a backbone model for various \emph{SplitP} (i.e., chunk-and-aggregate) 
approaches diminished  models' ability  to process and understand long contexts,
we carried out a focused comparison of several backbone models, including ELECTRA \cite{ELECTRA2020} and DEBERTA \cite{he2021debertav3}.

To this end, we used two methods: PARADE \cite{Parade2020} Attention and \emph{MaxP}.
PARADE Attention model achieved the largest average gain over \emph{FirstP} in our main
experiments (see Table~\ref{tab:main_results_real_short}),
whereas \emph{MaxP} models were extensively benchmarked in the past \cite{Parade2020,DaiC19,DBLP:conf/ecir/ZhangYL21}.

Although prior work found ELECTRA to be a better backbone model in terms of \emph{absolute} accuracy \cite{Parade2020,DBLP:conf/ecir/ZhangYL21},
we found no systematic evaluation of the relationship between a backbone model and gains achievable over \emph{FirstP}. 

Results in Tables \ref{tab:main_results_real_short} and \ref{tab:first_p_backbone_ablation}  confirm the overall superiority of both ELECTRA and DEBERTA over BERT-base. In that, DEBERTA models are nearly always more effective compared to ELECTRA models with biggest differences on Robust04.
However, their \emph{relative} effectiveness with respect to their respective \emph{FirstP} baselines does not exceed that of BERT-base.

The same is true for \emph{LongP} models. Except Longformer, they performed equally or worse compared to \emph{FirstP} on 8 test sets out of 18.
Moreover, all \emph{LongP} models achieved lower average gains over \emph{FirstP} (see the last column in Table~\ref{tab:main_results_real_short}).
We conclude that to measure capabilities of chunk-and-aggregate or long-context models to understand and incorporate long documents, 
it appears to be  \emph{beneficial} to use BERT-base instead of \mbox{ELECTRA} or DEBERTA.

Finally, we would like to note that  on standard benchmarks  Big-Bird's \cite{BigBird2020}  \emph{FirstP}
version \emph{always} outperformed its \emph{LongP} version (sometimes by as much as 10-15\%), which seems to be puzzling.
We noticed, however, that for shorter inputs, the model turns off sparse attention and prints the respective warning. Thus, we hypothesize that it is the use of sparse attention that causes this degradation. In contrast, the sparse attention implementation of the Longformer \cite{longformer2020} does not exhibit such a degradation (although with Longformer, not all attention is sparse: query-to-document attention is full).
Although Big-Bird underperforms on standard benchmarks, after fine-tuning, it still matches the Longformer accuracy on MS MARCO FarRelevant (see Table~\ref{tab:main_results_synthetic}).

\subsubsection{Comparison to Prior Art}
We also use Table~\ref{tab:first_p_backbone_ablation} to compare with prior art.
We generally reproduce prior art, in particular, experiments by \citealt{Parade2020}, who invented PARADE models. 
Our ELECTRA-based models achieve higher NDCG@10 on TREC DL 2019-2020 and PARADE Attention models
come very close, but they are about 3-5\% worse compared to their PARADE Transformer on Robust04.
At the same time, our DEBERTA-based PARADE Attention model achieves similar NDCG@20 scores.

Note that one should not expect identical results due to differences in training regimes and candidate generators. In particular, in the case of Robust04, \citealt{Parade2020} used RM3,
which is BM25 with pseudo-relevance feedback   \cite{RM3}.
On Robust04, it is more  effective than BM25 \cite{Robertson2004}, which we used as the first-stage retriever.

Another important comparison point  is Robust04 results by \citealt{DBLP:conf/ecir/ZhangYL21} who were able to reproduce
original \emph{MaxP} results by \citet{DaiC19},
who used BERT-base as a backbone. 
In addition, they experimented with ELECTRA models ``pre-finetuned'' on MS MARCO. 
When comparing results for the BERT-base backbone, \citet{DBLP:conf/ecir/ZhangYL21} have
the maximum relative gain of 6.6\% compared to ours 3.3\%. However, 
in absolute terms we got higher NDCG@20 for both \emph{FirstP} and \emph{MaxP}.
Their MaxP (ELECTRA) has comparable performance to ours on TREC DL 2019-2020, but it is 2-4\%
better on Robust04. In turn, our MaxP (DEBERTA) is better by 2-6\%.
Although we do not always match prior art using the same backbone models,
we generally match or outperform prior results, which, we believe, 
boosts the trustworthiness of our experiments.

\begin{figure}[!htbp]
    \centering 
        \includegraphics[width=0.5\textwidth]{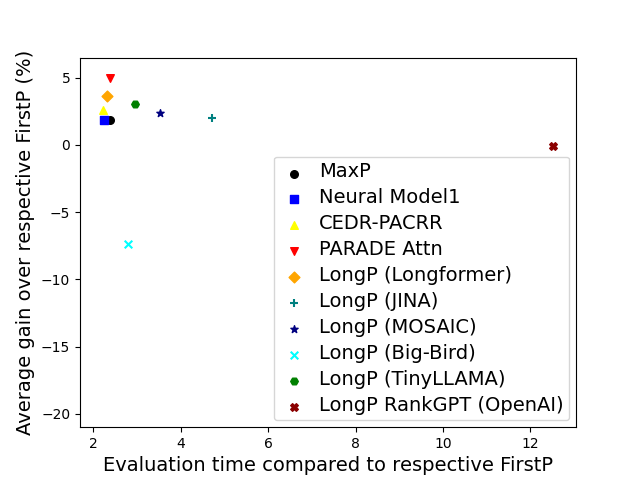}
    \caption{
    \revision{Efficiency of long-document models vs respective (truncation) \emph{FirstP} baselines.}
    The figure shows an average relative gain (in \%) vs. relative increase in run-time compared to
    \emph{respective} \emph{FirstP} baselines on MS MARCO, TREC DL 2019-2021, and Robust04 (for a representative subset of models). 
    Except LongP RankGPT, \emph{LongP} models truncate documents to be at most 1431 tokens. There is no truncation for RankGPT.
    \label{fig:tradeoff_plot_rel}}
\end{figure}

\subsection{Miscellaneous Ablations\label{sec:misc_ablations}}
\subsubsection{Varying the Number of Chunks}\label{sec:chunk_ablation}

To understand if truncating input to at most 1431 BERT tokens biased
 our experimental outcomes, we carried out an ablation study where one
of the top-performing models was trained and evaluated using inputs 
of varying maximum lengths.
To this end, we used PARADE Attention with the number of input chunks varying from
one to six. 
For each setting, the same number of chunks was used during training and evaluation,
i.e., we had to carry out six experiments.
Similar to our main experiments, we trained each model using 
three seeds. We carried out this ablation experiment using our MS MARCO
and Robust04 datasets.

The results are presented in Table~\ref{tab:chunk_ablation}.
We can see that---depending on the dataset---three or four input chunks are optimal.
However, compared to using three chunks,
the additional gains over the \emph{FirstP} baseline are at most 0.6\%
when averaged over all test sets.

\citealt{luyu2022moresplus} carried out a similar ablation using ClueWeb09.
Increasing the number of input chunks from three to six lead to only about 2.3\%
relative improvement in NDCG@20. 
However, even this modest gain could have been slightly inflated 
due to model not being trained \emph{directly} on shorter inputs.
Indeed, truncation of an input
for a six-chunk model to one chunk is potentially less effective than training
and evaluating the model using one-chunk data.

\subsubsection{Efficiency Evaluation}\label{sec:efficiency}

We conducted an efficiency evaluation because long-document models, despite their potential to capture more context, 
often incur substantial computational overhead compared to their \emph{FirstP} counterparts, raising questions about their practical utility.

From the efficiency-effectiveness plot in Fig.~\ref{fig:tradeoff_plot_rel}, we can see that, indeed,  all long-document models are at least 2$\times$ slower than respective \emph{FirstP} baselines. The biggest average gain of merely 5\% is achieved by the PARADE Attn model 
(with a BERT-base backbone) at the cost of being 2.5$\times$ slower than its \emph{FirstP} baseline.

All \emph{LongP} models are even slower and show less improvement. 
Given such small benefits at the cost of a substantial slow-down, 
one could question practicality of such models and suggest using \emph{FirstP} variants instead.

\section{Additional Dataset Details}
\subsection{Summary Dataset Statistics}
\label{sec:data_details}

\begin{table}[!tbp]
\centering \tablecustomsmall
\setlength{\tabcolsep}{0.25em}
\begin{tabular}{@{}lccc@{}}
\toprule
 & \multicolumn{1}{l}{\# of queries} & 
 \multicolumn{1}{l}{\begin{tabular}[c]{@{}l@{}}avg. \# of \\ BERT tokens\end{tabular}} & 
 \multicolumn{1}{l}{\begin{tabular}[c]{@{}l@{}}avg. \# of \\ pos. judgements\end{tabular}} \\ \midrule
\multicolumn{4}{c}{MS MARCO doc. v1} \\ \midrule
MS MARCO doc. train & 352K & 7 & 1 \\
MS MARCO doc. dev & 5193 & 7 & 1 \\
TREC DL 2019 & 43 & 7 & 153.4 \\
TREC DL 2020 & 45 & 7.4 & 39.3 \\\midrule
\multicolumn{4}{c}{MS MARCO v2} \\\midrule
TREC DL 2021 & 57 & 9.8 & 143.9 \\\midrule
\multicolumn{4}{c}{Robust04} \\\midrule
title & 250 & 3.6 & 69.6 \\
description & 250 & 18.7 & 69.6 \\ \midrule
\multicolumn{4}{c}{MS MARCO FarRelevant}\\\midrule
train & 50K & 7.0 & 1 \\
test & 1K & 7.0 & 1 \\\midrule
\multicolumn{4}{c}{LongEmbed}\\\midrule
Needle & 800 & 13.0 & 1 \\
Passkey & 800 & 9.7 & 1 \\\midrule
\multicolumn{4}{c}{MS MARCO pass. v1} \\ \midrule
TREC DL 2019 & 43 & 7 & 95.4 \\
TREC DL 2020 & 54 & 7.2 & 66.8 \\\midrule
\multicolumn{4}{c}{BEIR}\\\midrule
Natural Questions & 3452 & 9.9 & 1.2 \\
Touche & 49 & 19 & 7.9 \\
TREC COVID & 50 & 493.5 & 16 \\
NFCorpus (NFC) & 323 & 38.2 & 5 \\
DBP & 400 & 38.2 & 6.5 \\
SciFact & 300 & 1.1 & 20.8 \\
SciDocs & 1000 & 4.9 & 13.1 \\
\bottomrule
\end{tabular}
\caption{Query Statistics}\label{tab:queries}
\end{table}

\begin{table}[!htbp]
\centering \tablecustomsmall
\begin{threeparttable}
\begin{tabular}{@{}lll@{}}
\toprule
data set & \# of documents & average \# of  \\ 
         &                 & BERT tokens \\
         &                 & per document \\\midrule
\multicolumn{3}{c}{Long-document collections} \\\midrule
MS MARCO doc. v1 & 3.2M & 1.4K  \\
MS MARCO doc. v2 & 12M & 2K \\
Robust04 & 0.5M & 0.6K \\
MS MARCO FarRelevant & 0.53M & 1.1K \\
Needle (LongEmbed) & 0.8K &  variable-length  \\
Passkey (LongEmbed) & 0.8K & variable-length  \\\midrule
\multicolumn{3}{c}{Short-document collections} \\\midrule
MS MARCO pass. v1        & 8.8M & 75  \\\midrule
\multicolumn{3}{c}{Short-document collections: BEIR subset} \\\midrule
Natural Questions  & 2.7M & 107 \\
Touche & 382.5K & 382 \\
TREC COVID & 171.3K & 243 \\
NFCorpus (NFC) & 3.6K & 355 \\
DBP & 4.6M & 71 \\
SciFact & 5.2K & 335 \\
SciDocs & 25.7K & 234 \\
\bottomrule
\end{tabular}
  \begin{tablenotes}
        \tablecustomsmall
    \textsc{LongEmbed} subsets each have 16 subsets of documents
    whose lengths vary from (approximately) 256 to 32768 tokens.
  \end{tablenotes}
\end{threeparttable}
\caption{Document Statistics}\label{tab:docs}
\end{table}

Summary query and dataset statistics are shown in Tables~\ref{tab:queries} and \ref{tab:docs}.
In the case of BEIR, we use seven subsets, which include: 
\begin{itemize}
    \item  Natural Questions \cite{NaturalQuestions2019};
    \item Touch{\'e} 2020 \cite{touche2020};
    \item  TREC COVID \cite{treccovid};
    \item NFCorpus  \cite{nfc};
    \item DBPedia-Entity-v2 \cite{DBLP:conf/sigir/HasibiNXBBKC17};
    \item SciFact \cite{scifact}; 
    \item SciDocs \cite{scidocs}.
\end{itemize}
Please, note that in the case of MS MARCO FarRelevant, 
we created about 500K training and 7K testing queries,
but to reduce experimentation cost we ended up using only 50K and 1K queries, respectively.

\begin{table*}
\tablecustomsmall
\begin{threeparttable}
\setlength\columnsep{0.5em}
\begin{tabular}{l|l|lll}
 \toprule
\multicolumn{1}{c|}{Model} & \multicolumn{1}{|c|}{MS MARCO}  & \multicolumn{3}{|c}{TREC DL }  \\
 & \multicolumn{1}{|c|}{dev} &  \multicolumn{3}{|c}{2019-2021} \\\midrule
  & \multicolumn{1}{|l|}{MRR} & \multicolumn{1}{|l}{NDCG@10} &   \multicolumn{1}{l}{P@10} &   \multicolumn{1}{l}{MAP} \\\midrule
\revision{BM25} &
\revision{0.274} &
\revision{0.545} &
\revision{0.636} &
\revision{0.282 } 
\\
Retriever & 0.312 & 0.629 & 0.720 & 0.321 \\\midrule
FirstP (BERT) & 0.394 & 0.632 & 0.712 & 0.311 \\
FirstP (Longformer) & 0.404 & 0.643 & 0.722 & 0.317 \\
FirstP (ELECTRA) & 0.417 & 0.662 & 0.734 & 0.320 \\
FirstP (DEBERTA) & 0.415 & 0.672 & 0.741 & 0.327 \\
FirstP (Big-Bird) & 0.408 & 0.656 & 0.727 & 0.321 \\
FirstP (JINA) & 0.422 & 0.654 & 0.728 & 0.320 \\
FirstP (MOSAIC) & 0.423 & 0.643 & 0.726 & 0.316 \\
FirstP (TinyLLAMA) & 0.395 & 0.615 & 0.692 & 0.301 \\
FirstP (E5) & 0.380 & 0.641 & 0.722 & 0.317 \\
FirstP RankGPT (OpenAI) & -- & \textbf{0.708} & \textbf{0.790} & \textbf{0.352} \\
FirstP RankGPT (Anthropic) & -- & 0.703 & 0.776 & 0.347 \\\midrule
AvgP & 0.389 $(- 1.3\%)$ & 0.642 $(+ 1.5\%)$ & 0.717 $(+ 0.7\%)$ & 0.317$^{a}$ $(+ 2.0\%)$ \\\midrule
MaxP & 0.392 $(- 0.4\%)$ & 0.644$^{a}$ $(+ 1.9\%)$ & 0.723 $(+ 1.5\%)$ & 0.322$^{a}$ $(+ 3.7\%)$ \\
MaxP (ELECTRA) & 0.414 $(- 0.6\%)$ & 0.659 $(- 0.5\%)$ & 0.745 $(+ 1.5\%)$ & 0.326 $(+ 2.1\%)$ \\
MaxP (DEBERTA) & 0.402$^{a}$ $(- 3.2\%)$ & 0.671 $(- 0.1\%)$ & 0.746 $(+ 0.7\%)$ & 0.335$^{a}$ $(+ 2.5\%)$ \\
SumP & 0.390 $(- 1.0\%)$ & 0.639 $(+ 1.0\%)$ & 0.715 $(+ 0.4\%)$ & 0.319$^{a}$ $(+ 2.6\%)$ \\\midrule
CEDR-DRMM & 0.385$^{a}$ $(- 2.3\%)$ & 0.629 $(- 0.5\%)$ & 0.708 $(- 0.5\%)$ & 0.313 $(+ 0.6\%)$ \\
CEDR-KNRM & 0.379$^{a}$ $(- 3.8\%)$ & 0.630 $(- 0.3\%)$ & 0.711 $(- 0.1\%)$ & 0.313 $(+ 0.8\%)$ \\
CEDR-PACRR & 0.395 $(+ 0.3\%)$ & 0.643$^{a}$ $(+ 1.6\%)$ & 0.719 $(+ 0.9\%)$ & 0.320$^{a}$ $(+ 2.9\%)$ \\\midrule
Neural Model1 & 0.398 $(+ 0.9\%)$ & 0.650$^{a}$ $(+ 2.8\%)$ & 0.723$^{a}$ $(+ 1.5\%)$ & 0.323$^{a}$ $(+ 3.9\%)$ \\\midrule
PARADE Attn & 0.416$^{a}$ $(+ 5.5\%)$ & 0.652$^{a}$ $(+ 3.1\%)$ & 0.728$^{a}$ $(+ 2.2\%)$ & 0.324$^{a}$ $(+ 4.2\%)$ \\
PARADE Attn (ELECTRA) & 0.431$^{a}$ $(+ 3.3\%)$ & 0.680$^{a}$ $(+ 2.7\%)$ & 0.763$^{a}$ $(+ 3.9\%)$ & 0.335$^{a}$ $(+ 4.9\%)$ \\
PARADE Attn (DEBERTA) & 0.422$^{a}$ $(+ 1.6\%)$ & 0.688$^{a}$ $(+ 2.4\%)$ & 0.763$^{a}$ $(+ 3.0\%)$ & 0.339$^{a}$ $(+ 3.9\%)$ \\\midrule
PARADE Avg & 0.392 $(- 0.6\%)$ & 0.646$^{a}$ $(+ 2.1\%)$ & 0.715 $(+ 0.4\%)$ & 0.317$^{a}$ $(+ 2.1\%)$ \\
PARADE Max & 0.405$^{a}$ $(+ 2.7\%)$ & 0.655$^{a}$ $(+ 3.5\%)$ & 0.733$^{a}$ $(+ 2.9\%)$ & 0.324$^{a}$ $(+ 4.1\%)$ \\\midrule
PARADE Transf-RAND-L2 & 0.419$^{a}$ $(+ 6.3\%)$ & 0.655$^{a}$ $(+ 3.6\%)$ & 0.734$^{a}$ $(+ 3.1\%)$ & 0.326$^{a}$ $(+ 5.0\%)$ \\
PARADE Transf-RAND-L2 (ELECTRA) & \textbf{0.433$^{a}$ $(+ 3.9\%)$} & 0.670 $(+ 1.2\%)$ & 0.747 $(+ 1.8\%)$ & 0.327 $(+ 2.2\%)$ \\
PARADE Transf-PRETR-L6 & 0.402$^{a}$ $(+ 1.9\%)$ & 0.643 $(+ 1.6\%)$ & 0.717 $(+ 0.8\%)$ & 0.322$^{a}$ $(+ 3.6\%)$ \\
PARADE Transf-PRETR-LATEIR-L6 & 0.398 $(+ 1.1\%)$ & 0.626 $(- 0.9\%)$ & 0.707 $(- 0.7\%)$ & 0.307 $(- 1.1\%)$ \\\midrule
LongP (Longformer) & 0.412$^{a}$ $(+ 1.9\%)$ & 0.668$^{a}$ $(+ 3.9\%)$ & 0.752$^{a}$ $(+ 4.1\%)$ & 0.333$^{a}$ $(+ 5.1\%)$ \\
LongP (Big-Bird) & 0.397$^{a}$ $(- 2.9\%)$ & 0.651 $(- 0.7\%)$ & 0.726 $(- 0.2\%)$ & 0.322 $(+ 0.3\%)$ \\
LongP (JINA) & 0.416$^{a}$ $(- 1.5\%)$ & 0.665$^{a}$ $(+ 1.7\%)$ & 0.742$^{a}$ $(+ 2.0\%)$ & 0.328$^{a}$ $(+ 2.4\%)$ \\
LongP (MOSAIC) & 0.421 $(- 0.4\%)$ & 0.664$^{a}$ $(+ 3.3\%)$ & 0.740$^{a}$ $(+ 1.9\%)$ & 0.327$^{a}$ $(+ 3.7\%)$ \\
LongP (TinyLLAMA) & 0.402$^{a}$ $(+ 1.7\%)$ & 0.608 $(- 1.1\%)$ & 0.692 $(+ 0.0\%)$ & 0.306 $(+ 1.6\%)$ \\
LongP (E5) & 0.353$^{a}$ $(- 7.1\%)$ & 0.649 $(+ 1.3\%)$ & 0.724 $(+ 0.3\%)$ & 0.323 $(+ 1.8\%)$ \\
LongP RankGPT (OpenAI) & -- & 0.706 $(- 0.3\%)$ & 0.783 $(- 1.0\%)$ & 0.350 $(- 0.7\%)$ \\
LongP RankGPT (Anthropic) & -- & 0.707 $(+ 0.5\%)$ & 0.780 $(+ 0.5\%)$ & 0.348 $(+ 0.4\%)$ \\
\bottomrule
\end{tabular}   
  \begin{tablenotes}
        \tablecustomsmall
  
  In each column we show a relative gain with respect model's respective \emph{FirstP} baseline: The last column shows the average relative gain of \emph{FirstP}. Best numbers are in \textbf{bold}: Results are averaged over three seeds. Unless specified explicitly, the backbone is \textbf{BERT-base}.  \\
  Statistical significant differences with respect to this baseline are denoted using the superscript superscript \textbf{a}. $p$-value threshold is 0.01
  for an MS MARCO development collection and 0.05 otherwise. \\
  E5-models were used only in the zero-shot model, i.e., without fine-tuning. \\
 
    \end{tablenotes}
\end{threeparttable}
\caption{
\revision{Comparison between long-document models and respective FirstP (truncation) baselines. Results for MS MARCO and TREC DL.}
\label{tab:msmarco}}
\end{table*}

\hspace{-1em}\begin{table*}
\scriptsize
\setlength{\tabcolsep}{2pt}
\begin{threeparttable}
\begin{tabular}{l|lll|lll}
 \toprule
\multicolumn{1}{c|}{Model}   & \multicolumn{1}{|l}{NDCG@20} &   \multicolumn{1}{l}{P@20} &   \multicolumn{1}{l}{MAP} & \multicolumn{1}{|l}{NDCG@20} &   \multicolumn{1}{l}{P@20} &   \multicolumn{1}{l}{MAP}\\\midrule
\revision{ Retriever (BM25)} & 0.428 & 0.365 & 0.255 & 0.402 & 0.334 & 0.240 \\\midrule
FirstP (BERT) & 0.475 & 0.405 & 0.277 & 0.527 & 0.447 & 0.303 \\
FirstP (Longformer) & 0.483 & 0.413 & 0.277 & 0.540 & 0.454 & 0.307 \\
FirstP (ELECTRA) & 0.492 & 0.424 & 0.294 & 0.552 & 0.465 & 0.320 \\
FirstP (DEBERTA) & 0.534 & 0.459 & 0.319 & 0.596 & 0.503 & 0.350 \\
FirstP (Big-Bird) & 0.507 & 0.435 & 0.300 & 0.560 & 0.473 & 0.325 \\
FirstP (JINA) & 0.488 & 0.421 & 0.287 & 0.532 & 0.450 & 0.305 \\
FirstP (MOSAIC) & 0.453 & 0.390 & 0.266 & 0.538 & 0.455 & 0.310 \\
FirstP (TinyLLAMA) & 0.431 & 0.370 & 0.246 & 0.473 & 0.398 & 0.262 \\
FirstP (E5-4K) & 0.438 & 0.371 & 0.247 & 0.429 & 0.355 & 0.234 \\
FirstP RankGPT (OpenAI) & -- & -- & -- & 0.562 & 0.456 & 0.280 \\
FirstP RankGPT (Anthropic) & -- & -- & -- & 0.541 & 0.446 & 0.268 \\
\midrule
AvgP & 0.478 $(+ 0.5\%)$ & 0.411 $(+ 1.6\%)$ & 0.292$^{a}$ $(+ 5.4\%)$ & 0.531 $(+ 0.9\%)$ & 0.451 $(+ 1.0\%)$ & 0.324$^{a}$ $(+ 6.7\%)$ \\\midrule
MaxP & 0.488$^{a}$ $(+ 2.6\%)$ & 0.425$^{a}$ $(+ 5.1\%)$ & 0.306$^{a}$ $(+10.6\%)$ & 0.544$^{a}$ $(+ 3.3\%)$ & 0.467$^{a}$ $(+ 4.5\%)$ & 0.338$^{a}$ $(+11.5\%)$ \\
MaxP (ELECTRA) & 0.502 $(+ 2.0\%)$ & 0.441$^{a}$ $(+ 3.9\%)$ & 0.319$^{a}$ $(+ 8.3\%)$ & 0.563 $(+ 2.1\%)$ & 0.483$^{a}$ $(+ 4.0\%)$ & 0.350$^{a}$ $(+ 9.3\%)$ \\
MaxP (DEBERTA) & 0.535 $(+ 0.2\%)$ & 0.464 $(+ 1.2\%)$ & 0.340$^{a}$ $(+ 6.7\%)$ & 0.609$^{a}$ $(+ 2.2\%)$ & 0.519$^{a}$ $(+ 3.2\%)$ & 0.378$^{a}$ $(+ 7.9\%)$ \\
SumP & 0.486 $(+ 2.2\%)$ & 0.418$^{a}$ $(+ 3.4\%)$ & 0.305$^{a}$ $(+10.2\%)$ & 0.538 $(+ 2.1\%)$ & 0.461$^{a}$ $(+ 3.1\%)$ & 0.337$^{a}$ $(+11.1\%)$ \\\midrule
CEDR-DRMM & 0.466 $(- 2.0\%)$ & 0.403 $(- 0.4\%)$ & 0.287$^{a}$ $(+ 3.8\%)$ & 0.533 $(+ 1.3\%)$ & 0.458$^{a}$ $(+ 2.5\%)$ & 0.326$^{a}$ $(+ 7.6\%)$ \\
CEDR-KNRM & 0.483 $(+ 1.7\%)$ & 0.413 $(+ 1.9\%)$ & 0.291$^{a}$ $(+ 5.1\%)$ & 0.535 $(+ 1.7\%)$ & 0.456 $(+ 2.0\%)$ & 0.324$^{a}$ $(+ 6.8\%)$ \\
CEDR-PACRR & 0.496$^{a}$ $(+ 4.3\%)$ & 0.426$^{a}$ $(+ 5.3\%)$ & 0.307$^{a}$ $(+11.0\%)$ & 0.549$^{a}$ $(+ 4.2\%)$ & 0.466$^{a}$ $(+ 4.4\%)$ & 0.337$^{a}$ $(+11.2\%)$ \\\midrule
Neural Model1 & 0.484 $(+ 1.8\%)$ & 0.417$^{a}$ $(+ 3.1\%)$ & 0.298$^{a}$ $(+ 7.7\%)$ & 0.537 $(+ 1.9\%)$ & 0.459$^{a}$ $(+ 2.6\%)$ & 0.330$^{a}$ $(+ 8.8\%)$ \\\midrule
PARADE Attn & 0.503$^{a}$ $(+ 5.7\%)$ & 0.433$^{a}$ $(+ 6.9\%)$ & 0.311$^{a}$ $(+12.4\%)$ & 0.556$^{a}$ $(+ 5.6\%)$ & 0.476$^{a}$ $(+ 6.5\%)$ & 0.344$^{a}$ $(+13.3\%)$ \\
PARADE Attn (ELECTRA) & 0.523$^{a}$ $(+ 6.4\%)$ & 0.456$^{a}$ $(+ 7.4\%)$ & 0.329$^{a}$ $(+11.7\%)$ & 0.581$^{a}$ $(+ 5.3\%)$ & 0.495$^{a}$ $(+ 6.5\%)$ & 0.358$^{a}$ $(+11.9\%)$ \\
PARADE Attn (DEBERTA) & \textbf{0.549$^{a}$ $(+ 2.9\%)$} & \textbf{0.475$^{a}$ $(+ 3.6\%)$} & \textbf{0.346$^{a}$ $(+ 8.7\%)$} & \textbf{0.615$^{a}$ $(+ 3.2\%)$} & \textbf{0.522$^{a}$ $(+ 3.8\%)$} & \textbf{0.383$^{a}$ $(+ 9.4\%)$} \\
PARADE Avg & 0.483 $(+ 1.5\%)$ & 0.412 $(+ 1.8\%)$ & 0.291$^{a}$ $(+ 5.2\%)$ & 0.534 $(+ 1.5\%)$ & 0.457 $(+ 2.4\%)$ & 0.318$^{a}$ $(+ 4.7\%)$ \\
PARADE Max & 0.489$^{a}$ $(+ 2.8\%)$ & 0.420$^{a}$ $(+ 3.8\%)$ & 0.306$^{a}$ $(+10.8\%)$ & 0.548$^{a}$ $(+ 4.0\%)$ & 0.470$^{a}$ $(+ 5.3\%)$ & 0.337$^{a}$ $(+11.0\%)$ \\\midrule
PARADE Transf-RAND-L2 & 0.488$^{a}$ $(+ 2.8\%)$ & 0.423$^{a}$ $(+ 4.6\%)$ & 0.303$^{a}$ $(+ 9.7\%)$ & 0.548$^{a}$ $(+ 4.1\%)$ & 0.469$^{a}$ $(+ 5.0\%)$ & 0.338$^{a}$ $(+11.6\%)$ \\
PAR. Transf-RAND-L2 (ELECTRA) & 0.523$^{a}$ $(+ 6.3\%)$ & 0.454$^{a}$ $(+ 6.9\%)$ & 0.330$^{a}$ $(+12.2\%)$ & 0.574$^{a}$ $(+ 3.9\%)$ & 0.488$^{a}$ $(+ 5.0\%)$ & 0.354$^{a}$ $(+10.6\%)$ \\
PARADE Transf-PRETR-L6 & 0.494$^{a}$ $(+ 4.0\%)$ & 0.426$^{a}$ $(+ 5.3\%)$ & 0.308$^{a}$ $(+11.5\%)$ & 0.554$^{a}$ $(+ 5.1\%)$ & 0.474$^{a}$ $(+ 6.1\%)$ & 0.346$^{a}$ $(+14.1\%)$ \\
PAR. Transf-PRETR-LATEIR-L6 & 0.450$^{a}$ $(- 5.2\%)$ & 0.389$^{a}$ $(- 3.9\%)$ & 0.277 $(+ 0.3\%)$ & 0.501$^{a}$ $(- 4.9\%)$ & 0.423$^{a}$ $(- 5.3\%)$ & 0.302 $(- 0.5\%)$ \\\midrule
LongP (Longformer) & 0.500$^{a}$ $(+ 3.6\%)$ & 0.435$^{a}$ $(+ 5.3\%)$ & 0.309$^{a}$ $(+11.5\%)$ & 0.568$^{a}$ $(+ 5.1\%)$ & 0.482$^{a}$ $(+ 6.1\%)$ & 0.347$^{a}$ $(+12.9\%)$ \\
LongP (Big-Bird) & 0.452$^{a}$ $(-10.9\%)$ & 0.389$^{a}$ $(-10.7\%)$ & 0.274$^{a}$ $(- 8.8\%)$ & 0.477$^{a}$ $(-14.9\%)$ & 0.400$^{a}$ $(-15.5\%)$ & 0.279$^{a}$ $(-14.2\%)$ \\
LongP (JINA) & 0.503$^{a}$ $(+ 2.9\%)$ & 0.434$^{a}$ $(+ 3.1\%)$ & 0.309$^{a}$ $(+ 7.5\%)$ & 0.558$^{a}$ $(+ 4.9\%)$ & 0.473$^{a}$ $(+ 5.2\%)$ & 0.335$^{a}$ $(+ 9.7\%)$ \\
LongP (MOSAIC) & 0.456 $(+ 0.6\%)$ & 0.393 $(+ 0.8\%)$ & 0.280$^{a}$ $(+ 5.3\%)$ & 0.570$^{a}$ $(+ 6.0\%)$ & 0.484$^{a}$ $(+ 6.3\%)$ & 0.350$^{a}$ $(+13.0\%)$ \\
LongP (TinyLLAMA) & 0.452$^{a}$ $(+ 4.8\%)$ & 0.396$^{a}$ $(+ 6.9\%)$ & 0.267$^{a}$ $(+ 8.7\%)$ & 0.505$^{a}$ $(+ 6.7\%)$ & 0.428$^{a}$ $(+ 7.6\%)$ & 0.297$^{a}$ $(+13.3\%)$ \\
LongP (E5-4K) & 0.439 $(+ 0.1\%)$ & 0.375 $(+ 1.0\%)$ & 0.250 $(+ 1.3\%)$ & 0.434 $(+ 1.1\%)$ & 0.360 $(+ 1.6\%)$ & 0.241$^{a}$ $(+ 2.9\%)$ \\
LongP RankGPT (OpenAI) & -- & -- & -- & 0.562 $(+ 0.0\%)$ & 0.456 $(- 0.0\%)$ & 0.281 $(+ 0.3\%)$ \\
LongP RankGPT (Anthropic) & -- & -- & -- & 0.538 $(- 0.6\%)$ & 0.445 $(- 0.2\%)$ & 0.268 $(- 0.0\%)$ \\
\bottomrule
\end{tabular}   
  \begin{tablenotes}
        \tablecustomsmall
  
  In each column we show a relative gain with respect model's respective \emph{FirstP} baseline: The last column shows the average relative gain of \emph{FirstP}. Best numbers are in \textbf{bold}: Results are averaged over three seeds. Unless specified explicitly, the backbone is \textbf{BERT-base}.  \\
  Statistical significant differences with respect to this baseline are denoted using the superscript superscript \textbf{a}. $p$-value threshold is 0.05. \\
 E5-models were used only in the zero-shot model, i.e., without fine-tuning. \\
    \end{tablenotes}
\end{threeparttable}
\caption{
\revision{Comparison between long-document models and respective FirstP (truncation) baselines. Results for Robust04.}
\label{tab:robust04}}
\end{table*}

\subsection{MS MARCO FarRelevant Creation Algorithm}\label{sec:farrelevant_algo}
The algorithm iterates over the set of relevant passages in the original MS MARCO Passages collection.
Each iteration is as follows:
\begin{enumerate}
    
    \item Assume that $C_t$ is the number of tokens in the selected relevant passage. Select randomly a document length $D_{max}$ between $512 + C_t$ and 1431;
    \item Using rejection sampling, obtain $K_1$ non-relevant passages such that their \emph{total} length is $\ge 512$,
    but the length of $K_1 -1$ first passages is $<512$.
    \item Using the same approach, sample another $K_2$ non-relevant passages such that the total length of $K_1 + K_2 - 1$ 
    non-relevant passages is at most $D_{max} - C_t$, but the total length of $K_1 + K_2$ passages is $\ge D_{max} - C_t$.
    \item Randomly mix these $K_2$ non-relevant passages with a single relevant passage.
    \item Finally, append the resulting string to the concatenation of the first $K_1$ non-relevant passages.
\end{enumerate}
Initially, we planned to limit documents to have at most 1431 tokens. However,
step 3 in the above algorithm has a bug: It samples one extra non-relevant passage.
As a result, only about 5\% of generated documents have more than 1431 tokens and 
less than 1\% have more than 1500.

\begin{figure*}
    \centering
    \begin{textenv}

Aaron Swartz created a scraped feed of the essays page. November 2021(This essay is derived from a talk at the Cambridge Union. )When I was a kid, I'd have said there wasn't. My father told me so. Some people like some things, and other people like other things, and who's to say who's right?It seemed so obvious that there was no such thing as good taste that it was only through indirect evidence that I realized my father was wrong. And that's what I'm going to give you here: a proof by reductio ad absurdum. If we start from the premise that there's no such thing as good taste, we end up with conclusions that are obviously false, and therefore the premise must be wrong. We'd better start by saying what good taste is. There's a narrow sense in which it refers to aesthetic judgements and a broader one in which it refers to preferences of any kind. The strongest proof would be to show that taste exists in the narrowest sense, so I'm going to talk about taste in art. You have better taste than me if the art you like is better than the art I like. If there's no such thing as good taste, then there's no such thing as good art. Because if there is such a thing as good art, it's easy to tell which of two people has better taste. Show them a lot of works by artists they've never seen before and ask them to choose the best, and whoever chooses the better art has better taste. So if you want to discard the concept of good taste, you also have to discard the concept of good art. And that means you have to discard the possibility of people being good at making it. Which means there's no way for artists to be good at their jobs. And not just visual artists, but anyone who is in any sense an artist. You can't have good actors, or novelists, or composers, or dancers either. You can have popular novelists, but not good ones. We don't realize how far we'd have to go if we discarded the concept of good taste, because we don't even debate the most obvious cases. But it doesn't just mean we can't say which of two famous painters is better. It means we can't say that any painter is better than a randomly chosen eight year old. That was how I realized my father was wrong. I started studying painting. And it was just like other kinds of work I'd done: you could do it well, or badly, and if you tried hard, you could get better at it. And it was obvious that Leonardo and Bellini were much better at it than me. That gap between us was not imaginary. They were so good. And if they could be good, then art could be good, and there was such a thing as good taste after all. Now that I've explained how to show there is such a thing as good taste, I should also explain why people think there isn't. There are two reasons. One is that there's always so much disagreement about taste. Most people's response to art is a tangle of unexamined impulses. Is the artist famous? Is the subject attractive? Is this the sort of art they're supposed to like? Is it hanging in a famous museum, or reproduced in a big, expensive book? In practice most people's response to art is dominated by such extraneous factors.

\textbf{The Terracotta Army is a collection of terracotta sculptures depicting the armies of Qin Shi Huang, the first Emperor of China.}

And the people who do claim to have good taste are so often mistaken. The paintings admired by the so-called experts in one generation are often so different from those admired a few generations later. It's easy to conclude there's nothing real there at all. It's only when you isolate this force, for example by trying to paint and comparing your work to Bellini's, that you can see that it does in fact exist. The other reason people doubt that art can be good is that there doesn't seem to be any room in the art for this goodness. The argument goes like this. Imagine several people looking at a work of art and judging how good it is. If being good art really is a property of objects, it should be in the object somehow. But it doesn't seem to be; it seems to be something happening in the heads of each of the observers. And if they disagree, how do

\end{textenv}
    \caption{A sample relevant document for the Needle collection. The query/question is: ``What is the Terracotta Army?''.
    The answer-bearing sentence is marked by bold font.}
    \label{fig:sample_needle}
\end{figure*}

\begin{figure*}
    \centering
    \begin{textenv}
The grass is green. The sky is blue. The sun is yellow. Here we go. There and back again. The grass is green. The sky is blue. The sun is yellow. Here we go. There and back again. The grass is green. The sky is blue. The sun is yellow. Here we go. There and back again. The grass is green. The sky is blue. The sun is yellow. Here we go. There and back again. The grass is green. The sky is blue. The sun is yellow. Here we go. There and back again. The grass is green. The sky is blue. The sun is yellow. Here we go. There and back again. The grass is green. The sky is blue. The sun is yellow. Here we go. There and back again. The grass is green. The sky is blue. The sun is yellow. Here we go. There and back again. The grass is green. The sky is blue. The sun is yellow. Here we go. There and back again. The grass is green. The sky is blue. The sun is yellow. Here we go. There and back again. The grass is green. The sky is blue. The sun is yellow. Here we go. There and back again. The grass is green. The sky is blue. The sun is yellow. Here we go. There and back again. The grass is green. The sky is blue. The sun is yellow. Here we go. There and back again. The grass is green. The sky is blue. The sun is yellow. Here we go. There and back again. The grass is green. The sky is blue. The sun is yellow. Here we go. There and back again. The grass is green. The sky is blue. The sun is yellow. Here we go. There and back again. The grass is green. The sky is blue. The sun is yellow. Here we go. There and back again. The grass is green. The sky is blue. The sun is yellow. Here we go. There and back again. The grass is green. The sky is blue. The sun is yellow. Here we go. There and back again. The grass is green. The sky is blue. The sun is yellow. Here we go. There and back again. The grass is green. The sky is blue. The sun is yellow. Here we go. There and back again. The grass is green. The sky is blue. The sun is yellow. Here we go. There and back again. The grass is green. The sky is blue. The sun is yellow. Here we go. There and back again. The grass is green. The sky is blue. The sun is yellow. Here we go. There and back again. The grass is green. The sky is blue. The sun is yellow. Here we go. There and back again. The grass is green. The sky is blue. The sun is yellow. Here we go. There and back again. The grass is green. The sky is blue. The sun is yellow. Here we go. There and back again. The grass is green. The sky is blue. The sun is yellow. Here we go. There and back again. The grass is green.

\textbf{Jimmy Moses's pass key is 39566. Remember it. 39566 is the pass key for Jimmy Moses.}

The sky is blue. The sun is yellow. Here we go. There and back again. The grass is green. The sky is blue. The sun is yellow. Here we go. There and back again. The grass is green. The sky is blue. The sun is yellow. Here we go. There and back again. The grass is green. The sky is blue. The sun is yellow. Here we go. There and back again. The grass is green. The sky is blue. The sun is yellow. Here we go. There and back again. The grass is green. The sky is blue. The sun is yellow. Here we go. There and back again. The grass is green. The sky is blue. The sun is yellow. Here we go. There and back again. The grass is green. The sky is blue. The sun is yellow. Here we go. There and back again. The grass is green. The sky is blue. The sun is yellow. Here we go. There and back again. The grass is green. The sky is blue. The sun is yellow. Here we go. There and back again. The grass is green. The sky is blue. The sun is yellow. Here we go. There and back again. The grass
    \end{textenv}
    \caption{A sample relevant document for the Passkey collection. The query/question is: ``what is the passkey for Jimmy Moses?''.
    The answer-bearing sentence is marked by bold font.}
    \label{fig:sample_passkey}
\end{figure*}

\begin{figure*}
    \centering
    \begin{textenv}
Andhra Pradesh Airports make an easy access for tourists visiting the state. This huge state has many airports, which serves the needs of both tourists and residents commuting to different parts in its large expanse. However, Hyderabad Airport is the major as well as the only international airport of Andhra Pradesh. Hyderabad, a major IT hub of India, boasts of the sixth busiest airport in India. Keeping the rush of passengers in mind, the Government is planning to establish another airport in Hyderabad.

In contrast, traditional English Longbow shooters step into the bow, exerting force with both the bow arm and the string hand arm simultaneously, especially when using bows having draw weights from 100 lbs to over 175 lbs. Heavily stacked traditional bows (recurves, long bows, and the like) are released immediately upon reaching full draw at maximum weight, whereas compound bows reach their maximum weight around the last inch and a half, dropping holding weight significantly at full draw.
The Oreo Biscuit was first developed and produced by the National Biscuit Company (today known as Nabisco) in 1912 at its Chelsea, Manhattan factory in the current-day Chelsea Market complex, located on Ninth Avenue between 15th and 16th Streets. Today, this same block of Ninth Avenue is known as Oreo Way..

It's possible to take a day trip to the Bahamas by ferry. In some cases, less than it would cost to fly. The high-speed Balearia Bahamas Express travels from Fort Lauderdale to the city of Freeport on Grand Bahama, one of many Bahamian islands with British roots.It's roughly the distance from Philadelphia to New York.Prepare for a long day.n some cases, less than it would cost to fly. The high-speed Balearia Bahamas Express travels from Fort Lauderdale to the city of Freeport on Grand Bahama, one of many Bahamian islands with British roots. It's roughly the distance from Philadelphia to New York.

I was 17 when I took Bactrim for a UTI, I was a month away from turning 18 and was given this because another antibiotic would have more side effects I was told. I took it for 5 days and became horribly sick from a nasty cold and was told to stop Bactrim and to take Z Pac instead for 5 days.

Cases When Medicare Does NOT Automatically Start for You Medicare will NOT automatically start when you turn 65 if you're not receiving Social Security Benefits or Railroad Retirement Benefits for at least 4 months prior to your 65th birthday.
You wanted to know how you can feel on your belly that you're pregnant. And actually this a pretty hard thing to do.There are better ways to find out if you're pregnant or not. For example, if you ever miss a period, the best thing to do is to take a home pregnancy test, because that's the first sign of pregnancy.And if it's positive, ...here are better ways to find out if you're pregnant or not. For example, if you ever miss a period, the best thing to do is to take a home pregnancy test, because that's the first sign of pregnancy.

1 Whisk light soy sauce, dark soy sauce, red wine vinegar, chili oil, ginger, sugar, garlic, and green onion together in a bowl; pour into a sealable container, seal, and refrigerate 1 hour.  See how to make a simple sweet-and-sour peach sauce. See how easy and delicious it is to make horseradish sauce from scratch.

Here you'll find a number of Kentucky facts including the state history at a glance; Kentucky state facts such as the location of the state capital, city populations, geography and natural resources; Information on Kentucky' government, symbols and traditions; and even a list of famous Kentuckians.

Confidence votes 193. Replacing a car window usually cost around \$300 give or take based upon where you live, what options your car glass needs and what type of car you drive. Additionally, you can save money if you can repair your car glass instead of completely replacing it. However, if the window is completely shattered this will not be an option.

\textbf{Flying time from Chicago, IL to Cairo, Egypt. The total flight duration from Chicago, IL to Cairo, Egypt is 12 hours, 47 minutes. This assumes an average flight speed for a commercial airliner of 500 mph, which is equivalent to 805 km/h or 434 knots. It also adds an extra 30 minutes for take-off and landing. Your exact time may vary depending on wind speeds.}
\end{textenv}
    \caption{A sample relevant document for the MS MARCO FarRelevant collection. The query/question is: ``how long is the flight from chicago to cairo''. The answer-bearing passage is marked by bold font.}
    \label{fig:sample_far_relevant}
\end{figure*}

\subsection{Comparison of FarRelevant and Synthetic Data from LongEmbed}\label{sec:comp_synthetic}
Our synthetic data consists of two subsets ({Needle} and {Passkey}) from the {LongEmbed} collection \cite{E52024} and our newly created {MS MARCO FarRelevant} dataset. 
All these datasets can be seen a variant of a needle-in-the-haystack benchmark \cite{LoCo2024,E52024}.
They share a common limitation: the resulting documents are constructed by combining pieces of text in a purely mechanical fashion and do not represent natural documents. In this section, we describe {Needle} and {Passkey} in more detail and compare them to our proposed collection {MS MARCO FarRelevant}, which---we believe---offers several practical advantages despite also being synthetic.

The {Needle} subset was created by taking a single document (a Paul Graham essay on taste\footnote{\url{https://www.paulgraham.com/goodtaste.html}}), truncating it to generate 16 text pieces of varying lengths (from 256 to 32{,}768 tokens), 
and inserting a single answer-bearing sentence at a random location. 
An example query-document pair can be found in  Figure~\ref{fig:sample_needle}.
While this design ensures precise control over document length, it introduces several problems:
\begin{enumerate}
\item \textbf{Extremely low document diversity}: All examples are truncated variants of the same original text, which severely limits the variability in document style and content.
\item \textbf{Artificial signal separation}: The inserted answer sentence  differs substantially from
the background text, making it easy for models to identify it.
\end{enumerate}

The {Passkey} subset is similar in structure and also uses length-bucketed documents, but instead of a single sentence, it inserts a three-sentence passkey definition into a synthetic background context  (see Figure~\ref{fig:sample_passkey}). 
The background is even less natural than {Needle}'s subset background, being composed of unrelated or nonsensical declarative statements (e.g., ``The sky is blue. The sun is yellow. Here we go. There and back again.''). This leads to even greater distributional mismatch between signal and context.

Critically, both {Needle} and {Passkey} were designed primarily to test \emph{answer extraction} or retrieval of small, highly localized answer-bearing spans. They are not well suited to studying retrieval or ranking of entire passages or documents, especially when relevance is more distributed or contextual.

Our {MS MARCO FarRelevant} (see Figure~\ref{fig:sample_far_relevant} for an example) 
is designed to be textually similar to MS MARCO Documents but with different positional biases for relevant passages. We believe it offers a more robust testbed for long-context \emph{document ranking}. 
While it is also synthetic in construction, it avoids many of the pitfalls noted above. Each document is created by concatenating multiple passages, which are typically \emph{meaningful} and \emph{complete}, 
only one of which is relevant to the query. The remaining passages serve as distractors but are independently coherent. 
Although these documents do not exist in the wild, they are much more diverse in content and style than those in {Needle} or {Passkey} despite being typically much shorter. Furthermore, each individual passage is semantically complete and belongs to a real corpus, namely, {MS MARCO Passages}. 

Crucially, one may argue that it is natural for the models to fail on randomly concatenated passages
because they do not represent a complete ``natural'' document (this criticism  applies to the synthetic subsets of LongEmbed as well).
Although this is possible for  \emph{LongP} methods---where longer document lengths are ``natively'' supported by models---nearly 
all models in our study are \emph{SplitP} models, which chunk documents and process each chunk separately. 
Hence, these models do not ``care'' whether the full document is coherent or not. 
Furthermore, the strong performance of all the ranking models after fine-tuning on MS MARCO FarRelevant demonstrates that models do not inherently fail on unnatural documents.

\subsection{Positional Bias Identification}\label{sec:positional_bias_indent}
\begin{figure*}
    \centering
    \begin{subfigure}{0.28\textwidth}
        \centering
        \includegraphics[width=\linewidth]{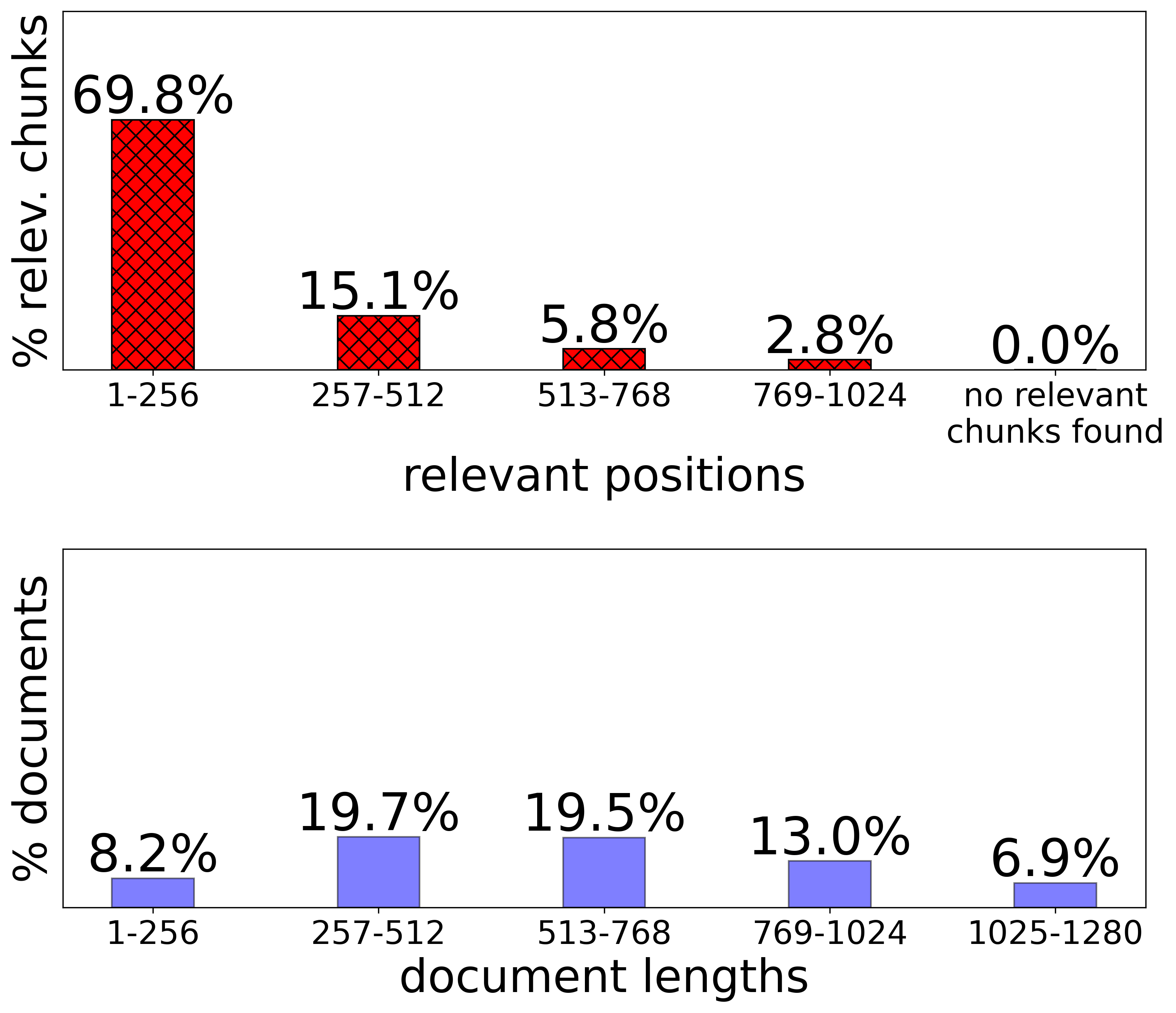}
        \caption{MS MARCO train}
    \end{subfigure}
    \begin{subfigure}{0.28\textwidth}
        \centering
        \includegraphics[width=\linewidth]{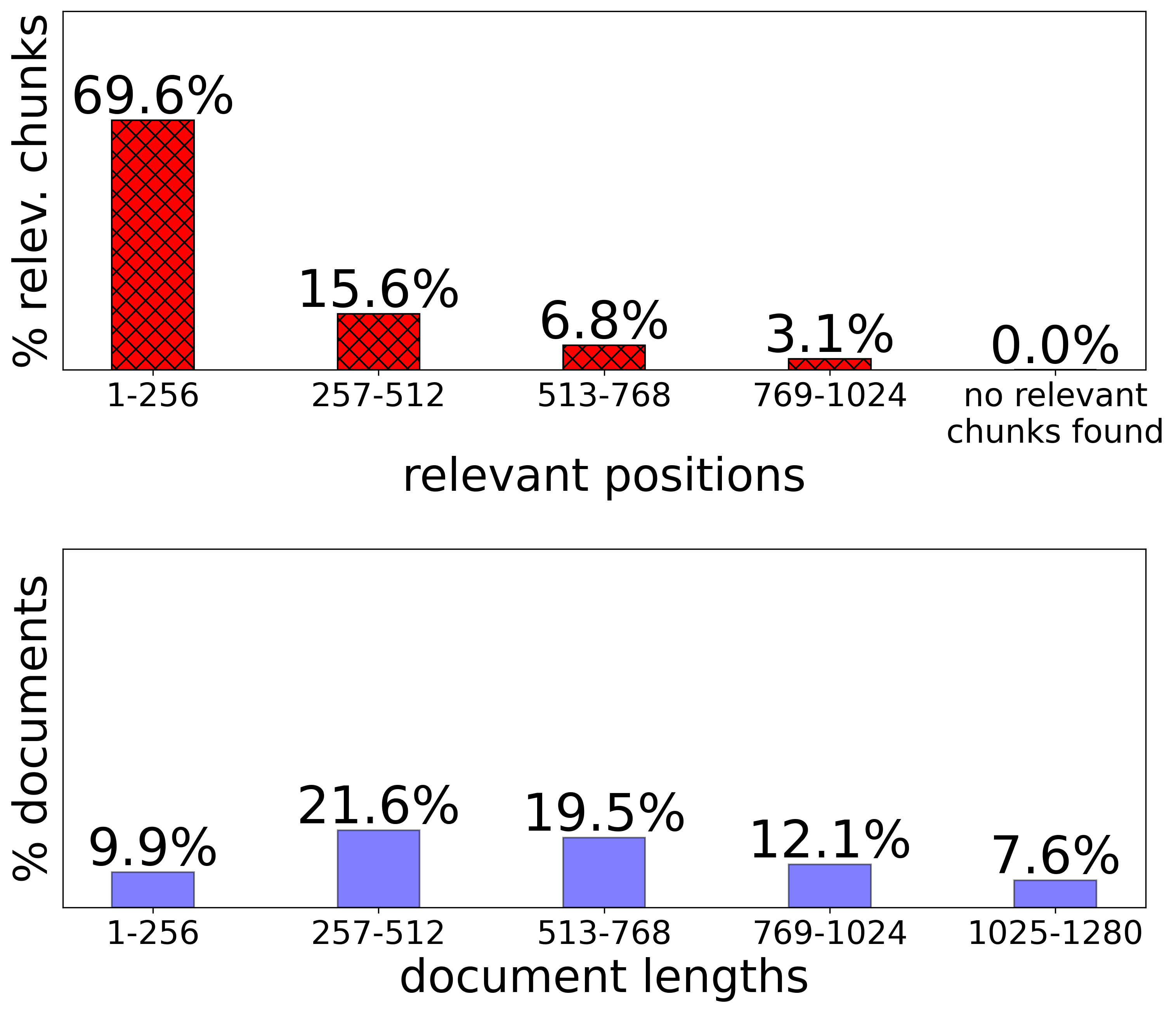}
        \caption{MS MARCO dev}
    \end{subfigure}
    \begin{subfigure}{0.28\textwidth}
        \centering
        \includegraphics[width=\linewidth]{relev_match_figures/test2019-2020.png}
        \caption{TREC DL 2019-2021 (combined)}
    \end{subfigure}
    
    \vspace{0.5cm} 
    
    \begin{subfigure}{0.28\textwidth}
        \centering
        \includegraphics[width=\linewidth]{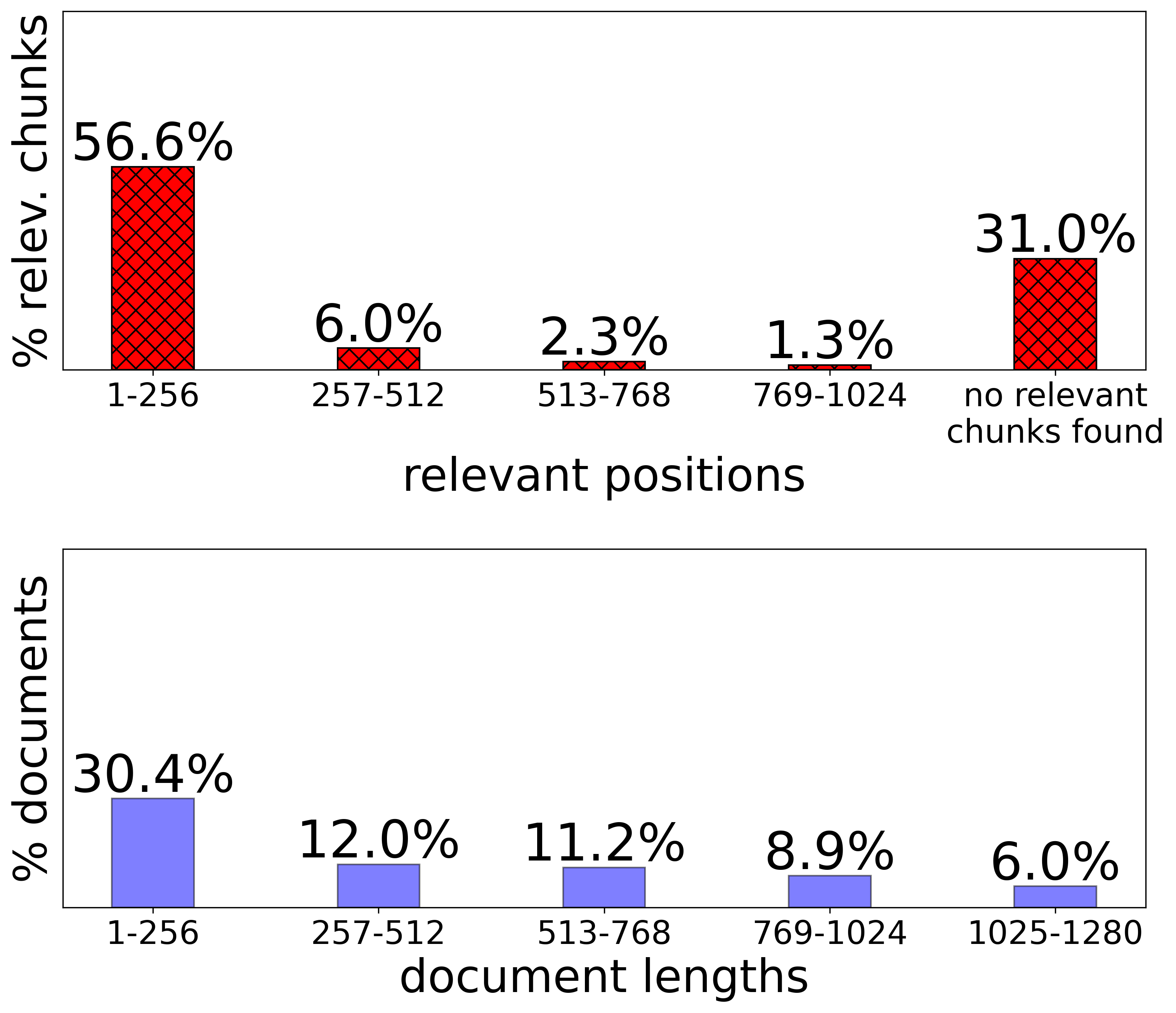}
        \caption{Gov2 (1MQ)}
    \end{subfigure}
    \begin{subfigure}{0.28\textwidth}
        \centering
        \includegraphics[width=\linewidth]{relev_match_figures/clueweb12.png}
        \caption{ClueWeb12 (WebTrack)}
    \end{subfigure}
    \begin{subfigure}{0.28\textwidth}
        \centering
        \includegraphics[width=\linewidth]{relev_match_figures/robust04.png}
        \caption{Robust04}
    \end{subfigure}
    
    \caption{\revision{Illustration of positional relevance bias for four 
    long-document collections and six query sets.
    We show a distribution of first relevant passage positions (red bars) vs. relevant document lengths (blue bars).}
    Positions and lengths are measured in the number of subword tokens (BERT-base tokenizer). Best viewed in color.}
    \label{fig:relev_match_full_plot}
\end{figure*}
\begin{figure*}
    \centering
    \begin{subfigure}{0.28\textwidth}
        \centering
        \includegraphics[width=\linewidth]{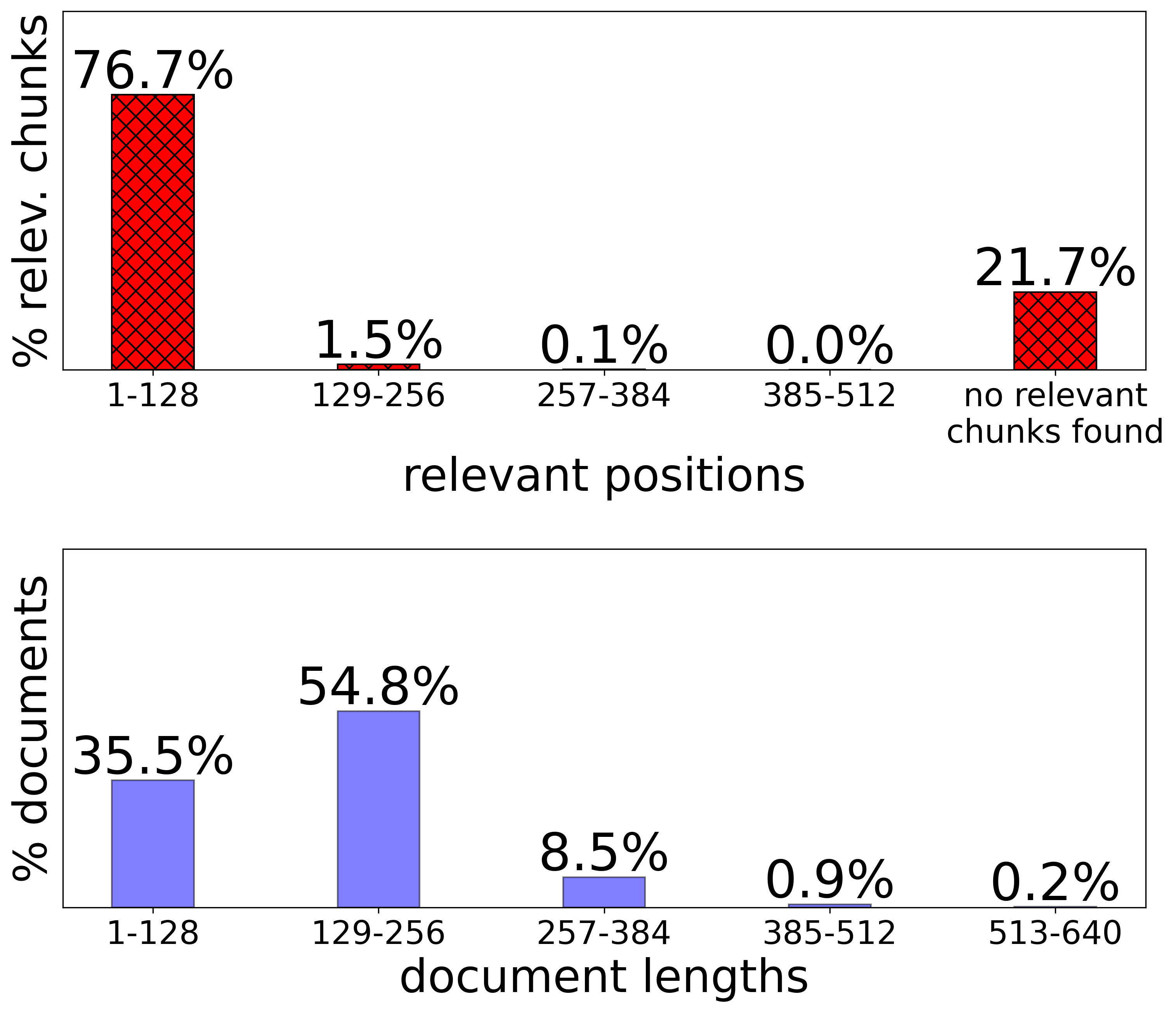 }
        \caption{Natural Questions (NQ)}
    \end{subfigure}
    \begin{subfigure}{0.28\textwidth}
        \centering
        \includegraphics[width=\linewidth]{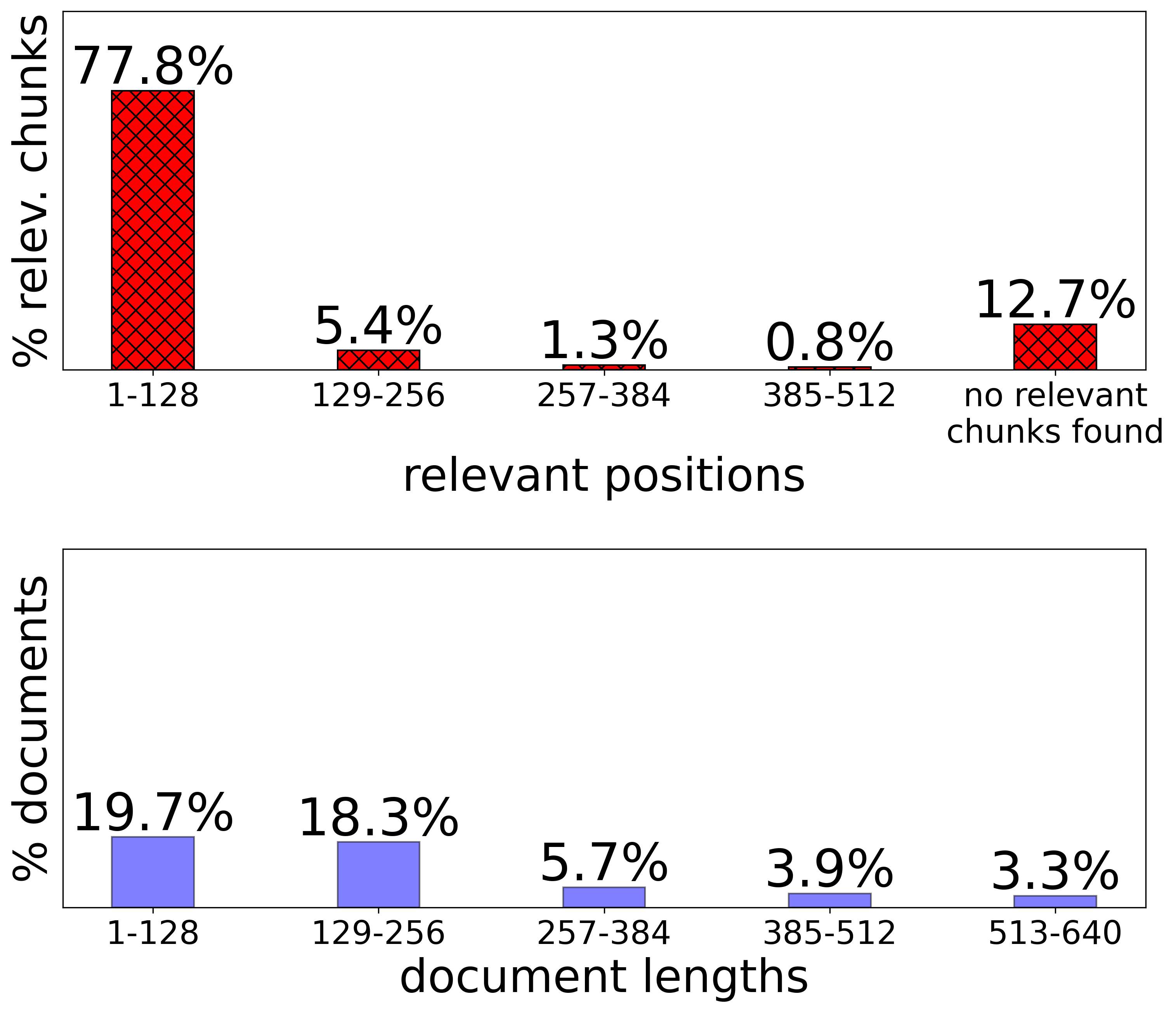}
        \caption{Touche}
    \end{subfigure}
    \begin{subfigure}{0.28\textwidth}
        \centering
        \includegraphics[width=\linewidth]{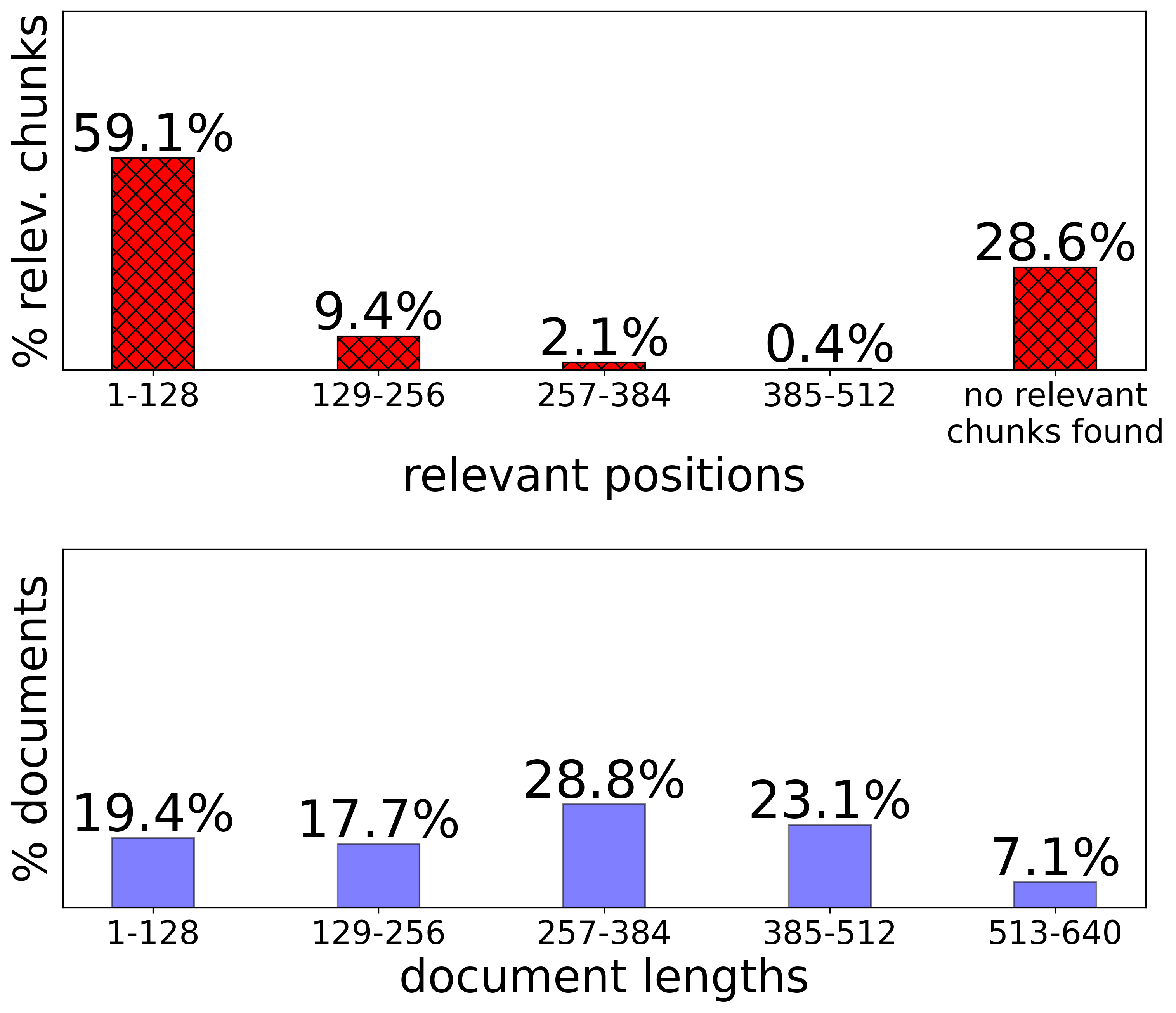}
        \caption{TREC COVID}
    \end{subfigure}
    
    \vspace{0.5cm} 

    \begin{subfigure}{0.28\textwidth}
        \centering
        \includegraphics[width=\linewidth]{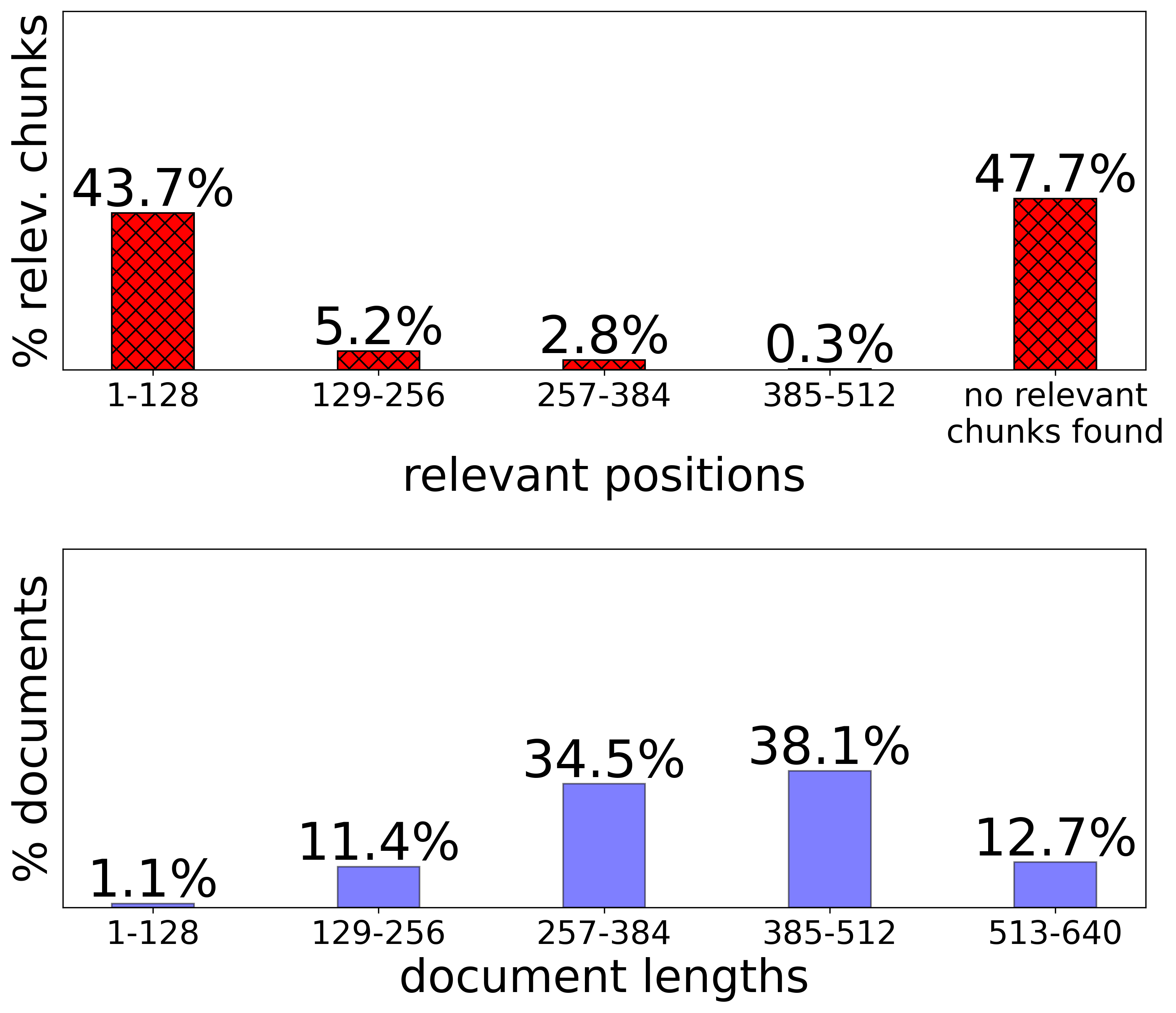}
        \caption{NFCorpus (NFC)}
    \end{subfigure}
    \begin{subfigure}{0.28\textwidth}
        \centering
        \includegraphics[width=\linewidth]{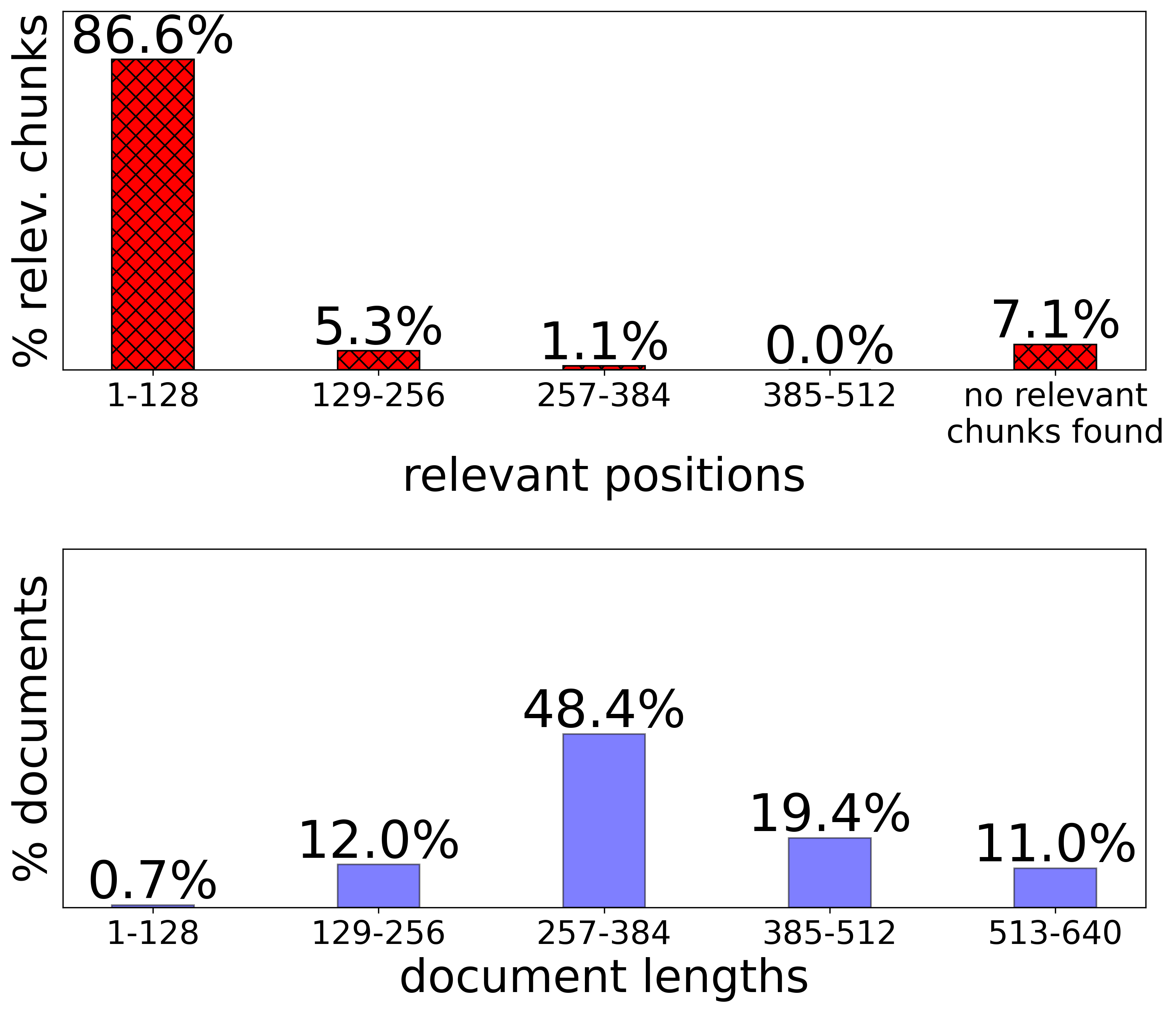}
        \caption{SciFact}
    \end{subfigure}
    \begin{subfigure}{0.28\textwidth}
        \centering
        \includegraphics[width=\linewidth]{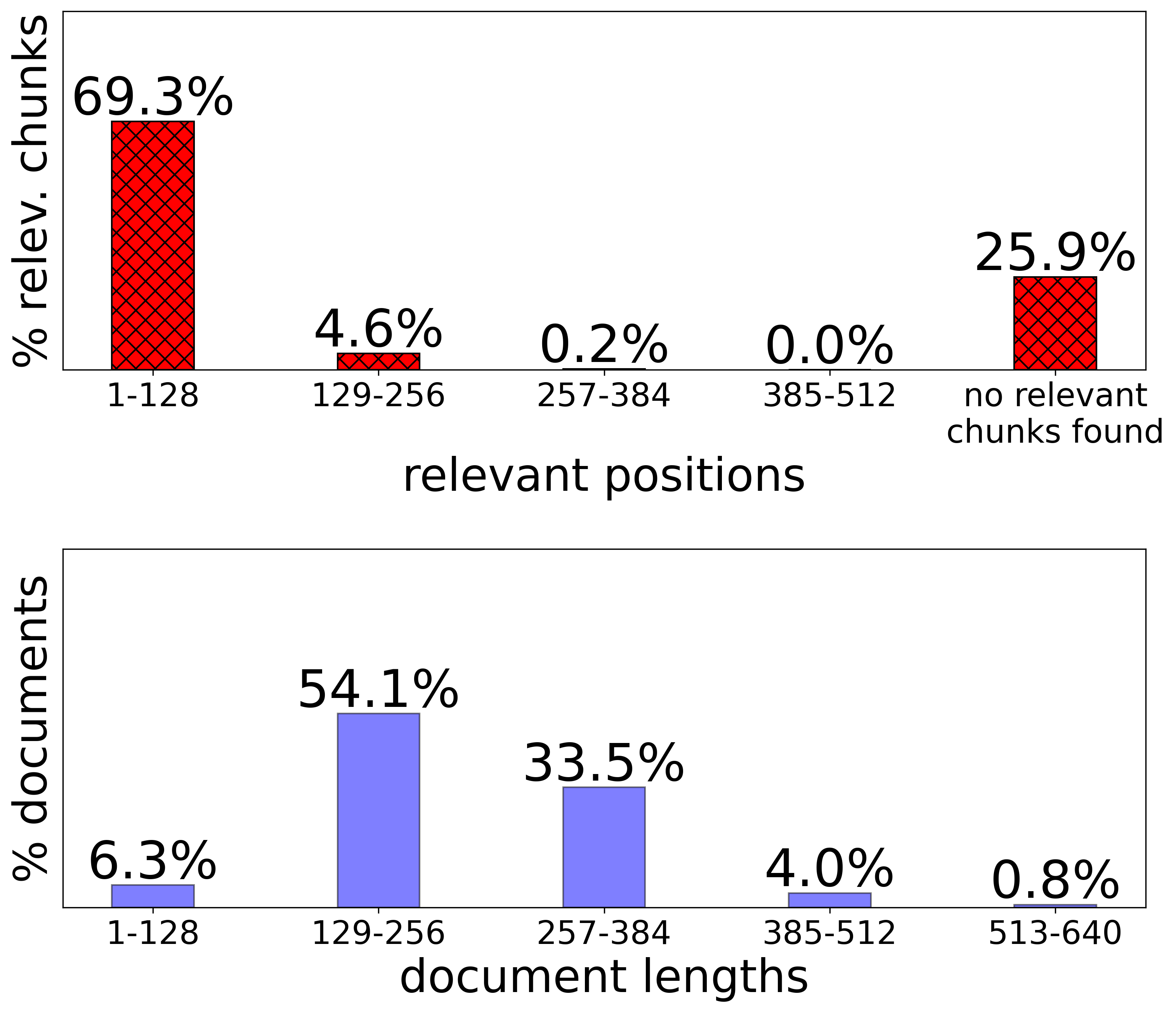}
        \caption{SciDocs}
    \end{subfigure}

    \caption{\revision{Illustration of positional relevance bias for six BEIR collections, all of which are typically categorized as 
    \emph{short-document} datasets.
    We show a distribution of first relevant passage positions (red bars) vs. relevant document lengths (blue bars).
    Positions and lengths are measured in the number of subword tokens (BERT-base tokenizer). Best viewed in color.}}
    \label{fig:relev_match_full_beir_plot}
\end{figure*}
To assess positional bias, we used a combination of approximate string matching and LLM-based judging, 
which was recently shown to highly correlate with human judgments \cite{DBLP:journals/corr/abs-2406-06519,arabzadeh2025benchmarking}.
The resulting distributions can be found in Figures~\ref{fig:relev_match_full_plot} and \ref{fig:relev_match_full_beir_plot}.

Approximate string matching was used for MS MARCO training and development (\emph{dev}) sets, both of which have sparse labels.
Although initially MS MARCO Passages were exact substrings of MS MARCO Documents, 
document and passage texts  were collected at different times. 
This led to some content divergence \cite{trecdl2020overview},
which made exact mapping  impossible in most cases.
There were prior attempts to recover initial positions, but only with a limited success (see a discussion below).

Our approximate string matching combines approximate substring matching with longest-substring matching and incorporates efficiency heuristics to identify initial candidate sets. 
Candidate sets were constructed using two approaches: selecting all relevant passages and documents (for a given query) and 
retrieving top-5 documents using relevant passages as queries.
To assess reliability, we manually inspected a subset of the matched passages and found the procedure to be sufficiently accurate. 
We then applied this approach to two sets of queries: 
\begin{itemize}
    \item A set of all 5193  queries from the \emph{devevelopment} set;
    \item A random and \emph{uniform} sample of 5000 training queries.
\end{itemize}
In both cases, we were able to find matches for about 85\% of the queries.

For queries sets with ``dense'' relevance judgments---produced by TREC NIST assessors---we used an LLM judge. 
This included TREC 2019, 2020, 2021 TREC DL queries~\cite{trecdl2021overview},
Robust04~\cite{Terabyte2004}, Gov2 with 2007, 2008 Million Query Track queries \cite{DBLP:conf/trec/AllanAPKC08},
and ClueWeb12 with 2012 and 2013 Web Track queries \cite{DBLP:conf/trec/Collins-Thompson13}.
We also used an LLM judge for BEIR datasets \cite{BEIR}.

In contrast, \citet{FIRA2020}  employed crowd-workers to identify the distribution of relevance chunks.
They found similar evidence of the relevance position bias,
but their study was limited only to the TREC DL 2019 query set.
For sparsely-judged training queries, \citet{FIRA2020} used matching on answer words---rather than passage text itself---and were able to match only 32\%  of the passages.
In particular, \cite{DBLP:conf/acl/CoelhoMMCX24} used exact matching and found only about one thousand matching passages.
Both of these attempts match only a small-to-modest fraction of passages while being subject to biases.
Finally, MS MARCO v2 authors \cite{trecdl2021overview} provided explicit mappings of relevant passages inside documents,
but how these mappings were obtained was not disclosed.

To avoid potential positional biases in LLM-judging, we divided each document into non-overlapping chunks and judged each chunk separately.
Chunking preserved sentence boundary while ensuring each chunk size was close to 256 tokens in length
for long-document collections and 128 tokens for BEIR collections \cite{BEIR}. 

For efficiency reasons, we
only considered at most 36 chunks (about 9K tokens) and 500-2000 positive query-document pairs per query set. 
A chunk was considered to be relevant if it received a positive grade from an LLM-judge.
The LLM-judge was a GPT4-mini-based UMBRELA \cite{openaiGPT2023,DBLP:journals/corr/abs-2406-06519}.

For most collections, there was only a small fraction of documents where the LLM-judge found no relevant query-document pairs. 
Among long-document datasets, one exception was the Gov2 collection with 2007, 2008 Million Query Track queries where this happened in about 30\% of the cases. 

Despite this uncertainty, Gov2 still had a substantial positional bias with about 57\% of the cases where the first chunk was deemed to be relevant (see Fig.~\ref{fig:relev_match_full_plot}).
Among BEIR datasets, five datasets also had a prominent positional bias (see Fig.~\ref{fig:relev_match_full_beir_plot}). 
However, we are less confident about the NFCorpus, where we found no relevant chunks for nearly 50\% of the query-document pairs.

\begin{table}[!htpb]
\centering
\small
\begin{tabular}{l|c}
\toprule
  Model family           &  \# of  \\ 
                         & params. \\ \midrule
  PARADE Transformer &  132-148M             \\
  Longformer             &  149M             \\
  BigBird                &  127M             \\
  JINA                   &  137M             \\
  MOSAIC                 &  137M             \\
  DEBERTA-based models         &  184M             \\
  TinyLLAMA-based models       &   1034M              \\
  Other BERT- and ELECTRA-based models& $\approx$110 M \\ \bottomrule
\end{tabular}
\caption{\revision{Number of model parameters (we do not include models behind cloud APIs)}\label{tab:model_params}}
\end{table}

\section{Ranking with Cross-Encoding Long-Document Models}
\label{sec:methods_detailed}
In this section, we describe long-document cross-encoding models in more details.
With an exception of TinyLLAMA \cite{TinyLLAMA2024} all models employ relatively small 
bidirectional encoder-only Transformers \cite{vaswani2017attention} with 
100-200M parameters in total (see Table~\ref{tab:model_params}). 
TinyLLAMA is a so-called LLM-ranker: a ``causal''
decoder-only Transformer that has about 1B parameters.
Moreover, we focus exclusively on pure Transformer architectures, leaving hybrid approaches such as RankMamba \cite{xu2024rankmamba} for future work.

We assume that an input text is split into small chunks of texts called \emph{tokens}.
Although tokens can be  complete English words, 
Transformer models usually split text into sub-word units \cite{WuSCLNMKCGMKSJL16}.

The length of a document $d$---denoted as $|d|$---is measured in the number of tokens.
Because neural networks cannot operate directly on text,
a sequence of tokens $t_1 t_2 \ldots t_n$ is first converted to a sequences 
of $d$-dimensional embedding vectors $w_1 w_2 \ldots w_n$ by an \emph{embedding} network.
These embeddings are context-independent, 
i.e., each token is always mapped to the same vector \cite{collobert2011natural,MikolovSCCD13}.

For a detailed description of Transformer models,
please see the  annotated Transformer guide \cite{rush2018annotated} as well 
as the recent survey by Lin et al.~\cite{lin2019neural}, 
which focuses on the use of BERT-style cross-encoding models for ranking and retrieval. 
For this paper, it is necessary to know only the following basic facts:
\begin{itemize}
    \item BERT is an encoder-only model,  which converts a sequence of tokens $t_1 t_2 \ldots t_n$ to a sequence 
        of $d$-dimensional vectors $w_1 w_2 \ldots w_n$.
        These vectors---which are token representations from the \emph{last} model layer---are commonly referred 
        to as contextualized token embeddings \cite{peters2018deep};
    \item BERT operates on word pieces \cite{WuSCLNMKCGMKSJL16} rather than on complete words;
    \item The vocabulary includes two special tokens: \clstok\  (an aggregator) and \septok\  (a separator);
    \item Using a \emph{pooled} representation of token vectors $w_1 w_2 \ldots w_n$, a linear layer is used to produce a ranking score. A nearly universal pooling approach in cross-encoding rankers is to use the vector of the \clstok\  token,
    i.e., $w_1$. However, we learned that some models (e.g., JINA \cite{gunther2023jina}) converge well \emph{only} with mean pooling, i.e., they use ${1 \over n}\sum_{i=1}^n w_i$.
\end{itemize}

A ``vanilla'' BERT ranker---dubbed as  monoBERT by \citet{lin2019neural}---uses a single fully-connect layer $F$ as a prediction head,
which converts the last-layer representation of the \clstok\  token (i.e., a contextualized embedding of \clstok) into 
a scalar \cite{nogueira2019passage}.
It makes a prediction based on the following  sequence of tokens:  
\clstok\  $q$ \septok\ $d$ \septok,
where $q$ is a query and $d$ is a document.

An alternative approach is to aggregate contextualized embeddings of regular tokens using a shallow neural network 
\cite{CEDR2019,BoytsovK21,KhattabZ20},
optionally combining these embeddings with the contextualized embedding of \clstok.

This  was first proposed by \citet{CEDR2019} who also found that incorporating \clstok\  improved performance.
However, \citet{BoytsovK21} proposed a shallow aggregating network that did not use the output
of the \clstok\  token  and achieved the same accuracy on MS MARCO datasets,
thus questioning the necessity of using \clstok\  for aggregation.

Replacing the standard BERT model in the vanilla BERT ranker with a BERT variant
that ``natively'' supports longer documents is conceptually the simplest way to handle long documents.
We collectively refer to these models as \emph{LongP} models.
For many older BERT models with limited context size,
long documents and queries still need to be split or truncated so that the total number of tokens does not exceed 512.
In the \emph{FirstP} mode, we process only the first chunk and ignore the truncated text.
In the \emph{SplitP} mode, each chunk is processed separately and the results are aggregated.
In the remaining of this section, we discuss \emph{LongP} and \emph{SplitP} approaches in detail.
We conclude the discussion with a description of several miscellaneous models, most of which we were unable to implement due to incomplete information or missing code.

\subsection{LongP models}
In our work, we benchmark both sparse-attention and full-attention models. 
Sparse attention LongP models include two popular variants: 
Longformer \cite{longformer2020} and Big-Bird \cite{BigBird2020}.
In this setting, we use the same approach to score documents as with the vanilla BERT ranker,
namely, concatenating queries with documents and making a prediction based 
on the contextualized embedding of the \clstok\   token \cite{nogueira2019passage}.
Both Big-Bird and Longformer use a combination of the local, ``scattered'' (our terminology), and global attention.
The local attention utilizes a sliding window of a constant length where each token attends to each
other token within this window.
In the case of the global attention, certain tokens can attend to \emph{all} other tokens and vice-versa,
In Big-Bird, only special tokens such as \clstok\  can attend globally. 
In Longformer, the user have to select such tokens explicitly. 
Following Beltagy et al.~\cite{longformer2020}, who applied this technique to question-answering,
we designate global attention only for query tokens.
Unlike the global attention, 
the scattered attention is limited to restricted sub-sets of tokens, 
but these subsets do not necessarily have locality.
In Big-Bird, the scattered attention uses randomly selected tokens,
whereas  Longformer uses a dilated sliding-window attention with layer- and head-specific dilation.

Full-attention models include JINABert \cite{gunther2023jina},
TinyLLAMA \cite{TinyLLAMA2024}, and MosaicBERT~\cite{portes2023mosaicbert}, henceforth, simply JINA, TinyLLAMA and MOSAIC. 
All these models use a recently proposed FlashAttention \cite{FlashAttention2022} 
to efficiently process long-contexts as well as special  positional embeddings 
 that can generalize to document lengths not seen during training.
 In particular, JINA and MOSAIC use AliBi \cite{press2022train}, while 
 TinyLLAM uses ROPE embeddings \cite{su2023roformer}.
 JINA and MOSAIC are bidirectional encoder-only Transformer model whereas TinyLLAMA 
 is a unidirectional (sometimes called causal) decoder-only Transformer model \cite{vaswani2017attention}.
 
Beyond architectural variations, models also vary in their pretraining strategies.
MOSAIC relies primarily on the masked language (MLM) objective without next sentence prediction (NSP). 
JINA uses this approach as a first step
and fine-tuning on retrieval and classification tasks with mean-pooled representations as a second step (following a RoBERTa pretraining strategy \cite{Roberta2019}). 
TinyLLAMA was trained using  an autoregressive language modeling objective \cite{TinyLLAMA2024}.
We found that JINA lost an ability to effectively pool on the \clstok\ token and we used mean-pooling instead. We also use mean pooling for TinyLLAMA.
For MOSAIC we used pooling on \clstok.

\subsection{SplitP models}\label{sec:split_p_desc}
SplitP models differ in partitioning and aggregation approaches.
Documents can be split into either disjoint or overlapping chunks.
In the first case, documents are split in a greedy fashion so that each document chunk---except possibly the last one---is exactly 512 tokens long (after being concatenated with a (padded) query  and three special tokens). 
Because we pad queries to be 32 tokens long, each document chunk contains at most $512 - 32 - 3 = 477$ tokens.

In the second case, we use a sliding window approach with a window size and stride that are not tied to the maximum
length of BERT input.
Because our primary focus is accuracy and we aim to understand the limits of long-document models,
we exclude from evaluation several \emph{SplitP} models,
which achieve better efficiency-effectiveness trade-offs by pre-selecting certain document parts and feeding only selected parts into a BERT ranker. 
This includes, but is not limited to, papers by \citet{HofstatterMZCH21,ZouZCMCWSCY21}.

\paragraph{Greedy partitioning into disjoint chunks}
CEDR models~\cite{CEDR2019} and the Neural Model 1~\cite{BoytsovK21} use the first method,
which involves:
\begin{itemize} 
    \item tokenizing the document $d$;
    \item greedily splitting a tokenized document $d$ into $m$ disjoint chunks: $d=d_1 d_2 \ldots d_m$; 
    \item generating $m$ token sequences \clstok\  $q$ \septok\ $d_i$ \septok\  by
          concatenating the query with document chunks;
    \item processing each sequence with a BERT model to generate contextualized embeddings 
          for regular tokens as well as for \clstok.
\end{itemize}
The outcome of this procedure is $m$ \clstok-vectors $cls_i$ and $n$ contextualized vectors $w_1 w_2 \ldots w_n$
(one for \emph{each} document token $t_i$)
that are aggregated in a model-specific way.

\citet{CEDR2019} use contextualized embeddings as a direct replacement of context-free 
embeddings in the following neural architectures: KNRM \cite{XiongDCLP17}, PACRR \cite{HuiYBM18}, and DRMM \cite{GuoFAC16}.
To boost performance, they incorporate \clstok\ -vectors  in a model-specific way.
We call the respective models as \emph{CEDR-KNRM}, \emph{CEDR-PACRR}, and \emph{CEDR-DRMM}.

They also proposed an extension of the vanilla BERT ranker that makes a prediction
using the average \clstok\  token: $\frac{1}{m}\sum_{i=1}^{m} cls_i$
by passing it through a linear projection layer.
We call this method \emph{AvgP}.

The Neural Model 1 \cite{BoytsovK21} uses the same greedy partitioning approach as CEDR, 
but a different aggregator network, which does not use the embeddings of the \clstok\  token.
This network is a neural parametrization of the classic Model 1 \cite{berger1999information,brown1993mathematics}.

\paragraph{Sliding window approach}

The BERT MaxP/SumP \cite{DaiC19} and PARADE \cite{Parade2020} models use a sliding window approach.
Assume $w$ is the size of the window and $s$ is the stride.
Then the processing can be summarized as follows:

\begin{itemize} 
    \item tokenizing, the document $d$ into sub-words $t_1 t_2 \ldots t_n$ ;
    \item splitting a tokenized document $d$ into $m$ possibly overlapping chunks $d_i = t_{i \cdot s} t_{i \cdot s + 1} \ldots t_{i \cdot s + w - 1}$, where $s$ is a stride and $w$ is a window size. Trailing chunks may have fewer than $w$ tokens.
    \item generating $m$ token sequences \clstok\  $q$ \septok\ $d_i$ \septok\  by
          concatenating the query with document chunks;
    \item processing each sequence with a BERT model to generate a last-layer output
          for each sequence \clstok\  token.
\end{itemize}
The outcome of this procedure is $m$ \clstok-vectors $cls_i$, which are subsequently aggregated in a model-specific way.
Note that  PARADE and MaxP/SumP models do not use contextualized embeddings of regular tokens.

\paragraph{BERT MaxP/SumP} These models use a linear layer $F$ to 
produce $m$ chunk-relevance scores $F(cls_i)$.
Document scores are computed 
as $\max_{i=1}^m F(cls_i)$ and $\sum_{i=1}^m F(cls_i)$
for  MaxP and SumP models, respectively  \cite{DaiC19}.

\paragraph{PARADE} These models \cite{Parade2020} can be divided into two groups.
The first group includes PARADE average, PARADE max,
and PARADE attention models, which all use simple approaches to
produce an aggregated representation of $m$ \clstok-vectors $cls_i$.
To compute a document-relevance score these aggregated representations are passed
through a linear layer $F$.

In particular, PARADE average and PARADE max 
combine $cls_i$ using  averaging and the element-wise maximum operation, respectively, 
to generate aggregated representation of $m$
\clstok\  tokens $cls_i$.\footnote{Note that both PARADE average and AvgP vanilla ranker use
the same approach to aggregate contextualized embeddings of \clstok\ tokens,
but they differ in their approach to select document chunks.
In particular, AvgP uses non-overlapping chunks while
PARADE average relies on the sliding window approach.}
The PARADE attention model uses a learnable attention \cite{BahdanauCB14} vector $C$ 
to compute a scalar weight $w_i$ of each $cls_i$ as follows: 
$w_1 w_2 \ldots w_m = softmax (C\cdot cls_1, C \cdot  cls_2, \ldots , C \cdot cls_m)$.
These weights are used to compute the aggregated representation as
$\sum_{i=1}^{m} w_i cls_i$

PARADE Transformer models aggregates \clstok-vectors  $cls_i$ using an additional 
 transformer model $AggregTransf()$, i.e., an \emph{aggregator} model. 
The input of the aggregator Transformer is a sequence of $cls_i$ vectors 
prepended with a learnable vector $C$, which plays a role
of a \clstok\  embedding for $AggregTransf()$. 
Optionally, one can use learnable positional embedding vectors together with C and $\{cls_i\}$.
The last-layer representation of the first vector
is passed through a linear layer $F$ to produce a relevance score:
\begin{equation}\label{eq:aggreg}
F(AggregTransf(C, cls_1, cls_2, \ldots , cls_m)[0]) 
\end{equation}

An aggregator Transformer can be either pretrained or randomly initialized.
In the case of a pretrained transformer, we completely discard the embedding layer.
Furthermore, if the dimensionality of $cls_i$ vectors is different from the dimensionality of input embeddings in $AggregTransf$, 
we project $cls_i$ using a linear transformation.

\paragraph{Miscellaneous models}
We attempted to implement additional state-of-the-art models \cite{luyu2022moresplus,Chengzhen2022gnn}. 
Gao and Callan \cite{luyu2022moresplus} introduced a  late-interaction model MORES+,
which is a modular long document reranker that uses a sequence-to-sequence transformer in a 
non-auto-regressive mode.
In MORES+ document chunks are first encoded using the encoder-only Transformer model.
Then they use a modified decoder Transformer for joint query-to-all-document-chunk cross-attention:
This modification changes a causal Transformer into an encoder-only bidirectional Transformer model.
As of the moment of writing, the MORES+ model holds the first position on a competitive MS MARCO document leaderboard.\footnote{\url{https://microsoft.github.io/MSMARCO-Document-Ranking-Submissions/leaderboard/}}.
However, the authors provide only incomplete implementation which does not fully match the description in the paper (i.e., crucial details are missing). We reimplemented this model to the best of our understanding, but our implementation failed
to outperform even BM25.

Inspired by this approach, we managed to implement a late-interaction variant of the PARADE model, which we denoted
as PARADE-LATEIR. Similar to the original PARADE model, it splits documents into overlapping chunks. However, it then
encodes chunks and queries independently. Next, it uses an interaction Transformer to (1) mix these representations,
and (2) combine output using an aggregator Transformer.
In total, the model uses three backbone encoder-only Transformers: 
All of these Transformers are initialized using pretrained models.

\citet{Chengzhen2022gnn} proposed a multi-view  interactions-based ranking model (MIR). 
They implement inter-passage interactions via a multi-view attention mechanism, 
which enables information propagation at token, sentence, and passage levels. 
Due to the computational complexity, they restrict these interactions to a set of salient/pivot tokens.
However, the paper does not provide enough details regarding the choices of these tokens.
There is no software available and authors did not respond to our clarification requests.
Thus, this model is also excluded from our evaluation.


\begin{thebibliography}{103}
\expandafter\ifx\csname natexlab\endcsname\relax\def\natexlab#1{#1}\fi

\bibitem[{Allan et~al.(2008)Allan, Aslam, Pavlu, Kanoulas, and
  Carterette}]{DBLP:conf/trec/AllanAPKC08}
James Allan, Javed~A. Aslam, Virgil Pavlu, Evangelos Kanoulas, and Ben
  Carterette. 2008.
\newblock Million query track 2008 overview.
\newblock In \emph{{TREC}}, volume 500-277 of \emph{{NIST} Special
  Publication}. National Institute of Standards and Technology {(NIST)}.

\bibitem[{An et~al.(2024)An, Ma, Lin, Zheng, Lou, and
  Chen}]{DBLP:conf/nips/AnML0LC24}
Shengnan An, Zexiong Ma, Zeqi Lin, Nanning Zheng, Jian{-}Guang Lou, and Weizhu
  Chen. 2024.
\newblock Make your {LLM} fully utilize the context.
\newblock In \emph{NeurIPS}.

\bibitem[{Anthropic(2024)}]{anthropic2024claude3}
Anthropic. 2024.
\newblock \href {https://www.anthropic.com/news/claude-3-family} {The claude 3
  model family: Opus, sonnet, haiku}.

\bibitem[{Arabzadeh and Clarke(2025)}]{arabzadeh2025benchmarking}
Negar Arabzadeh and Charles~LA Clarke. 2025.
\newblock Benchmarking llm-based relevance judgment methods.
\newblock \emph{arXiv preprint arXiv:2504.12558}.

\bibitem[{Bahdanau et~al.(2015)Bahdanau, Cho, and Bengio}]{BahdanauCB14}
Dzmitry Bahdanau, Kyunghyun Cho, and Yoshua Bengio. 2015.
\newblock Neural machine translation by jointly learning to align and
  translate.
\newblock In \emph{3rd International Conference on Learning Representations,
  {ICLR} 2015}.

\bibitem[{Bajaj et~al.(2016)Bajaj, Campos, Craswell, Deng, Gao, Liu, Majumder,
  McNamara, Mitra, Nguyen et~al.}]{msmarco}
Payal Bajaj, Daniel Campos, Nick Craswell, Li~Deng, Jianfeng Gao, Xiaodong Liu,
  Rangan Majumder, Andrew McNamara, Bhaskar Mitra, Tri Nguyen, et~al. 2016.
\newblock {MS MARCO}: A human generated machine reading comprehension dataset.
\newblock \emph{arXiv preprint arXiv:1611.09268}.

\bibitem[{Beltagy et~al.(2020)Beltagy, Peters, and Cohan}]{longformer2020}
Iz~Beltagy, Matthew~E. Peters, and Arman Cohan. 2020.
\newblock Longformer: The long-document transformer.
\newblock \emph{CoRR}, abs/2004.05150.

\bibitem[{Berger and Lafferty(1999)}]{berger1999information}
Adam Berger and John Lafferty. 1999.
\newblock Information retrieval as statistical translation.
\newblock In \emph{Proceedings of the 22nd annual international ACM SIGIR
  conference on Research and development in information retrieval}, pages
  222--229.

\bibitem[{Bondarenko et~al.(2020)Bondarenko, Fr{\"{o}}be, Beloucif, Gienapp,
  Ajjour, Panchenko, Biemann, Stein, Wachsmuth, Potthast, and
  Hagen}]{touche2020}
Alexander Bondarenko, Maik Fr{\"{o}}be, Meriem Beloucif, Lukas Gienapp, Yamen
  Ajjour, Alexander Panchenko, Chris Biemann, Benno Stein, Henning Wachsmuth,
  Martin Potthast, and Matthias Hagen. 2020.
\newblock Overview of touch{\'{e}} 2020: Argument retrieval - extended
  abstract.
\newblock In \emph{{CLEF}}, volume 12260 of \emph{Lecture Notes in Computer
  Science}, pages 384--395. Springer.

\bibitem[{Boteva et~al.(2016)Boteva, Gholipour, Sokolov, and Riezler}]{nfc}
Vera Boteva, Demian Gholipour, Artem Sokolov, and Stefan Riezler. 2016.
\newblock A full-text learning to rank dataset for medical information
  retrieval.
\newblock In \emph{Advances in Information Retrieval: 38th European Conference
  on IR Research, ECIR 2016, Padua, Italy, March 20--23, 2016. Proceedings 38},
  pages 716--722. Springer.

\bibitem[{Boytsov and Kolter(2021)}]{BoytsovK21}
Leonid Boytsov and Zico Kolter. 2021.
\newblock Exploring classic and neural lexical translation models for
  information retrieval: Interpretability, effectiveness, and efficiency
  benefits.
\newblock In \emph{{ECIR} {(1)}}, volume 12656 of \emph{Lecture Notes in
  Computer Science}, pages 63--78. Springer.

\bibitem[{Boytsov et~al.(2022)Boytsov, Lin, Gao, Zhao, Huang, and
  Nyberg}]{boytsov2022understanding}
Leonid Boytsov, Tianyi Lin, Fangwei Gao, Yutian Zhao, Jeffrey Huang, and Eric
  Nyberg. 2022.
\newblock \href {https://arxiv.org/abs/2207.01262v1} {Understanding performance
  of long-document ranking models through comprehensive evaluation and
  leaderboarding}.
\newblock \emph{CoRR}, abs/2207.01262v1.

\bibitem[{Boytsov and Naidan(2013)}]{boytsov2013engineering}
Leonid Boytsov and Bilegsaikhan Naidan. 2013.
\newblock Engineering efficient and effective non-metric space library.
\newblock In \emph{International Conference on Similarity Search and
  Applications}, pages 280--293. Springer.

\bibitem[{Boytsov and Nyberg(2020)}]{FlexNeuART}
Leonid Boytsov and Eric Nyberg. 2020.
\newblock Flexible retrieval with {NMSLIB} and {FlexNeuART}.
\newblock In \emph{Proceedings of Second Workshop for NLP Open Source Software
  (NLP-OSS)}, pages 32--43.

\bibitem[{Brown et~al.(1993)Brown, Pietra, Pietra, and
  Mercer}]{brown1993mathematics}
Peter~F. Brown, Stephen~Della Pietra, Vincent J.~Della Pietra, and Robert~L.
  Mercer. 1993.
\newblock The mathematics of statistical machine translation: Parameter
  estimation.
\newblock \emph{Computational Linguistics}, 19(2):263--311.

\bibitem[{Clark et~al.(2020)Clark, Luong, Le, and Manning}]{ELECTRA2020}
Kevin Clark, Minh{-}Thang Luong, Quoc~V. Le, and Christopher~D. Manning. 2020.
\newblock {ELECTRA:} pre-training text encoders as discriminators rather than
  generators.
\newblock In \emph{{ICLR}}. OpenReview.net.

\bibitem[{Clarke et~al.(2004)Clarke, Craswell, and Soboroff}]{Terabyte2004}
Charles L.~A. Clarke, Nick Craswell, and Ian Soboroff. 2004.
\newblock Overview of the {TREC} 2004 terabyte track.
\newblock In \emph{{TREC}}, volume 500-261 of \emph{{NIST} Special
  Publication}. National Institute of Standards and Technology {(NIST)}.

\bibitem[{Coelho et~al.(2024)Coelho, Martins, Magalh{\~{a}}es, Callan, and
  Xiong}]{DBLP:conf/acl/CoelhoMMCX24}
Jo{\~{a}}o Coelho, Bruno Martins, Jo{\~{a}}o Magalh{\~{a}}es, Jamie Callan, and
  Chenyan Xiong. 2024.
\newblock Dwell in the beginning: How language models embed long documents for
  dense retrieval.
\newblock In \emph{{ACL} (Short Papers)}, pages 370--377. Association for
  Computational Linguistics.

\bibitem[{Cohan et~al.(2020)Cohan, Feldman, Beltagy, Downey, and
  Weld}]{scidocs}
Arman Cohan, Sergey Feldman, Iz~Beltagy, Doug Downey, and Daniel~S Weld. 2020.
\newblock Specter: Document-level representation learning using
  citation-informed transformers.
\newblock In \emph{Proceedings of the 58th Annual Meeting of the Association
  for Computational Linguistics}, pages 2270--2282.

\bibitem[{Collins{-}Thompson et~al.(2013{\natexlab{a}})Collins{-}Thompson,
  Bennett, Diaz, Clarke, and Voorhees}]{Web2012}
Kevyn Collins{-}Thompson, Paul~N. Bennett, Fernando Diaz, Charlie Clarke, and
  Ellen~M. Voorhees. 2013{\natexlab{a}}.
\newblock {TREC} 2013 web track overview.
\newblock In \emph{{TREC}}, volume 500-302 of \emph{{NIST} Special
  Publication}. National Institute of Standards and Technology {(NIST)}.

\bibitem[{Collins{-}Thompson et~al.(2013{\natexlab{b}})Collins{-}Thompson,
  Bennett, Diaz, Clarke, and Voorhees}]{DBLP:conf/trec/Collins-Thompson13}
Kevyn Collins{-}Thompson, Paul~N. Bennett, Fernando Diaz, Charlie Clarke, and
  Ellen~M. Voorhees. 2013{\natexlab{b}}.
\newblock {TREC} 2013 web track overview.
\newblock In \emph{{TREC}}, volume 500-302 of \emph{{NIST} Special
  Publication}. National Institute of Standards and Technology {(NIST)}.

\bibitem[{Collobert et~al.(2011)Collobert, Weston, Bottou, Karlen, Kavukcuoglu,
  and Kuksa}]{collobert2011natural}
Ronan Collobert, Jason Weston, L{\'e}on Bottou, Michael Karlen, Koray
  Kavukcuoglu, and Pavel Kuksa. 2011.
\newblock Natural language processing (almost) from scratch.
\newblock \emph{J. Mach. Learn. Res.}, 12:2493--2537.

\bibitem[{Craswell et~al.(2021{\natexlab{a}})Craswell, Mitra, Yilmaz, and
  Campos}]{trecdl2020overview}
Nick Craswell, Bhaskar Mitra, Emine Yilmaz, and Daniel Campos.
  2021{\natexlab{a}}.
\newblock Overview of the {TREC} 2020 deep learning track.
\newblock \emph{CoRR}, abs/2102.07662.

\bibitem[{Craswell et~al.(2021{\natexlab{b}})Craswell, Mitra, Yilmaz, Campos,
  and Lin}]{trecdl2021overview}
Nick Craswell, Bhaskar Mitra, Emine Yilmaz, Daniel Campos, and Jimmy Lin.
  2021{\natexlab{b}}.
\newblock Overview of the {TREC} 2021 deep learning track.

\bibitem[{Craswell et~al.(2022)Craswell, Mitra, Yilmaz, Campos, Lin, Voorhees,
  and Soboroff}]{trecdl2022overview}
Nick Craswell, Bhaskar Mitra, Emine Yilmaz, Daniel Campos, Jimmy Lin, Ellen~M.
  Voorhees, and Ian Soboroff. 2022.
\newblock \href {https://trec.nist.gov/pubs/trec31/papers/Overview\_deep.pdf}
  {Overview of the {TREC} 2022 deep learning track}.
\newblock In \emph{Proceedings of the Thirty-First Text REtrieval Conference,
  {TREC} 2022, online, November 15-19, 2022}, volume 500-338 of \emph{{NIST}
  Special Publication}. National Institute of Standards and Technology
  {(NIST)}.

\bibitem[{Craswell et~al.(2020)Craswell, Mitra, Yilmaz, Campos, and
  Voorhees}]{trecdl2019overview}
Nick Craswell, Bhaskar Mitra, Emine Yilmaz, Daniel Campos, and Ellen~M.
  Voorhees. 2020.
\newblock Overview of the {TREC} 2019 deep learning track.
\newblock \emph{CoRR}, abs/2003.07820.

\bibitem[{Craswell et~al.(2023)Craswell, Mitra, Yilmaz, Rahmani, Campos, Lin,
  Voorhees, and Soboroff}]{trecdl2023overview}
Nick Craswell, Bhaskar Mitra, Emine Yilmaz, Hossein~A. Rahmani, Daniel Campos,
  Jimmy Lin, Ellen~M. Voorhees, and Ian Soboroff. 2023.
\newblock Overview of the {TREC} 2023 deep learning track.
\newblock In \emph{{TREC}}, volume 500-xxx of \emph{{NIST} Special
  Publication}. National Institute of Standards and Technology {(NIST)}.

\bibitem[{Dai and Callan(2019)}]{DaiC19}
Zhuyun Dai and Jamie Callan. 2019.
\newblock Deeper text understanding for {IR} with contextual neural language
  modeling.
\newblock In \emph{{SIGIR}}, pages 985--988. {ACM}.

\bibitem[{Dao et~al.(2022)Dao, Fu, Ermon, Rudra, and
  R{\'{e}}}]{FlashAttention2022}
Tri Dao, Daniel~Y. Fu, Stefano Ermon, Atri Rudra, and Christopher R{\'{e}}.
  2022.
\newblock Flashattention: Fast and memory-efficient exact attention with
  io-awareness.
\newblock In \emph{NeurIPS}.

\bibitem[{Devlin et~al.(2019)Devlin, Chang, Lee, and
  Toutanova}]{devlin2018bert}
Jacob Devlin, Ming{-}Wei Chang, Kenton Lee, and Kristina Toutanova. 2019.
\newblock {BERT:} pre-training of deep bidirectional transformers for language
  understanding.
\newblock pages 4171--4186.

\bibitem[{Formal et~al.(2021)Formal, Lassance, Piwowarski, and
  Clinchant}]{SPLADEv2}
Thibault Formal, Carlos Lassance, Benjamin Piwowarski, and St{\'{e}}phane
  Clinchant. 2021.
\newblock {SPLADE} v2: Sparse lexical and expansion model for information
  retrieval.
\newblock \emph{CoRR}, abs/2109.10086.

\bibitem[{Fu et~al.(2022)Fu, Hu, Feng, Dou, Jia, Chen, Yu, and
  Cao}]{Chengzhen2022gnn}
Chengzhen Fu, Enrui Hu, Letian Feng, Zhicheng Dou, Yantao Jia, Lei Chen, Fan
  Yu, and Zhao Cao. 2022.
\newblock Leveraging multi-view inter-passage interactions for neural document
  ranking.
\newblock In \emph{Proceedings of the Fifteenth ACM International Conference on
  Web Search and Data Mining}, WSDM '22, page 298–306, New York, NY, USA.
  Association for Computing Machinery.

\bibitem[{Gao and Callan(2022)}]{luyu2022moresplus}
Luyu Gao and Jamie Callan. 2022.
\newblock Long document re-ranking with modular re-ranker.
\newblock In \emph{Proceedings of the 45th International ACM SIGIR Conference
  on Research and Development in Information Retrieval}, SIGIR '22, page
  2371–2376, New York, NY, USA. Association for Computing Machinery.

\bibitem[{Guo et~al.(2016)Guo, Fan, Ai, and Croft}]{GuoFAC16}
Jiafeng Guo, Yixing Fan, Qingyao Ai, and W.~Bruce Croft. 2016.
\newblock A deep relevance matching model for ad-hoc retrieval.
\newblock In \emph{{CIKM}}, pages 55--64. {ACM}.

\bibitem[{Guo et~al.(2019)Guo, Fan, Pang, Yang, Ai, Zamani, Wu, Croft, and
  Cheng}]{guo2019deep}
Jiafeng Guo, Yixing Fan, Liang Pang, Liu Yang, Qingyao Ai, Hamed Zamani, Chen
  Wu, W~Bruce Croft, and Xueqi Cheng. 2019.
\newblock A deep look into neural ranking models for information retrieval.
\newblock \emph{Information Processing \& Management}, page 102067.

\bibitem[{Guo et~al.(2022)Guo, Ainslie, Uthus, Ontanon, Ni, Sung, and
  Yang}]{guo2022longt5}
Mandy Guo, Joshua Ainslie, David Uthus, Santiago Ontanon, Jianmo Ni, Yun-Hsuan
  Sung, and Yinfei Yang. 2022.
\newblock \href {https://doi.org/10.18653/v1/2022.findings-naacl.55}
  {{L}ong{T}5: {E}fficient text-to-text transformer for long sequences}.
\newblock In \emph{Findings of the Association for Computational Linguistics:
  NAACL 2022}, pages 724--736, Seattle, United States. Association for
  Computational Linguistics.

\bibitem[{Günther et~al.(2023)Günther, Ong, Mohr, Abdessalem, Abel, Akram,
  Guzman, Mastrapas, Sturua, Wang, Werk, Wang, and Xiao}]{gunther2023jina}
Michael Günther, Jackmin Ong, Isabelle Mohr, Alaeddine Abdessalem, Tanguy
  Abel, Mohammad~Kalim Akram, Susana Guzman, Georgios Mastrapas, Saba Sturua,
  Bo~Wang, Maximilian Werk, Nan Wang, and Han Xiao. 2023.
\newblock \href {http://arxiv.org/abs/2310.19923} {Jina embeddings 2:
  8192-token general-purpose text embeddings for long documents}.

\bibitem[{Hasibi et~al.(2017)Hasibi, Nikolaev, Xiong, Balog, Bratsberg, Kotov,
  and Callan}]{DBLP:conf/sigir/HasibiNXBBKC17}
Faegheh Hasibi, Fedor Nikolaev, Chenyan Xiong, Krisztian Balog, Svein~Erik
  Bratsberg, Alexander Kotov, and Jamie Callan. 2017.
\newblock Dbpedia-entity v2: {A} test collection for entity search.
\newblock In \emph{{SIGIR}}, pages 1265--1268. {ACM}.

\bibitem[{He et~al.(2021)He, Gao, and Chen}]{he2021debertav3}
Pengcheng He, Jianfeng Gao, and Weizhu Chen. 2021.
\newblock \href {http://arxiv.org/abs/2111.09543} {Debertav3: Improving deberta
  using electra-style pre-training with gradient-disentangled embedding
  sharing}.

\bibitem[{Hofst{\"{a}}tter et~al.(2021{\natexlab{a}})Hofst{\"{a}}tter, Lipani,
  Althammer, Zlabinger, and Hanbury}]{DBLP:conf/ecir/HofstatterLAZH21}
Sebastian Hofst{\"{a}}tter, Aldo Lipani, Sophia Althammer, Markus Zlabinger,
  and Allan Hanbury. 2021{\natexlab{a}}.
\newblock Mitigating the position bias of transformer models in passage
  re-ranking.
\newblock In \emph{{ECIR} {(1)}}, volume 12656 of \emph{Lecture Notes in
  Computer Science}, pages 238--253. Springer.

\bibitem[{Hofst{\"{a}}tter et~al.(2021{\natexlab{b}})Hofst{\"{a}}tter, Mitra,
  Zamani, Craswell, and Hanbury}]{HofstatterMZCH21}
Sebastian Hofst{\"{a}}tter, Bhaskar Mitra, Hamed Zamani, Nick Craswell, and
  Allan Hanbury. 2021{\natexlab{b}}.
\newblock Intra-document cascading: Learning to select passages for neural
  document ranking.
\newblock In \emph{{SIGIR}}, pages 1349--1358. {ACM}.

\bibitem[{Hofst{\"{a}}tter et~al.(2020{\natexlab{a}})Hofst{\"{a}}tter,
  Zlabinger, and Hanbury}]{HofstatterZH20}
Sebastian Hofst{\"{a}}tter, Markus Zlabinger, and Allan Hanbury.
  2020{\natexlab{a}}.
\newblock Interpretable {\&} time-budget-constrained contextualization for
  re-ranking.
\newblock In \emph{{ECAI}}, volume 325 of \emph{Frontiers in Artificial
  Intelligence and Applications}, pages 513--520. {IOS} Press.

\bibitem[{Hofst{\"{a}}tter et~al.(2020{\natexlab{b}})Hofst{\"{a}}tter,
  Zlabinger, Sertkan, Schr{\"{o}}der, and Hanbury}]{FIRA2020}
Sebastian Hofst{\"{a}}tter, Markus Zlabinger, Mete Sertkan, Michael
  Schr{\"{o}}der, and Allan Hanbury. 2020{\natexlab{b}}.
\newblock Fine-grained relevance annotations for multi-task document ranking
  and question answering.
\newblock In \emph{{CIKM}}, pages 3031--3038. {ACM}.

\bibitem[{Hsieh et~al.(2024)Hsieh, Chuang, Li, Wang, Le, Kumar, Glass, Ratner,
  Lee, Krishna, and Pfister}]{DBLP:conf/acl/HsiehCL0LKGRLKP24}
Cheng{-}Yu Hsieh, Yung{-}Sung Chuang, Chun{-}Liang Li, Zifeng Wang, Long~T. Le,
  Abhishek Kumar, James~R. Glass, Alexander Ratner, Chen{-}Yu Lee, Ranjay
  Krishna, and Tomas Pfister. 2024.
\newblock Found in the middle: Calibrating positional attention bias improves
  long context utilization.
\newblock In \emph{{ACL} (Findings)}, pages 14982--14995. Association for
  Computational Linguistics.

\bibitem[{Hui et~al.(2018)Hui, Yates, Berberich, and de~Melo}]{HuiYBM18}
Kai Hui, Andrew Yates, Klaus Berberich, and Gerard de~Melo. 2018.
\newblock Co-pacrr: {A} context-aware neural {IR} model for ad-hoc retrieval.
\newblock In \emph{{WSDM}}, pages 279--287. {ACM}.

\bibitem[{Huston and Croft(2014)}]{huston2014comparison}
Samuel Huston and W~Bruce Croft. 2014.
\newblock A comparison of retrieval models using term dependencies.
\newblock In \emph{Proceedings of the 23rd ACM International Conference on
  Conference on Information and Knowledge Management}, pages 111--120.

\bibitem[{Jaleel et~al.(2004)Jaleel, Allan, Croft, Diaz, Larkey, Li, Smucker,
  and Wade}]{RM3}
Nasreen~Abdul Jaleel, James Allan, W.~Bruce Croft, Fernando Diaz, Leah~S.
  Larkey, Xiaoyan Li, Mark~D. Smucker, and Courtney Wade. 2004.
\newblock Umass at {TREC} 2004: Novelty and {HARD}.
\newblock In \emph{{TREC}}, volume 500-261 of \emph{{NIST} Special
  Publication}. National Institute of Standards and Technology {(NIST)}.

\bibitem[{J{\"{a}}rvelin and
  Kek{\"{a}}l{\"{a}}inen(2002)}]{DBLP:journals/tois/JarvelinK02}
Kalervo J{\"{a}}rvelin and Jaana Kek{\"{a}}l{\"{a}}inen. 2002.
\newblock Cumulated gain-based evaluation of {IR} techniques.
\newblock \emph{{ACM} Trans. Inf. Syst.}, 20(4):422--446.

\bibitem[{Karpukhin et~al.(2020)Karpukhin, Oguz, Min, Lewis, Wu, Edunov, Chen,
  and Yih}]{DPR2020}
Vladimir Karpukhin, Barlas Oguz, Sewon Min, Patrick S.~H. Lewis, Ledell Wu,
  Sergey Edunov, Danqi Chen, and Wen{-}tau Yih. 2020.
\newblock Dense passage retrieval for open-domain question answering.
\newblock In \emph{{EMNLP} {(1)}}, pages 6769--6781. Association for
  Computational Linguistics.

\bibitem[{Khattab and Zaharia(2020)}]{KhattabZ20}
Omar Khattab and Matei Zaharia. 2020.
\newblock Colbert: Efficient and effective passage search via contextualized
  late interaction over {BERT}.
\newblock In \emph{{SIGIR}}, pages 39--48. {ACM}.

\bibitem[{Kwiatkowski et~al.(2019)Kwiatkowski, Palomaki, Redfield, Collins,
  Parikh, Alberti, Epstein, Polosukhin, Devlin, Lee, Toutanova, Jones, Kelcey,
  Chang, Dai, Uszkoreit, Le, and Petrov}]{NaturalQuestions2019}
Tom Kwiatkowski, Jennimaria Palomaki, Olivia Redfield, Michael Collins,
  Ankur~P. Parikh, Chris Alberti, Danielle Epstein, Illia Polosukhin, Jacob
  Devlin, Kenton Lee, Kristina Toutanova, Llion Jones, Matthew Kelcey,
  Ming{-}Wei Chang, Andrew~M. Dai, Jakob Uszkoreit, Quoc Le, and Slav Petrov.
  2019.
\newblock \href {https://doi.org/10.1162/TACL\_A\_00276} {Natural questions: a
  benchmark for question answering research}.
\newblock \emph{Trans. Assoc. Comput. Linguistics}, 7:452--466.

\bibitem[{Li et~al.(2024)Li, Yates, MacAvaney, He, and Sun}]{Parade2020}
Canjia Li, Andrew Yates, Sean MacAvaney, Ben He, and Yingfei Sun. 2024.
\newblock \href {https://doi.org/10.1145/3600088} {{PARADE:} passage
  representation aggregation for document reranking}.
\newblock \emph{{ACM} Trans. Inf. Syst.}, 42(2):36:1--36:26.

\bibitem[{Lin(2019)}]{lin2019neural}
Jimmy Lin. 2019.
\newblock The neural hype and comparisons against weak baselines.
\newblock In \emph{ACM SIGIR Forum}, volume~52, pages 40--51. ACM New York, NY,
  USA.

\bibitem[{Lin et~al.(2021)Lin, Nogueira, and Yates}]{2021LinNY}
Jimmy Lin, Rodrigo Nogueira, and Andrew Yates. 2021.
\newblock \emph{Pretrained Transformers for Text Ranking: {BERT} and Beyond}.
\newblock Synthesis Lectures on Human Language Technologies. Morgan {\&}
  Claypool Publishers.

\bibitem[{Liu et~al.(2024)Liu, Lin, Hewitt, Paranjape, Bevilacqua, Petroni, and
  Liang}]{DBLP:journals/tacl/LiuLHPBPL24}
Nelson~F. Liu, Kevin Lin, John Hewitt, Ashwin Paranjape, Michele Bevilacqua,
  Fabio Petroni, and Percy Liang. 2024.
\newblock Lost in the middle: How language models use long contexts.
\newblock \emph{Trans. Assoc. Comput. Linguistics}, 12:157--173.

\bibitem[{Liu et~al.(2019)Liu, Ott, Goyal, Du, Joshi, Chen, Levy, Lewis,
  Zettlemoyer, and Stoyanov}]{Roberta2019}
Yinhan Liu, Myle Ott, Naman Goyal, Jingfei Du, Mandar Joshi, Danqi Chen, Omer
  Levy, Mike Lewis, Luke Zettlemoyer, and Veselin Stoyanov. 2019.
\newblock Roberta: {A} robustly optimized {BERT} pretraining approach.
\newblock \emph{CoRR}, abs/1907.11692.

\bibitem[{Loshchilov and Hutter(2017)}]{loshchilov2017decoupled}
Ilya Loshchilov and Frank Hutter. 2017.
\newblock Decoupled weight decay regularization.
\newblock \emph{arXiv preprint arXiv:1711.05101}.

\bibitem[{Ma et~al.(2023)Ma, Wang, Yang, Wei, and
  Lin}]{DBLP:journals/corr/abs-2310-08319}
Xueguang Ma, Liang Wang, Nan Yang, Furu Wei, and Jimmy Lin. 2023.
\newblock Fine-tuning llama for multi-stage text retrieval.
\newblock \emph{CoRR}, abs/2310.08319.

\bibitem[{MacAvaney et~al.(2022)MacAvaney, Feldman, Goharian, Downey, and
  Cohan}]{DBLP:journals/tacl/MacAvaneyFGDC22}
Sean MacAvaney, Sergey Feldman, Nazli Goharian, Doug Downey, and Arman Cohan.
  2022.
\newblock {ABNIRML:} analyzing the behavior of neural {IR} models.
\newblock \emph{Trans. Assoc. Comput. Linguistics}, 10:224--239.

\bibitem[{MacAvaney et~al.(2019)MacAvaney, Yates, Cohan, and
  Goharian}]{CEDR2019}
Sean MacAvaney, Andrew Yates, Arman Cohan, and Nazli Goharian. 2019.
\newblock {CEDR:} contextualized embeddings for document ranking.
\newblock In \emph{{SIGIR}}, pages 1101--1104. {ACM}.

\bibitem[{MacAvaney et~al.(2021)MacAvaney, Yates, Feldman, Downey, Cohan, and
  Goharian}]{irds2021}
Sean MacAvaney, Andrew Yates, Sergey Feldman, Doug Downey, Arman Cohan, and
  Nazli Goharian. 2021.
\newblock Simplified data wrangling with ir-datasets.
\newblock In \emph{SIGIR}.

\bibitem[{Mikolov et~al.(2013)Mikolov, Sutskever, Chen, Corrado, and
  Dean}]{MikolovSCCD13}
Tomas Mikolov, Ilya Sutskever, Kai Chen, Gregory~S. Corrado, and Jeffrey Dean.
  2013.
\newblock Distributed representations of words and phrases and their
  compositionality.
\newblock In \emph{{NIPS}}, pages 3111--3119.

\bibitem[{Mokrii et~al.(2021)Mokrii, Boytsov, and
  Braslavski}]{Boytsov_Pseudo_2021}
Iurii Mokrii, Leonid Boytsov, and Pavel Braslavski. 2021.
\newblock \href {https://doi.org/10.1145/3404835.3463093} {\emph{A Systematic
  Evaluation of Transfer Learning and Pseudo-Labeling with BERT-Based Ranking
  Models}}, page 2081–2085. Association for Computing Machinery, New York,
  NY, USA.

\bibitem[{Mosbach et~al.(2020)Mosbach, Andriushchenko, and
  Klakow}]{Mosbach2020-kn}
Marius Mosbach, Maksym Andriushchenko, and Dietrich Klakow. 2020.
\newblock On the stability of fine-tuning {BERT:} misconceptions, explanations,
  and strong baselines.
\newblock \emph{CoRR}, abs/2006.04884.

\bibitem[{Nogueira and Cho(2019)}]{nogueira2019passage}
Rodrigo Nogueira and Kyunghyun Cho. 2019.
\newblock Passage re-ranking with {BERT}.
\newblock \emph{CoRR}, abs/1901.04085.

\bibitem[{Nogueira and Lin(2019)}]{Nogueira2019FromDT}
Rodrigo Nogueira and Jimmy Lin. 2019.
\newblock From doc2query to {docTTTTTquery}.
\newblock \emph{MS MARCO passage retrieval task publication}.

\bibitem[{Nogueira et~al.(2019)Nogueira, Yang, Lin, and
  Cho}]{nogueira2019document}
Rodrigo Nogueira, Wei Yang, Jimmy Lin, and Kyunghyun Cho. 2019.
\newblock Document expansion by query prediction.
\newblock \emph{CoRR}, abs/1904.08375.

\bibitem[{OpenAI(2023)}]{openaiGPT2023}
OpenAI. 2023.
\newblock \href {https://doi.org/10.48550/ARXIV.2303.08774} {{GPT-4} technical
  report}.
\newblock \emph{CoRR}, abs/2303.08774.

\bibitem[{Ouyang et~al.(2022)Ouyang, Wu, Jiang, Almeida, Wainwright, Mishkin,
  Zhang, Agarwal, Slama, Ray, Schulman, Hilton, Kelton, Miller, Simens, Askell,
  Welinder, Christiano, Leike, and Lowe}]{InstructGPT2022}
Long Ouyang, Jeffrey Wu, Xu~Jiang, Diogo Almeida, Carroll~L. Wainwright, Pamela
  Mishkin, Chong Zhang, Sandhini Agarwal, Katarina Slama, Alex Ray, John
  Schulman, Jacob Hilton, Fraser Kelton, Luke Miller, Maddie Simens, Amanda
  Askell, Peter Welinder, Paul~F. Christiano, Jan Leike, and Ryan Lowe. 2022.
\newblock Training language models to follow instructions with human feedback.
\newblock In \emph{NeurIPS}.

\bibitem[{Paszke et~al.(2019)Paszke, Gross, Massa, Lerer, Bradbury, Chanan,
  Killeen, Lin, Gimelshein, Antiga et~al.}]{paszke2019pytorch}
Adam Paszke, Sam Gross, Francisco Massa, Adam Lerer, James Bradbury, Gregory
  Chanan, Trevor Killeen, Zeming Lin, Natalia Gimelshein, Luca Antiga, et~al.
  2019.
\newblock Pytorch: An imperative style, high-performance deep learning library.
\newblock In \emph{Advances in neural information processing systems}, pages
  8026--8037.

\bibitem[{Penha et~al.(2022)Penha, C{\^{a}}mara, and
  Hauff}]{DBLP:conf/ecir/PenhaCH22}
Gustavo Penha, Arthur C{\^{a}}mara, and Claudia Hauff. 2022.
\newblock Evaluating the robustness of retrieval pipelines with query variation
  generators.
\newblock In \emph{{ECIR} {(1)}}, volume 13185 of \emph{Lecture Notes in
  Computer Science}, pages 397--412. Springer.

\bibitem[{Peters et~al.(2018)Peters, Neumann, Iyyer, Gardner, Clark, Lee, and
  Zettlemoyer}]{peters2018deep}
Matthew~E Peters, Mark Neumann, Mohit Iyyer, Matt Gardner, Christopher Clark,
  Kenton Lee, and Luke Zettlemoyer. 2018.
\newblock Deep contextualized word representations.
\newblock In \emph{Proceedings of NAACL-HLT}, pages 2227--2237.

\bibitem[{Portes et~al.(2023)Portes, Trott, Havens, KING, Venigalla, Nadeem,
  Sardana, Khudia, and Frankle}]{portes2023mosaicbert}
Jacob Portes, Alexander~R Trott, Sam Havens, DANIEL KING, Abhinav Venigalla,
  Moin Nadeem, Nikhil Sardana, Daya Khudia, and Jonathan Frankle. 2023.
\newblock \href {https://openreview.net/forum?id=5zipcfLC2Z} {Mosaic{BERT}: A
  bidirectional encoder optimized for fast pretraining}.
\newblock In \emph{Thirty-seventh Conference on Neural Information Processing
  Systems}.

\bibitem[{Press et~al.(2022)Press, Smith, and Lewis}]{press2022train}
Ofir Press, Noah Smith, and Mike Lewis. 2022.
\newblock \href {https://openreview.net/forum?id=R8sQPpGCv0} {Train short, test
  long: Attention with linear biases enables input length extrapolation}.
\newblock In \emph{International Conference on Learning Representations}.

\bibitem[{Qu et~al.(2021)Qu, Ding, Liu, Liu, Ren, Zhao, Dong, Wu, and
  Wang}]{RocketQA2021}
Yingqi Qu, Yuchen Ding, Jing Liu, Kai Liu, Ruiyang Ren, Wayne~Xin Zhao, Daxiang
  Dong, Hua Wu, and Haifeng Wang. 2021.
\newblock Rocketqa: An optimized training approach to dense passage retrieval
  for open-domain question answering.
\newblock In \emph{{NAACL-HLT}}, pages 5835--5847. Association for
  Computational Linguistics.

\bibitem[{Rau et~al.(2024)Rau, Dehghani, and
  Kamps}]{DBLP:journals/tois/RauDK24}
David Rau, Mostafa Dehghani, and Jaap Kamps. 2024.
\newblock Revisiting bag of words document representations for efficient
  ranking with transformers.
\newblock \emph{{ACM} Trans. Inf. Syst.}, 42(5):114:1--114:27.

\bibitem[{Robertson(2004)}]{Robertson2004}
Stephen Robertson. 2004.
\newblock Understanding inverse document frequency: on theoretical arguments
  for {IDF}.
\newblock \emph{Journal of Documentation}, 60(5):503--520.

\bibitem[{Ruch et~al.(2006)Ruch, Tbahriti, Gobeill, and
  Aronson}]{DBLP:conf/acl/RuchTGA06}
Patrick Ruch, Imad Tbahriti, Julien Gobeill, and Alan~R. Aronson. 2006.
\newblock Argumentative feedback: {A} linguistically-motivated term expansion
  for information retrieval.
\newblock In \emph{{ACL}}. The Association for Computer Linguistics.

\bibitem[{Rush(2018)}]{rush2018annotated}
Alexander~M Rush. 2018.
\newblock The annotated transformer.
\newblock In \emph{Proceedings of workshop for NLP open source software
  (NLP-OSS)}, pages 52--60.

\bibitem[{Saad{-}Falcon et~al.(2024)Saad{-}Falcon, Fu, Arora, Guha, and
  R{\'{e}}}]{LoCo2024}
Jon Saad{-}Falcon, Daniel~Y. Fu, Simran Arora, Neel Guha, and Christopher
  R{\'{e}}. 2024.
\newblock Benchmarking and building long-context retrieval models with loco and
  {M2-BERT}.
\newblock \emph{CoRR}, abs/2402.07440.

\bibitem[{Su et~al.(2024)Su, Ahmed, Lu, Pan, Bo, and Liu}]{RoFORMER}
Jianlin Su, Murtadha H.~M. Ahmed, Yu~Lu, Shengfeng Pan, Wen Bo, and Yunfeng
  Liu. 2024.
\newblock Roformer: Enhanced transformer with rotary position embedding.
\newblock \emph{Neurocomputing}, 568:127063.

\bibitem[{Su et~al.(2023)Su, Lu, Pan, Murtadha, Wen, and Liu}]{su2023roformer}
Jianlin Su, Yu~Lu, Shengfeng Pan, Ahmed Murtadha, Bo~Wen, and Yunfeng Liu.
  2023.
\newblock \href {http://arxiv.org/abs/2104.09864} {Roformer: Enhanced
  transformer with rotary position embedding}.

\bibitem[{Sun et~al.(2023)Sun, Yan, Ma, Wang, Ren, Chen, Yin, and
  Ren}]{RankGPT2023}
Weiwei Sun, Lingyong Yan, Xinyu Ma, Shuaiqiang Wang, Pengjie Ren, Zhumin Chen,
  Dawei Yin, and Zhaochun Ren. 2023.
\newblock Is chatgpt good at search? investigating large language models as
  re-ranking agents.
\newblock In \emph{{EMNLP}}, pages 14918--14937. Association for Computational
  Linguistics.

\bibitem[{Tay et~al.(2020)Tay, Dehghani, Bahri, and Metzler}]{Tay2020}
Yi~Tay, Mostafa Dehghani, Dara Bahri, and Donald Metzler. 2020.
\newblock Efficient transformers: {A} survey.
\newblock \emph{CoRR}, abs/2009.06732.

\bibitem[{Thakur et~al.(2021)Thakur, Reimers, R{\"{u}}ckl{\'{e}}, Srivastava,
  and Gurevych}]{BEIR}
Nandan Thakur, Nils Reimers, Andreas R{\"{u}}ckl{\'{e}}, Abhishek Srivastava,
  and Iryna Gurevych. 2021.
\newblock {BEIR:} {A} heterogeneous benchmark for zero-shot evaluation of
  information retrieval models.
\newblock In \emph{NeurIPS Datasets and Benchmarks}.

\bibitem[{Upadhyay et~al.(2024)Upadhyay, Pradeep, Thakur, Craswell, and
  Lin}]{DBLP:journals/corr/abs-2406-06519}
Shivani Upadhyay, Ronak Pradeep, Nandan Thakur, Nick Craswell, and Jimmy Lin.
  2024.
\newblock {UMBRELA:} umbrela is the (open-source reproduction of the) bing
  relevance assessor.
\newblock \emph{CoRR}, abs/2406.06519.

\bibitem[{Vaswani et~al.(2017)Vaswani, Shazeer, Parmar, Uszkoreit, Jones,
  Gomez, Kaiser, and Polosukhin}]{vaswani2017attention}
Ashish Vaswani, Noam Shazeer, Niki Parmar, Jakob Uszkoreit, Llion Jones,
  Aidan~N Gomez, {\L}ukasz Kaiser, and Illia Polosukhin. 2017.
\newblock Attention is all you need.
\newblock In \emph{{NIPS}}, pages 5998--6008.

\bibitem[{Voorhees(2004)}]{Robust04}
Ellen Voorhees. 2004.
\newblock Overview of the trec 2004 robust retrieval track.
\newblock In \emph{TREC}.

\bibitem[{Voorhees et~al.(2021)Voorhees, Alam, Bedrick, Demner-Fushman, Hersh,
  Lo, Roberts, Soboroff, and Wang}]{treccovid}
Ellen Voorhees, Tasmeer Alam, Steven Bedrick, Dina Demner-Fushman, William~R
  Hersh, Kyle Lo, Kirk Roberts, Ian Soboroff, and Lucy~Lu Wang. 2021.
\newblock Trec-covid: constructing a pandemic information retrieval test
  collection.
\newblock In \emph{ACM SIGIR Forum}, volume~54, pages 1--12. ACM New York, NY,
  USA.

\bibitem[{Wadden et~al.(2020)Wadden, Lin, Lo, Wang, van Zuylen, Cohan, and
  Hajishirzi}]{scifact}
David Wadden, Shanchuan Lin, Kyle Lo, Lucy~Lu Wang, Madeleine van Zuylen, Arman
  Cohan, and Hannaneh Hajishirzi. 2020.
\newblock Fact or fiction: Verifying scientific claims.
\newblock In \emph{Proceedings of the 2020 Conference on Empirical Methods in
  Natural Language Processing (EMNLP)}, pages 7534--7550.

\bibitem[{Wei et~al.(2022)Wei, Bosma, Zhao, Guu, Yu, Lester, Du, Dai, and
  Le}]{FLAN22}
Jason Wei, Maarten Bosma, Vincent~Y. Zhao, Kelvin Guu, Adams~Wei Yu, Brian
  Lester, Nan Du, Andrew~M. Dai, and Quoc~V. Le. 2022.
\newblock Finetuned language models are zero-shot learners.
\newblock In \emph{{ICLR}}. OpenReview.net.

\bibitem[{Wolf et~al.(2019)Wolf, Debut, Sanh, Chaumond, Delangue, Moi, Cistac,
  Rault, Louf, Funtowicz, Davison, Shleifer, von Platen, Ma, Jernite, Plu, Xu,
  Scao, Gugger, Drame, Lhoest, and Rush}]{Wolf2019HuggingFacesTS}
Thomas Wolf, Lysandre Debut, Victor Sanh, Julien Chaumond, Clement Delangue,
  Anthony Moi, Pierric Cistac, Tim Rault, Rémi Louf, Morgan Funtowicz, Joe
  Davison, Sam Shleifer, Patrick von Platen, Clara Ma, Yacine Jernite, Julien
  Plu, Canwen Xu, Teven~Le Scao, Sylvain Gugger, Mariama Drame, Quentin Lhoest,
  and Alexander~M. Rush. 2019.
\newblock Huggingface's transformers: State-of-the-art natural language
  processing.
\newblock \emph{ArXiv}, abs/1910.03771.

\bibitem[{Wu et~al.(2016)Wu, Schuster, Chen, Le, Norouzi, Macherey, Krikun,
  Cao, Gao, Macherey, Klingner, Shah, Johnson, Liu, Kaiser, Gouws, Kato, Kudo,
  Kazawa, Stevens, Kurian, Patil, Wang, Young, Smith, Riesa, Rudnick, Vinyals,
  Corrado, Hughes, and Dean}]{WuSCLNMKCGMKSJL16}
Yonghui Wu, Mike Schuster, Zhifeng Chen, Quoc~V. Le, Mohammad Norouzi, Wolfgang
  Macherey, Maxim Krikun, Yuan Cao, Qin Gao, Klaus Macherey, Jeff Klingner,
  Apurva Shah, Melvin Johnson, Xiaobing Liu, Lukasz Kaiser, Stephan Gouws,
  Yoshikiyo Kato, Taku Kudo, Hideto Kazawa, Keith Stevens, George Kurian,
  Nishant Patil, Wei Wang, Cliff Young, Jason Smith, Jason Riesa, Alex Rudnick,
  Oriol Vinyals, Greg Corrado, Macduff Hughes, and Jeffrey Dean. 2016.
\newblock Google's neural machine translation system: Bridging the gap between
  human and machine translation.
\newblock \emph{CoRR}, abs/1609.08144.

\bibitem[{Xiong et~al.(2017)Xiong, Dai, Callan, Liu, and Power}]{XiongDCLP17}
Chenyan Xiong, Zhuyun Dai, Jamie Callan, Zhiyuan Liu, and Russell Power. 2017.
\newblock End-to-end neural ad-hoc ranking with kernel pooling.
\newblock In \emph{{SIGIR}}, pages 55--64. {ACM}.

\bibitem[{Xiong et~al.(2021)Xiong, Xiong, Li, Tang, Liu, Bennett, Ahmed, and
  Overwijk}]{ANCE2021}
Lee Xiong, Chenyan Xiong, Ye~Li, Kwok{-}Fung Tang, Jialin Liu, Paul~N. Bennett,
  Junaid Ahmed, and Arnold Overwijk. 2021.
\newblock Approximate nearest neighbor negative contrastive learning for dense
  text retrieval.
\newblock In \emph{{ICLR}}. OpenReview.net.

\bibitem[{Xu(2024)}]{xu2024rankmamba}
Zhichao Xu. 2024.
\newblock Rankmamba: Benchmarking mamba's document ranking performance in the
  era of transformers.
\newblock \emph{arXiv preprint arXiv:2403.18276}.

\bibitem[{Yilmaz et~al.(2019)Yilmaz, Wang, Yang, Zhang, and Lin}]{YilmazWYZL19}
Zeynep~Akkalyoncu Yilmaz, Shengjin Wang, Wei Yang, Haotian Zhang, and Jimmy
  Lin. 2019.
\newblock Applying {BERT} to document retrieval with birch.
\newblock In \emph{{EMNLP/IJCNLP} {(3)}}, pages 19--24. Association for
  Computational Linguistics.

\bibitem[{Zaheer et~al.(2020)Zaheer, Guruganesh, Dubey, Ainslie, Alberti,
  Onta{\~{n}}{\'{o}}n, Pham, Ravula, Wang, Yang, and Ahmed}]{BigBird2020}
Manzil Zaheer, Guru Guruganesh, Kumar~Avinava Dubey, Joshua Ainslie, Chris
  Alberti, Santiago Onta{\~{n}}{\'{o}}n, Philip Pham, Anirudh Ravula, Qifan
  Wang, Li~Yang, and Amr Ahmed. 2020.
\newblock Big bird: Transformers for longer sequences.
\newblock In \emph{NeurIPS}.

\bibitem[{Zerveas et~al.(2021)Zerveas, Rekabsaz, Cohen, and
  Eickhoff}]{Zerveas2021CODERAE}
George Zerveas, Navid Rekabsaz, Daniel Cohen, and Carsten Eickhoff. 2021.
\newblock Coder: An efficient framework for improving retrieval through
  contextualized document embedding reranking.
\newblock \emph{ArXiv}, abs/2112.08766.

\bibitem[{Zhang et~al.(2024)Zhang, Zeng, Wang, and Lu}]{TinyLLAMA2024}
Peiyuan Zhang, Guangtao Zeng, Tianduo Wang, and Wei Lu. 2024.
\newblock Tinyllama: An open-source small language model.
\newblock \emph{CoRR}, abs/2401.02385.

\bibitem[{Zhang et~al.(2021)Zhang, Yates, and Lin}]{DBLP:conf/ecir/ZhangYL21}
Xinyu Zhang, Andrew Yates, and Jimmy Lin. 2021.
\newblock Comparing score aggregation approaches for document retrieval with
  pretrained transformers.
\newblock In \emph{{ECIR} {(2)}}, volume 12657 of \emph{Lecture Notes in
  Computer Science}, pages 150--163. Springer.

\bibitem[{Zhu et~al.(2024)Zhu, Wang, Yang, Song, Wu, Wei, and Li}]{E52024}
Dawei Zhu, Liang Wang, Nan Yang, Yifan Song, Wenhao Wu, Furu Wei, and Sujian
  Li. 2024.
\newblock Longembed: Extending embedding models for long context retrieval.
\newblock pages 802--816.

\bibitem[{Zou et~al.(2021)Zou, Zhang, Cai, Ma, Cheng, Wang, Shi, Cheng, and
  Yin}]{ZouZCMCWSCY21}
Lixin Zou, Shengqiang Zhang, Hengyi Cai, Dehong Ma, Suqi Cheng, Shuaiqiang
  Wang, Daiting Shi, Zhicong Cheng, and Dawei Yin. 2021.
\newblock Pre-trained language model based ranking in baidu search.
\newblock In \emph{{KDD}}, pages 4014--4022. {ACM}.

\end{thebibliography}
\end{document}